\documentclass[reprint,aps,pre,amsmath,amssymb]{revtex4-2}

\usepackage[T1]{fontenc}
\usepackage{lmodern}
\usepackage{microtype}
\usepackage{amsmath,amssymb,bm}
\usepackage{xcolor}
\usepackage{booktabs}
\usepackage{tikz}
\usetikzlibrary{arrows.meta}
\definecolor{oiblue}{HTML}{0072B2}
\definecolor{oiverm}{HTML}{D55E00}
\usepackage{graphicx}
\newcommand{\dd}{\mathrm{d}}

\usepackage{hyperref}
\usepackage{orcidlink}

\begin{document}

\title{Classical heat exchanger fouling-resistance laws as first-moment identities of additive stochastic deposition}

\author{Simone Casolo\,\orcidlink{0000-0003-2945-8931}}
\affiliation{Cognite, J. Standrups vei 10, 1344 Lysaker, Norway}

\date{\today}

\begin{abstract}
Heat-exchanger fouling is monitored through the fouling resistance, whose
record is routinely fitted with the classical linear (Ebert--Panchal) or asymptotic
(Kern--Seaton) law, and the shape of the fit is often taken as evidence of the
underlying deposition or removal mechanism. We show that it cannot be. If the
deposit grows by non-negative, statistically stationary increments, the mean
fouling resistance is exactly linear without removal and exactly the
Kern--Seaton asymptote under first-order removal, whatever the microscopic
deposition law. Shear-driven erosion, per-element detachment, intermittent
sloughing and saturation of the surface all give the same mean curve, so a fit
identifies a net deposition rate and a relaxation time, and nothing else.
Modelling the deposit as a L\'evy subordinator, we show where the missing
information lies and what it costs to retrieve it from monitoring data.
Removal mechanisms separate in the variance: sloughing leaves a plateau
scatter that persists even for smooth deposition and can be read from a
record that saturates. Deposition laws separate only in higher cumulants,
which the measurement noise of a thermally reconstructed resistance
attenuates; the record needed to distinguish them grows as the fourth power
of the conductance error, which places it beyond reach at realistic
measurement accuracy. For the time to reach a cleaning limit, the deposition
law matters only near the threshold; farther away, the mean rate and its
uncertainty dominate the prediction.
\end{abstract}

\maketitle
\begin{table*}[tb]
\noindent\textbf{Nomenclature}\par\vspace{3pt}
\footnotesize
\begin{tabular*}{\textwidth}{@{\extracolsep{\fill}}l p{0.34\textwidth} l l p{0.34\textwidth}}
\toprule
\multicolumn{2}{l}{\emph{Latin symbols}} & & \multicolumn{2}{l}{\emph{Greek symbols}}\\
$A$ & heat-transfer area, m$^2$ & & $\alpha$ & shape exponent of the event spectrum, Eq.~\eqref{eq:tempered_stable}\\
$B$ & amplitude of the event spectrum, Eq.~\eqref{eq:tempered_stable} & & $\gamma_1,\gamma_2$ & standardized skewness and excess kurtosis of an increment\\
$c$ & coefficient of variation of the deposit over one sampling period, Eq.~\eqref{eq:cv_reference_interval} & & $\gamma_{n,k}$ & the same for measured increments over $k$ periods\\
$c(\Delta)$ & the same over a window $\Delta$ & & $\Delta$ & window length, s\\
$c_p$ & specific heat, J\,kg$^{-1}$\,K$^{-1}$ & & $\Delta T,\ \Delta T_{\mathrm{lm}}$ & temperature change of a stream, log-mean temperature difference, K\\
$D(t)$ & cumulative deposition, m$^2$\,K\,W$^{-1}$ & & $\epsilon$ & relative random error of the inferred conductance\\
$d'$ & separability index & & $\varepsilon_j$ & measurement error of reading $j$, m$^2$\,K\,W$^{-1}$\\
$d_p$ & particle diameter, m & & $\zeta$ & relative conductance change per sampling period\\
$E_a$ & activation energy, J\,mol$^{-1}$ & & $\eta$ & fraction of the deposit removed per sloughing event\\
$g(R)$ & removal law, m$^2$\,K\,W$^{-1}$\,s$^{-1}$ & & $\theta$ & fractional surface coverage\\
$J_1$ & resistance added by one particle, m$^2$\,K\,W$^{-1}$ & & $\kappa_n$ & $n$th cumulant\\
$k$ & window length in sampling periods & & $\lambda$ & tempering rate of the event spectrum, m$^{-2}$\,K$^{-1}$\,W\\
$k_d,\ k_r$ & deposition and removal constants & & $\Lambda(t)$ & operational clock, Eq.~\eqref{eq:operational_clock}\\
$k_f$ & deposit thermal conductivity, W\,m$^{-1}$\,K$^{-1}$ & & $\mu$ & mean fouling resistance, m$^2$\,K\,W$^{-1}$\\
$\ell$ & size-biased event size, Eq.~\eqref{eq:ell}, m$^2$\,K\,W$^{-1}$ & & $\nu(\mathrm dx)$ & L\'evy measure (event spectrum), s$^{-1}$\\
$m$ & mean deposition rate, m$^2$\,K\,W$^{-1}$\,s$^{-1}$ & & $\rho$ & noise-to-signal ratio, Eq.~\eqref{eq:rho}\\
$\dot m$ & mass flow rate, kg\,s$^{-1}$ & & $\sigma$ & standard deviation of the measurement error, m$^2$\,K\,W$^{-1}$\\
$N$ & record length in sampling periods & & $\tau$ & time lag, s\\
$N_{\mathrm{eff}},\ N_\infty$ & size-biased events in a window, on the plateau & & $\tau_0$ & sampling period, s\\
$n_w$ & number of independent windows & & $\tau_w$ & wall shear stress, Pa\\
$n_x$ & distance to the cleaning limit in mean increments, Eq.~\eqref{eq:threshold_distance} & & $\phi_s$ & solid fraction of the deposit\\
$Q$ & heat duty, W & & $\chi$ & relative deposition intensity\\
$q_n$ & $n$th cumulant rate, Eq.~\eqref{eq:cumulant_rates} & & $\psi$ & removal rate constant, s$^{-1}$\\
$R$ & fouling resistance, m$^2$\,K\,W$^{-1}$ & & & \\
$R_0,\ R_\infty$ & initial and plateau resistance, m$^2$\,K\,W$^{-1}$ & & \multicolumn{2}{l}{\emph{Subscripts and abbreviations}}\\
$R_g$ & universal gas constant, J\,mol$^{-1}$\,K$^{-1}$ & & clean & clean exchanger\\
$r$ & activity (total event rate), s$^{-1}$ & & $\infty$ & plateau or stationary value\\
$r_s$ & sloughing rate, s$^{-1}$ & & $k$ & increment over $k$ sampling periods\\
$s^2$ & variance rate, m$^4$\,K$^2$\,W$^{-2}$\,s$^{-1}$ & & CV & coefficient of variation\\
$T_f$ & film temperature, K & & SNR & signal-to-noise ratio\\
$T_x$ & time to reach the cleaning limit $x$, s & & & \\
$t,\ t_c,\ t_i$ & time; $t_c=s^2/m^2$; induction time, s & & & \\
$U$ & overall heat-transfer coefficient, W\,m$^{-2}$\,K$^{-1}$ & & & \\
$u$ & flow velocity, m\,s$^{-1}$ & & & \\
$V$ & variance of $R$, m$^4$\,K$^2$\,W$^{-2}$ & & & \\
$w(x)$ & mass-weighted event spectrum, Eq.~\eqref{eq:mass_spectrum} & & & \\
$x$ & cleaning limit, m$^2$\,K\,W$^{-1}$ & & & \\
$Y_j$ & measured resistance at reading $j$, m$^2$\,K\,W$^{-1}$ & & & \\
\bottomrule
\end{tabular*}
\end{table*}

\section{Introduction}
\label{sec:introduction}

\begin{figure}[tb]
\centering
\resizebox{0.75\columnwidth}{!}{%
\begin{tikzpicture}[scale=1.0, every node/.style={font=\large}]
  \draw[-{Latex[length=2mm]}] (0,0) -- (7.6,0) node[right] {$t$};
  \draw[-{Latex[length=2mm]}] (0,0) -- (0,5.2) node[above] {$R$};

  \def\Rinf{4.2}
  \def\tauKS{1.6}
  \def\ti{0.8}
  \draw[thick, oiblue] (0,0) -- (\ti,0);
  \draw[thick, oiblue, domain=\ti:7.4, samples=120, smooth]
    plot (\x, {\Rinf*(1-exp(-(\x-\ti)/\tauKS))});
  \draw[dashed, oiblue!50] (0,\Rinf) -- (7.6,\Rinf);
  \node[oiblue, anchor=west] at (7.5,\Rinf+0.15) {$R_\infty$};

  \def\REP{0.15}         
  \def\slopeEP{0.62}
  \draw[thick, oiverm] (0,\REP) -- (\ti,\REP);
  \draw[thick, oiverm, domain=\ti:7.4, samples=2]
    plot (\x, {\REP + \slopeEP*(\x-\ti)});
  \draw[densely dotted, gray] (\ti,0) -- (\ti,\REP+0.05);
  \node[gray, below] at (\ti,0) {$t_i$};

  \draw[thick, oiblue] (4.3,1.1) -- (4.9,1.1);
  \node[anchor=west] at (4.95,1.1) {Kern--Seaton (asymptotic)};
  \draw[thick, oiverm] (4.3,0.55) -- (4.9,0.55);
  \node[anchor=west] at (4.95,0.55) {Ebert--Panchal (threshold)};
\end{tikzpicture}%
}
\caption{Prototypical fouling-resistance trajectories. Both laws show
an induction period $t_i$ before the resistance grows, 
after which the Kern--Seaton law saturates to a plateau
$R_\infty$ at a rate set by the removal process, while the Ebert--Panchal
law grows approximately linearly once the deposition term dominates.}
\label{fig:models}
\end{figure}
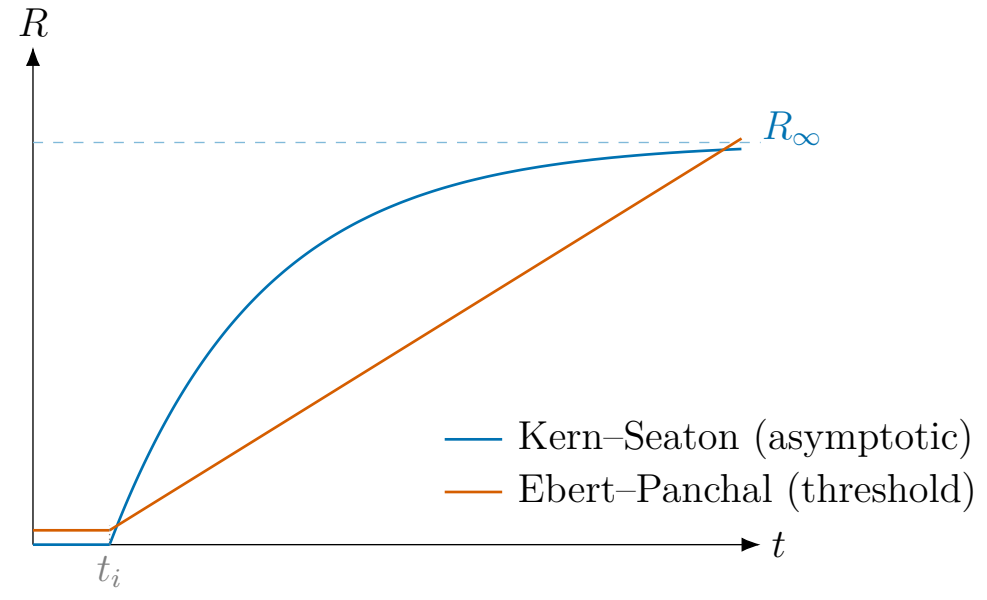

Fouling, the slow accumulation of material at the interface between a
flowing phase and a solid, is monitored in heat exchangers through the fouling
resistance. Its growth costs fuel, design margin and
production lost to cleaning~\cite{Epstein1983,Bott1995,MullerSteinhagen2011},
and the operator's question is when it will reach the 
limit at which cleaning becomes 
necessary~\cite{IshiyamaCleaning2011,Pogiatzis2012}. In operation the
resistance is not measured but reconstructed from temperatures and flow rates
through the exchanger's energy balance~\cite{Tatara2011,Shen2015}, so that
its change from one sampling period to the next is typically smaller than its
measurement error. Decades of work have
identified the contributing processes in detail, such as
particle transport and attachment, crystallization, chemical reactions and ageing~\cite{Taborek1972a,Somerscales1990,MullerSteinhagen2011,AugustinMeloWilson2025}.
Fouling records are often still read against two main models, shown in Fig.~\ref{fig:models}.
Kern and Seaton
assume deposition at a constant rate and removal proportional to the deposit,
giving an asymptotic rise to a plateau~\cite{KernSeaton1959}; Ebert and
Panchal set a thermally activated deposition rate against a shear-dependent
term, giving a threshold above which the resistance grows
linearly~\cite{EbertPanchal1997}.
Records are routinely sorted as linear, falling-rate or
asymptotic~\cite{Epstein1983,Somerscales1990,Bott1995,MullerSteinhagen2011},
while subsequent work has concentrated on what sets the \emph{rate}: particle
transport and
attachment~\cite{Yao1971,Epstein1988,Bohnet1987,Kostoglou1998,Brahim2003,Han2019},
crystallization and wax deposition~\cite{Maddahi2019,Akbarzadeh2008},
chemical-reaction fouling~\cite{WatkinsonWilson1997,EP6,EP4,EP3,EP2,EP1,Wilson24052017,Geddert2009,Mayer2013,YangCFD2020},
ageing~\cite{EP5,Ishiyama2011,IshiyamaAgeing2013,ColettiMacchietto2011,DiazBejarano2015,ColettiCrittenden2015,Diaz-Bejarano24052017}
and parameter estimation or data-driven prediction from plant
data~\cite{LoyolaFuentes2019,LoyolaFuentesSmith2019,Markowski2025,Sundar2020}.
Replicate fouling runs under identical conditions scatter
widely~\cite{Zubair1997CaCO3}, and probabilistic treatments attach this scatter
to the deterministic laws as random growth parameters, from which they derive
time-to-cleaning distributions and maintenance
policies~\cite{Zubair1992,Zubair1997,Zubair2000,Sheikh2001}; other stochastic
treatments~\cite{Vatistas1988,Flora1993,Wessling2001,Liu2020} are mostly
simulations. None derives the deterministic laws from the
stochastic deposition, or asks what the distribution of the fouling measure
allows a record to identify.

If fouling reflects many coupled processes, each with its own strong
dependence on operating conditions, why do measured trajectories repeatedly
collapse onto a small number of elementary shapes? The usual reading is that
the shape identifies the dominant mechanism. In heat
exchangers an asymptote is read as the signature of removal~\cite{Shaikh2023,Fguiri2020}, as the
Kern--Seaton derivation suggests, and the saw-tooth pattern and scatter of
plant data as evidence of it~\cite{Crittenden1992}; yet Ebert and Panchal's
threshold concerns only the initial rate, and linear growth is not evidence
that removal is absent~\cite{EbertPanchal1997,Wilson24052017}.
The mechanistic readings have been debated---asymptotic data need not imply
removal~\cite{Epstein1983}, suppression and removal are indistinguishable
during growth~\cite{Diaz-Bejarano24052017}, and likelihood comparison can
select the wrong increment law~\cite{ZhangRevie2016}---but mostly through
counterexamples and case studies, not a general statement of what a record
can identify.

We show in this study that, under very general conditions, 
the classical laws are universal first-moment consequences of additive deposition processes.
If deposition increments are stationary,
independent and non-negative, the deposit is a L\'evy subordinator and the two
classical laws are first-moment identities: the mean depends only on the net
deposition and removal rates, for every additive mechanism. The statement
follows directly from the linearity of expectation, but its consequences are
not obvious. The information absent
from the mean lies in the fluctuations and is unequally accessible: removal
mechanisms separate in a plateau variance that survives as the deposition
becomes smooth, whereas the increment law separates only in higher cumulants
that measurement noise attenuates, turning the noise-free sample size of
model selection~\cite{ZhangRevie2016} into a record length that grows as the
fourth power of the noise. The dependence of the rate on operating conditions
remains the proper object of mechanistic modelling, but it cannot be inferred
from a single record at a single operating point.
Section~\ref{sec:implications} translates these results into what an
operator can and cannot conclude from a monitored fouling resistance.

\section{Framework and deterministic reference models}
\label{sec:theory}

Fouling denotes the accumulation or growth of unwanted material on a
surface \cite{Epstein1983,Bott1995,AugustinMeloWilson2025}. We describe the
state of a fouled surface by a nonnegative fouling measure $R(t)$, which is
kept generic throughout the theory. Its principal application here is the
thermal fouling resistance of a heat exchanger,
$R=1/U-1/U_{\mathrm{clean}}$ 
where $U$ is the overall heat-transfer coefficient. 
For a homogeneous deposit of thickness $\delta$
and thermal conductivity $k_f$ this is proportional to the deposit 
thickness $R=\delta/k_f$. 

This framework describes deposition under approximately stationary operating
conditions in which the deposit does not feed back on the process that creates
it. Specifically, we assume that (i)~the variables that set the deposition rate
(flow, wall shear stress, film temperature, foulant concentration) are constant
in time and uniform over the surface, so that $R$ is an area-averaged scalar;
(ii)~the deposit does not appreciably change these variables, i.e.\ the loss of
flow cross-section and the resulting changes in wall shear stress and interface
temperature are negligible, such feedback being one route to an
asymptotic record~\cite{Rehman2023}; (iii)~$R$ is additive in the deposited material, so
that ageing, consolidation, and roughness effects that change $R$ at fixed
deposited mass are neglected \cite{Crittenden1988,Ishiyama2011}; and (iv)~removal, where present, is a
deterministic function of the state.
Removal \emph{triggered by the deposit state itself} (e.g.\ thickness-dependent
spalling) is not treated; Poisson-timed removal of a fixed fraction
is (Sec.~\ref{sec:removal_sloughing}). 
Additivity~(iii) and the absence of feedback~(ii) are what make the mean
independent of mechanism; when (i)--(iv) fail, the net rate depends on the
state and the mean laws of Table~\ref{tab:removal} no longer hold in general,
as shown for nonlinear removal in Sec.~\ref{sec:removal_nonlinear}.

\subsection{Deterministic fouling models}
\label{sec:deterministic_models}

Fouling is conventionally modeled as a balance between the deposition and
removal of material,
\begin{equation}
  \frac{dR}{dt}=\dot R_{\mathrm d}-\dot R_{\mathrm r},
  \label{eq:deterministic_mass_balance}
\end{equation}
where $\dot R_{\mathrm d}$ and $\dot R_{\mathrm r}$ are the deposition and
removal rates in units of $R$ per unit time. Depending on the relative
importance of deposition, adhesion, erosion, and detachment, observed
fouling curves may be approximately linear, asymptotic, falling-rate, or irregular 
\cite{Bott1995, AugustinMeloWilson2025}. 
Two prototypical 
deterministic models used in the industry may 
provide useful reference limits: the asymptotic
(Kern--Seaton) model \cite{KernSeaton1959} 
and the linear (Ebert--Panchal) model
\cite{EbertPanchal1997}.

\paragraph*{Kern--Seaton.}
Deposition is proportional to the fluid velocity
$u$ and to the foulant concentration $C$, 
whereas removal grows with the wall
shear stress $\tau_w$ and with the amount of material already deposited. For particulate fouling, this reads as
\begin{equation}
  \dot R_{\mathrm d}=k_d\,uC,\qquad \dot R_{\mathrm r}=k_r\,\tau_w R,
  \label{eq:kern_seaton_rates}
\end{equation}
where $k_d$ and $k_r$ are effective rate coefficients. The clean-surface
deposition rate $m=k_d\,uC$ may be interpreted as a mass-transfer-controlled
capture rate of classical filtration theory under particulate-fouling
assumptions, multiplied by an attachment efficiency
\cite{Yao1971,Epstein1988}; deposition--removal models of this form for
particulate and crystallisation fouling are developed in 
Refs.~\cite{Bohnet1987,Kostoglou1998} and have been a staple of fouling modeling 
in the industry.
The removal law
assumes that $R$ is proportional to the deposited mass, and in fully
turbulent flow the wall shear stress is approximately proportional to $u^{2}$,
so that under stationary conditions 
\begin{equation}
  \frac{dR}{dt}=m-\psi R,
\end{equation}
with $m=k_d\,uC$ and
$\psi=k_r'u^{2}$, where $k_r'$ is $k_r$ multiplied by the constant of
proportionality between $\tau_w$ and $u^{2}$. The fouling resistance then rises monotonically to the
plateau $R_\infty=m/\psi$ with time constant
$\psi^{-1}$~\cite{KernSeaton1959,MullerSteinhagen2011}, and because removal is
proportional to $R$ the state cannot become negative.

\paragraph*{Ebert--Panchal.} 
In this family of models the net fouling rate is the 
difference between
a thermally activated deposition contribution and a shear-dependent 
removal contribution \cite{Wilson24052017}. 
Following the original derivation
\begin{equation}
  \frac{dR}{dt}=\underbrace{a\,\mathrm{Re}^{b}\exp\!\Bigl(-\frac{E_a}{R_gT_f}\Bigr)}_{\displaystyle m}
  -\gamma\tau_w\;\equiv\;K_{\mathrm{EP}} .
  \label{eq:ebert_panchal_rate}
\end{equation}
Here $E_a$ is the activation energy per mole, $R_g$ the gas constant,
$T_f$ the film temperature, and $a$, $b$, $\gamma$ are empirical 
parameters.
The second term is interpreted as either a removal or suppression of the
deposition rate, and it reduces the rate of sticking, \emph{i.e.} 
the increase of $R$ even without any removal mechanism 
\cite{EP5,Diaz-Bejarano24052017}. 
The rate is accordingly clipped at zero,
$dR/dt=\max(K_{\mathrm{EP}},0)$, so that $R$ never decreases.
In the stochastic description of Sec.~\ref{sec:stochastic_formulation}
the two readings of the second term are no longer equivalent. Read as
\emph{suppression}, $\gamma\tau_w$ lowers the fraction of arriving material
that sticks; this is a thinning of the deposition process
(Sec.~\ref{sec:removal_summary}, Eq.~\eqref{eq:thinning}).
Read as \emph{erosion}, it is a constant removal rate
$g(R)=\gamma\tau_w$ acting between deposition events, so that the path
decreases during quiescent stretches and can reach $R=0$, where it must be
held. The two readings share the linear mean but not the variance, and the
erosion reading is exact only away from the wall
(Table~\ref{tab:removal}).

Several further models have been built on this basis
\cite{EP4,EP6,EP3,EP2,EP1}, and in all of them $m$ and the suppression rate
$\gamma\tau_w$ are time-independent under stationary conditions, so that $R$
grows linearly at the net rate $K_{\mathrm{EP}}$; the condition
$K_{\mathrm{EP}}>0$ defines the net-fouling regime, and below it no net
accumulation is predicted. Both models are instances of $dR/dt=m-g(R)$ with
the same deposition rate $m$ and different removal laws $g$, and both mean
trajectories are collected in Table~\ref{tab:removal}.

\begin{table*}[tb]
\caption{Deterministic fouling laws, and the
first two cumulants of the fouling state $R(t)$ when the deposition is a
subordinator with mean rate $m$ and variance rate $s^2=c^2m^2\tau_0$, starting
from $R(0)=R_0$.}
\label{tab:removal}
\begin{ruledtabular}
\begin{tabular}{lllll}
Removal law $g(R)$ & Deterministic law &  $\mathbb E[R(t)]$ &  $\mathrm{Var}[R(t)]$ & Closure \\
\hline
$0$ & pure deposition & $R_0+mt$ & $s^2\,t$ & exact \\
$\gamma\tau_w$ (constant) & Ebert--Panchal & $R_0+(m-\gamma\tau_w)\,t$ & $s^2\,t$ & see note\footnotemark[1] \\
$\psi R$ (linear) & Kern--Seaton & $R_0e^{-\psi t}+R_\infty\bigl(1-e^{-\psi t}\bigr)$ & $\dfrac{s^2}{2\psi}\bigl(1-e^{-2\psi t}\bigr)$ & exact \\
\end{tabular}
\end{ruledtabular}
\footnotetext[1]{Erosion reading, $g(R)=\gamma\tau_w$ for $R>0$; a
reflection at $R=0$ keeps $R\ge0$. The tabulated cumulants are exact while
$t<R_0/(\gamma\tau_w)$ and, in the net-fouling regime $m>\gamma\tau_w$,
acquire only bounded corrections once the wall is reached. Under the
suppression (thinning) reading, $g=0$ and $p=1-\gamma\tau_w/m$ in
Eq.~\eqref{eq:thinning}, the mean is the same and exact, and the variance is
$p\,s^2\,t$.}
\end{table*}

\section{Stochastic formulation of linear fouling models}
\label{sec:stochastic_formulation}

The deterministic models describe mean trajectories under fixed operating
conditions. They do not determine the variability among nominally identical
campaigns, the intermittency of deposition events, or the uncertainty in the
time needed to reach a prescribed fouling threshold. The same mean
trajectory may arise from frequent small deposition events or from rare,
comparatively large bursts.

To represent this unresolved variability we keep the accumulation--removal
structure of Eq.~\eqref{eq:deterministic_mass_balance} but replace the
smooth deposition increment by the increment of a non-negative stochastic
process. Let $D(t)$ be the cumulative deposition, with $D(0)=0$, and let
$g(R)$ be a deterministic, state-dependent removal rate. The system is then governed by the Langevin-type equation:
\begin{equation}
  dR(t)=dD(t)-g[R(t)]\,dt,\quad R(0)=R_0\ge0 .
  \label{eq:stochastic_balance}
\end{equation}

The deposition $dD(t)$ acts as a non-negative noise, made of the deposition
increments $D(t+\Delta)-D(t)$, i.e.\ the material accumulated in a window of
duration $\Delta$, counteracted by the deterministic, state-dependent removal
term $g[R(t)]$. Under stationary operating conditions we make three
assumptions:
\begin{enumerate}
  \item[(a)] \textit{Stationarity.} The distribution of an increment depends
    on the window length $\Delta$, but not on the time at which the window
    begins.
  \item[(b)] \textit{Independence.} Deposition increments over
    non-overlapping windows are independent.
  \item[(c)] \textit{Non-negativity.} Deposition increments are nonnegative,
    so $D(t)$ cannot decrease.
\end{enumerate}
Together with $D(0)=0$ and stochastic continuity, assumptions (a)--(c) make
$D(t)$ a L\'evy process. We take its
deterministic drift to be zero, so that all deposition occurs through events,
however small, making the process a \emph{subordinator}~\cite{Sato1999,Applebaum2009}.
Two kinds of subordinator exist, distinguished by the total mass of the
L\'evy measure, $\nu(\dd x)$, which we call the \emph{activity}:
\begin{itemize}
  \item \emph{finite activity}, $r=\int\nu(\dd x)<\infty$: a finite number of events per period, with possibly long quiet intervals in which nothing deposits (bursty, intermittent fouling);
  \item \emph{infinite activity}, $r=\int\nu(\dd x)=\infty$: infinitely many ever-smaller events in any interval, so the deposit strictly increases over every window almost surely; the sample paths remain discontinuous (pure-jump) despite this (gradual fouling).
\end{itemize}
The L\'evy measure fixes the whole process through the L\'evy--Khintchine
formula~\cite{Sato1999,Applebaum2009}; in particular, by Campbell's
theorem~\cite{Campbell1909}, the cumulants of $D(t)$ are proportional to $t$,
\begin{equation}
  \kappa_n[D(t)]=q_n\,t,\qquad q_n=\int_0^\infty x^n\,\nu(dx),
  \label{eq:cumulant_rates}
\end{equation}
with
\begin{align}
  \kappa_1[D(t)]=\mathbb E[D(t)]&=m\,t, \qquad m=q_1, \label{eq:deposition_mean}\\
  \kappa_2[D(t)]=\operatorname{Var}[D(t)]&=s^2\,t, \qquad s^2=q_2, \label{eq:deposition_variance_rate}
\end{align}
where $m$ is the mean deposition rate and $s^2$ the variance rate. Taking
the expectation of Eq.~\eqref{eq:stochastic_balance} then gives
$d\,\mathbb E[R]/dt=m-\mathbb E[g(R)]$, which for absent, constant or linear
$g$ closes into the mean laws of Table~\ref{tab:removal} whatever the L\'evy
measure. The
variance rate and the mean rate have different dimensions, so the relative
irregularity of the deposition must be referred to a time window. For a fixed
reference window $\tau_0$ we define the coefficient of variation ($c$ or CV)
of the accumulated deposition as
\begin{equation}
  c\equiv c(\tau_0),\qquad
  c(\Delta)^2=\frac{\operatorname{Var}[D(\Delta)]}{\mathbb E[D(\Delta)]^2}
  =\frac{s^2}{m^2\Delta},
  \label{eq:cv_reference_interval}
\end{equation}
so that
\begin{equation}
\begin{gathered}
  s^2=c^2m^2\tau_0,\qquad
  \operatorname{Var}[D(t)]=c^2m^2\tau_0\,t,\\[2pt]
  c(\Delta)=c\sqrt{\tau_0/\Delta}.
\end{gathered}
  \label{eq:deposition_variance}
\end{equation}
The CV therefore depends on the temporal resolution and decreases as windows
are aggregated. Operationally, $\tau_0$ is set by the record itself---the
native interval at which $R(t)$ is sampled or reported---while $\Delta=k\tau_0$
for integer $k$ is any coarser window obtained by aggregating $k$ successive
periods to suppress noise. The choice of $\tau_0$ is a convention; the physical content is
the time $t_c=s^2/m^2=c^2\tau_0$, which is independent of $\tau_0$ and
equals the mean time needed to deposit one size-biased event (Sec.~\ref{sec:activity}).
The parameter $m$ sets how fast fouling accumulates on average, whereas $c$
measures the relative irregularity of deposition over the chosen reference
window. The value $c=2$ used in most examples is illustrative, chosen
large enough to make the differences between the deposition laws visible.
Order-unity values require spatially coherent fluctuations of the deposition;
granular arrivals alone give far smaller values
(Sec.~\ref{sec:variance_test}). The deterministic models of Sec.~\ref{sec:theory} are recovered in
the limit $t_c\to0$ (equivalently $c\to0$ at fixed $m$), in which
fluctuations vanish relative to the mean and $R(t)$ reduces to the smooth
trajectories of Table~\ref{tab:removal}.
The third and fourth cumulants give the \emph{skewness} $\gamma_1=\kappa_3/\kappa_2^{3/2}$ and the
\emph{excess kurtosis} $\gamma_2=\kappa_4/\kappa_2^2$,
all referring to the deposit in one sampling period.
Since $\kappa_n[D(\Delta)]=q_n\Delta$, the standardized cumulant of order $n$
decays as $\Delta^{1-n/2}$ under aggregation, with an exponent that does not
depend on the L\'evy measure.


\subsection{Event spectrum and effective number of events}
\label{sec:activity}
 
Under assumptions (a)--(c) the deposition is a superposition of effective
events, each adding a positive amount of size $x$ to $R$ 
(the attachment of a particle of foulant), which need not
be resolved individually. 
Events of each size occur as independent Poisson
processes with rate density $\nu(x)$, so that a
size class $x$ contributes $x\,\nu(x)\,dx$ to the mean 
deposition rate and
$x^2\nu(x)\,dx$ to its variance rate. From 
Eq.~\eqref{eq:cumulant_rates} then follows that the fluctuations
(higher order cumulants) are dominated by the larger events.
 
The activity (Sec.~\ref{sec:stochastic_formulation})
separates two regimes. For finite $r$ the process is a compound
Poisson process (Supplemental Material,
Sec.~S2).
Deposition occurs at random times and with random magnitude.
It is bursty, with quiescent
periods that are in principle observable.
Only finitely many events occur in
a finite interval, and the increment $\Delta D=D(t+\Delta)-D(t)$ vanishes with
probability $\Pr(\Delta D=0)=e^{-r\Delta}$.
For $r=\infty$ instead, infinitely many
arbitrarily small events occur in every interval and
$\Delta D>0$ almost surely
for every $\Delta>0$. Deposition is then gradual, but it can still be intermittent,
because most of the deposited mass may be carried by a few large events. Since
$D$ is nondecreasing, its paths have finite variation in both cases, and the
model contains no Gaussian component.
The distribution of the deposited mass over event sizes is described by the
mass spectrum $w(x)=x\,\nu(x)/m$, the fraction of the deposit carried by events
of size $x$. Its mean,
\begin{equation}
  \ell=\int_0^\infty x\,w(x)\,dx=\frac{q_2}{q_1}=\frac{s^2}{m}=m\,t_c ,
  \label{eq:ell}
\end{equation}
is the size-biased event size, i.e.\ the mean size of the event that deposits a
given unit of $R$ (denoted $\ell$ to distinguish it from the process $D(t)$).
It has the dimension of $R$, does not depend on $\Delta$, and differs from the
mean event size $\int x\,\nu\,dx\big/\!\int\nu\,dx$, which is dominated by the
numerous small events. The squared coefficient of variation over a window $\Delta$
is then
\begin{equation}
  c(\Delta)^2=\frac{\ell}{m\Delta}=\frac{1}{N_{\mathrm{eff}}(\Delta)},\qquad
  N_{\mathrm{eff}}(\Delta)=\frac{m\Delta}{\ell},
  \label{eq:neff}
\end{equation}
where $N_{\mathrm{eff}}$ is the number of size-biased events contained in the
mean deposit $m\Delta$; equivalently $N_{\mathrm{eff}}=\Delta/t_c$, the time
$t_c$ named in Sec.~\ref{sec:stochastic_formulation} and now given
explicitly as $t_c=\ell/m$. The irregularity of the deposition is thus the inverse
of the effective number of events.
For $N_{\mathrm{eff}}\lesssim1$ it is lumpy, with quiescent periods if
$r$ is finite and with intermittent gradual growth if $r$ is infinite and a few large
events carry most of the mass.
Instead, for
$N_{\mathrm{eff}}\gg1$ the deposit in a window is Gaussian irrespective of the
activity, as a consequence of the central limit theorem: provided the
cumulant rate $q_2$ is finite, 
the standardized higher cumulants $\gamma_i$
tend to zero as $N_{\mathrm{eff}}$ increases.

\subsection{Tempered-stable family of event spectra}
\label{sec:tempered}

A pure-jump deposition process is specified by the spectrum of its events. 
We use a tempered-stable L\'evy subordinator form
\begin{equation}
  \nu(dx)=B\,x^{-1-\alpha}e^{-\lambda x}\,dx,\qquad x>0,\quad\alpha<1 .
    \label{eq:tempered_stable}
\end{equation}
The power-law factor controls the relative occurrence of small and large
events, and the exponential factor imposes a maximum event size of order
$\lambda^{-1}$, which makes all moments finite (see Ref.~\cite{Rosinski2007}
for the general theory of tempering). The cutoff may represent, for example, a
maximum cluster or flake size that can form or attach 
and in the following $\alpha$ is used as a statistical shape
parameter. The condition $\alpha<1$ ensures a finite mean deposition rate.

Given $m$ and $c$, the two scale parameters $B$ and $\lambda$ are fixed for
any $\alpha$; in particular the cutoff is $\lambda=(1-\alpha)/\ell$
(Supplemental Material, Sec.~S3). The family therefore has the same mean and
variance rates for all $\alpha$, and the standardized skewness and excess
kurtosis over a window $\Delta$ are
\begin{equation}
\begin{aligned}
  \gamma_1(\Delta)&=\frac{2-\alpha}{1-\alpha}\,c(\Delta), \\
  \gamma_2(\Delta)&=\frac{(2-\alpha)(3-\alpha)}{(1-\alpha)^2}\,c(\Delta)^2, \\
  c(\Delta)&=c\sqrt{\tau_0/\Delta}.
\end{aligned}
\label{eq:aggregation_scaling}
\end{equation}
The prefactor $(2-\alpha)/(1-\alpha)$ is the variance power of the
Tweedie exponential-dispersion family, to which these laws belong
\cite{Jorgensen1987,Hougaard1986}.

The parameter $\alpha$ has a direct physical reading. The mass-weighted
spectrum of the family $w(x)$
is a Gamma density of shape $1-\alpha$ and rate
$\lambda$,
\begin{equation}
  w(x)=\frac{\lambda^{1-\alpha}}{\Gamma(1-\alpha)}\,x^{-\alpha}e^{-\lambda x},
  \label{eq:mass_spectrum}
\end{equation}
with mean $\ell$ for every $\alpha$ and relative width $(1-\alpha)^{-1/2}$. At
fixed $m$ and $c$, $\alpha$ therefore controls how the deposit is distributed
over event sizes. For $\alpha\to-\infty$ all events have the same size
$\ell$, and deposition is a Poisson process of equal events; as $\alpha$
increases, the deposit is carried by an increasingly broad range of sizes,
with more of it in rare large events and, at the same time, in very small
ones. Since the mean and the variance are unchanged, this broadening shows up
only from the third cumulant onwards, and never lowers the
skewness below $c(\Delta)$, a bound that holds for every subordinator
(Supplemental Material, Sec.~S3).

\begin{table*}[tb]
\caption{The three deposition laws as members of the tempered-stable family,
Eq.~\eqref{eq:tempered_stable}, at common $m$, $c$ and $\tau_0$; $\ell=c^2m\tau_0$.
Skewness and excess kurtosis over a window $\Delta$ follow from
Eq.~\eqref{eq:aggregation_scaling}: $\gamma_1=(\gamma_1/c)\,c(\Delta)$ and
$\gamma_2=(\gamma_2/c^2)\,c(\Delta)^2$.}
\label{tab:three_laws}
\begin{ruledtabular}
\begin{tabular}{lllp{4.2cm}ccc}
$\alpha$ & Law & Activity & Event picture & $\Pr(\Delta D=0)$ & $\gamma_1/c$ & $\gamma_2/c^2$ \\
\hline
$-1$ & compound Poisson & finite & discrete events, exponential sizes of mean $\ell/2$ & $e^{-2\Delta/t_c}$ & $3/2$ & $3$ \\
$0$ & Gamma & infinite & dense small events, exponential mass spectrum & $0$ & $2$ & $6$ \\
$1/2$ & inverse Gaussian & infinite & dense small events, broader mass spectrum & $0$ & $3$ & $15$ \\
\end{tabular}
\end{ruledtabular}
\end{table*}

\subsection{The three deposition laws}
\label{sec:three_laws}

\begin{figure*}[t]
\centering
\includegraphics[width=\textwidth]{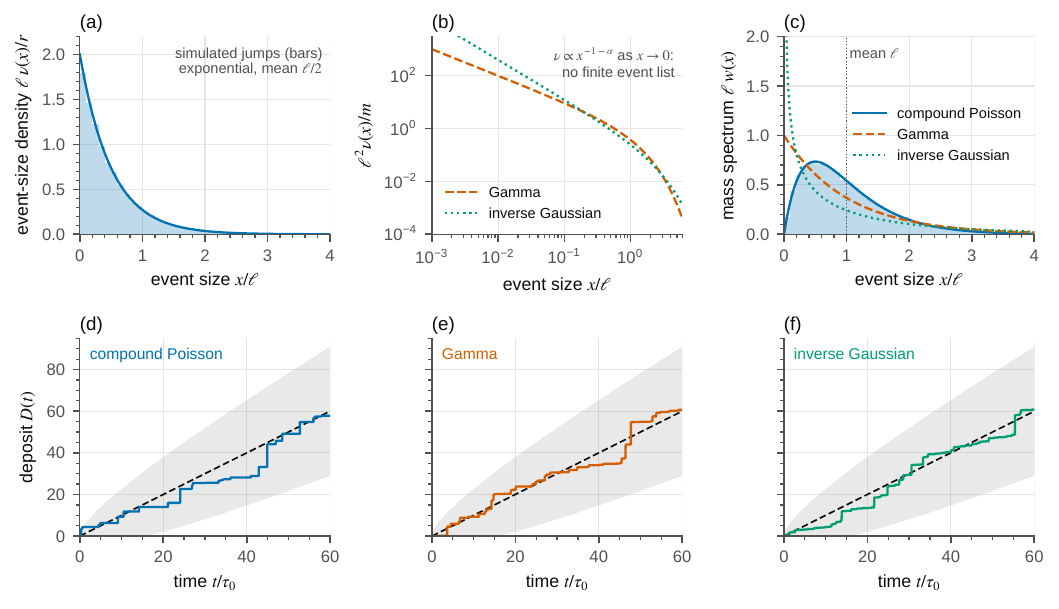}
\caption{Event spectrum, mass spectrum and sample paths for the three laws of
Table~\ref{tab:three_laws} at matched $m$, $c=2$ and $\tau_0$.
(a)~Event-count spectrum $\nu(x)$ of the compound-Poisson law, with its
exponential jump law of mean $\beta=\lambda^{-1}=\ell/2$. (b)~The same for the two
infinite-activity laws; not normalizable. 
(c)~Mass spectrum $w(x)=x\,\nu(x)/m$,
with the common size-biased mean $\ell$ (dashed). (d)--(f)~One path of each
law with the mean $mt$ and its $\pm2\sigma$ band; only compound Poisson~(d)
shows quiescent stretches.}
\label{fig:paths}
\end{figure*}

Three values of $\alpha$ reproduce the standard processes used to model
particulate deposition (Table~\ref{tab:three_laws}); their explicit increment
distributions are given in the Supplemental Material, Sec.~S3.
They differ only in
activity and in the breadth of the event-size spectrum. Compound Poisson
($\alpha=-1$) is the only finite-activity member, depositing discrete events of
mean size $\ell/2$ at rate $2/t_c$ and so admitting quiescent periods of
probability $e^{-r\Delta}$; any $\alpha<0$ is the same process with
Gamma-distributed event sizes (Supplemental Material, Sec.~S3). 
Gamma ($\alpha=0$) is its limit of increasingly
frequent and increasingly weak events, in which attachment of fine particles,
changes of contact area, pore filling and cluster growth cannot be told
apart~\cite{AbdelHameed1975,LawlessCrowder2004,vanNoortwijk2009}, and inverse
Gaussian ($\alpha=1/2$) differs from Gamma only in carrying a broader spectrum.
All three are standard reliability models of monotone degradation, originating
in shock-and-wear models~\cite{Esary1973}, developed in detail for the
inverse-Gaussian case~\cite{WangXu2010,YeChen2014,YeXie2015,Si2011,Kahle2016},
long since unified within the class of L\'evy degradation
processes~\cite{Cinlar1977,AbdelHameed1984,Shu2016}, and applied to heat-exchanger health~\cite{Collet2026}.

The three laws should be viewed as alternative mesoscopic descriptions of
deposition with identical mean and variance. As shown in
Fig.~\ref{fig:paths}, their event-size distributions differ, but the mean
and the variance of the deposit are the same by construction at every
window. Only the higher-order cumulants, skewness and excess kurtosis, are
sensitive to the event-size distribution, and their sample estimates
converge to the analytical values of Table~\ref{tab:three_laws}
(Supplemental Material, Sec.~S6).

\subsection{Aggregation and approach to Gaussian behavior}
\label{sec:aggregation}

Equation~\eqref{eq:aggregation_scaling} shows that aggregation acts on the
three laws in the same way: the decay of the skewness and of the kurtosis with
the window length has the same exponent for all of them, and $\alpha$ only
changes the prefactor. The crossover window to Gaussian behavior, defined by
$\gamma_1(\Delta_c)=1$, is
\begin{equation}
  \Delta_c=t_c\Bigl(\frac{2-\alpha}{1-\alpha}\Bigr)^{2}.
  \label{eq:crossover_window}
\end{equation}
It is a crossover, not a sharp threshold, and it grows without bound
as $\alpha\to1$: the broader the spectrum of event sizes, the more events are
needed before the deposit looks Gaussian. For $\Delta\gg\Delta_c$ the three
laws become increasingly difficult to distinguish on the basis of the
shape of the increments alone, and the deposition
is statistically a drift $m$ with Gaussian fluctuations of variance rate
$s^2$: this is the central limit theorem for the underlying random walk, 
with $N_{\mathrm{eff}}=\Delta/t_c$ playing the role of
the number of summands, and Eq.~\eqref{eq:aggregation_scaling} quantifies
the decay of the leading non-Gaussian cumulants.

\section{Removal: fluctuations, intermittency and the limits of closure}
\label{sec:removal}

Section~\ref{sec:stochastic_formulation} showed that, for absent, constant
or linear removal, the mean
fouling curve does not depend on the statistics of the deposition. Here we ask
what removal does to the \emph{fluctuations}, whether two removal mechanisms
with the same mean trajectory can be told apart, and when the closure of the
mean fails.

\subsection{Linear removal: a L\'evy-driven Ornstein--Uhlenbeck process}
\label{sec:removal_linear}

Solving Eq.~\eqref{eq:stochastic_balance} with $g=\psi R$
gives
\begin{equation}
  R(t)=R_0e^{-\psi t}+\int_0^t e^{-\psi(t-u)}\,dD(u),
  \label{eq:ou_solution}
\end{equation}
a superposition of deposition events, each decaying exponentially over the
removal time $\psi^{-1}$. This is a shot-noise process and, in the language of
stochastic processes, a L\'evy-driven Ornstein--Uhlenbeck process
\cite{Sato1999,BarndorffNielsenShephard2001}.
Since the removal kernel is
linear, the cumulants of $R$ follow from those of $D$:
\begin{equation}
  \kappa_n[R(t)]=q_n\,\frac{1-e^{-n\psi t}}{n\psi}\qquad(n\ge2),
  \label{eq:ou_cumulants}
\end{equation}
while the mean is given in Table~\ref{tab:removal}. All cumulants therefore
saturate, and the plateau is characterized by:
\begin{equation}
\begin{aligned}
  \mathbb E[R]_\infty&=R_\infty=\frac m\psi, &\qquad
  \operatorname{Var}[R]_\infty&=\frac{s^2}{2\psi},\\[2pt]
  \mathrm{CV}_\infty^2&=\frac{\psi t_c}{2}=\frac1{2N_\infty}, &\qquad
  N_\infty&=\frac{R_\infty}{\ell},
\end{aligned}
\label{eq:plateau}
\end{equation}
where $N_\infty$ is the number of size-biased events held in the plateau
deposit.

Per-element detachment is a different mechanism: each deposited element
survives an exponential time, then is removed \emph{entirely}---the usual
reading of the Kern--Seaton time constant as a mean residence
time~\cite{Shaikh2023}. It shares the mean $R_\infty=m/\psi$ but gives
\begin{equation}
  \kappa_n[R]_\infty=\frac{q_n}{\psi},
  \label{eq:mginf_cumulants}
\end{equation}
against $q_n/(n\psi)$ under continuous removal,
Eq.~\eqref{eq:ou_cumulants}. The difference lies in the fate of a single
event of size $x$ deposited at time $u$. Continuous removal erodes it
steadily, so that it contributes $x\,e^{-\psi(t-u)}$; per-element detachment
leaves it intact until it goes, so that it contributes $x$ with probability
$e^{-\psi(t-u)}$ and nothing otherwise. The two contributions have the same
mean, but their $n$th powers average to $x^ne^{-n\psi(t-u)}$ and
$x^ne^{-\psi(t-u)}$ respectively, and since the cumulants of a Poisson sum
add these powers, every cumulant of order $n$ is $n$ times larger under
detachment: the mean is unchanged and the variance doubled. Once standardized
the factors nearly cancel: the plateau excess kurtosis is the same,
$\gamma_2(\psi^{-1})$, and the skewness, $\gamma_1(\psi^{-1})$, exceeds the
continuous-removal value by only $3/(2\sqrt2)\approx1.06$ (Supplemental
Material, Sec.~S4). The factor $n$ requires each event to detach as
a single unit: an event made of many sub-units detaching independently loses
mass smoothly, as $x\,e^{-\psi(t-u)}$, and returns the continuous-removal
cumulants.

The plateau is thus built from $N_\infty=R_\infty/\ell$ size-biased events
(Sec.~\ref{sec:activity}), and its relative fluctuations are $1/\sqrt{2N_\infty}$:
a plateau that contains many events is regular, whereas one made of a few
large events is not. The transient variance, $(s^2/2\psi)(1-e^{-2\psi t})$,
grows as $s^2t$ (pure deposition) for $t\ll\psi^{-1}$ and saturates after a
time of order $(2\psi)^{-1}$.

For the tempered-stable family the stationary skewness and excess kurtosis
under continuous linear removal are
\begin{equation}
\begin{aligned}
  \gamma_{1,\infty}&=\frac43\,\frac{2-\alpha}{1-\alpha}\,\mathrm{CV}_\infty,\\
  \gamma_{2,\infty}&=2\,\frac{(2-\alpha)(3-\alpha)}{(1-\alpha)^2}\,\mathrm{CV}_\infty^2,
\end{aligned}
  \label{eq:plateau_shape}
\end{equation}
equivalently
\begin{equation}
\gamma_{1,\infty}=\tfrac{2\sqrt2}{3}\gamma_1(\psi^{-1}), \qquad
\gamma_{2,\infty}=\gamma_2(\psi^{-1}).
\end{equation}
The plateau has the shape of the
deposit accumulated over one removal time, $\Delta=\psi^{-1}$
(Supplemental Material, Sec.~S4).
The pattern of Sec.~\ref{sec:tempered} therefore repeats one level up:
$R_\infty$ and $\mathrm{Var}[R]_\infty$ are independent of $\alpha$, while
$\gamma_{1,\infty}$ and $\gamma_{2,\infty}$ are not.

What removal changes is how that distinguishing information behaves in time.
Without removal the standardized cumulants decay indefinitely with the
observation window, as $\Delta^{-1/2}$ and $\Delta^{-1}$, so the longer the
deposit accumulates before being inspected the harder the three laws are to
tell apart. Under linear removal $R(t)$ is built only from deposition of the
last $\psi^{-1}$ or so, older material having been removed, and the effective
window cannot grow past $\Delta\sim\psi^{-1}$ however long the process runs.
The exact finite-time cumulants of Eq.~\eqref{eq:ou_cumulants} show this
directly: $\gamma_1(t)$ and $\gamma_2(t)$ follow their no-removal decays while
$t\ll\psi^{-1}$, then saturate at the values of
Eq.~\eqref{eq:plateau_shape}. Removal does not erase the shape information the
way further aggregation does; it freezes it at the value reached after about
one relaxation time.
Figure~\ref{fig:removal}(a)--(b) illustrates this directly: sample paths
with the common mean and $\pm2\sigma$ band, and the exact skewness
$\gamma_1(t)$ for each law (solid), which tracks the no-removal
$t^{-1/2}$ decay (dashed) while $t\ll\psi^{-1}$ and bends over to the
law-dependent plateau $\gamma_{1,\infty}$ once removal takes effect.

\subsection{Intermittent removal: sloughing}
\label{sec:removal_sloughing}

Continuous removal is not the only way to produce a plateau. Fouling layers
may lose material in discrete events (sloughing, spalling, detachment of
flakes)~\cite{Stewart1993,Bohn2007}. At the scale of a single particle the same
structure arises from re-entrainment by turbulent bursts, each sweeping a fixed
fraction of the wall and removing material that has not yet
stabilised~\cite{Vatistas1988}. We model them as a Poisson process $N_s(t)$ of rate $r_s$, each event
removing a fixed fraction $\eta\in(0,1]$ of the deposit,
\begin{equation}
  dR(t)=dD(t)-\eta\,R(t^-)\,dN_s(t).
  \label{eq:sloughing}
\end{equation}
The rate and the fraction do not depend on the state of the deposit: removal is
triggered by time, not by the thickness reached. (Thickness-triggered spalling
is not treated here.) Averaging Eq.~\eqref{eq:sloughing} gives
$d\,\mathbb E[R]/dt=m-r_s\eta\,\mathbb E[R]$, the same mean equation as for
continuous removal, with $\psi=r_s\eta$. The mean trajectory, the plateau
$R_\infty$ and the relaxation time cannot discriminate between the two
mechanisms. The same holds for the two-point correlation: $\mathbb E[R(t+\tau)\,|\,R(t)]$ obeys the same linear
equation in both cases, so the autocorrelation is again $e^{-\psi\tau}$.

\begin{figure*}[t]
\centering
\includegraphics[width=\textwidth]{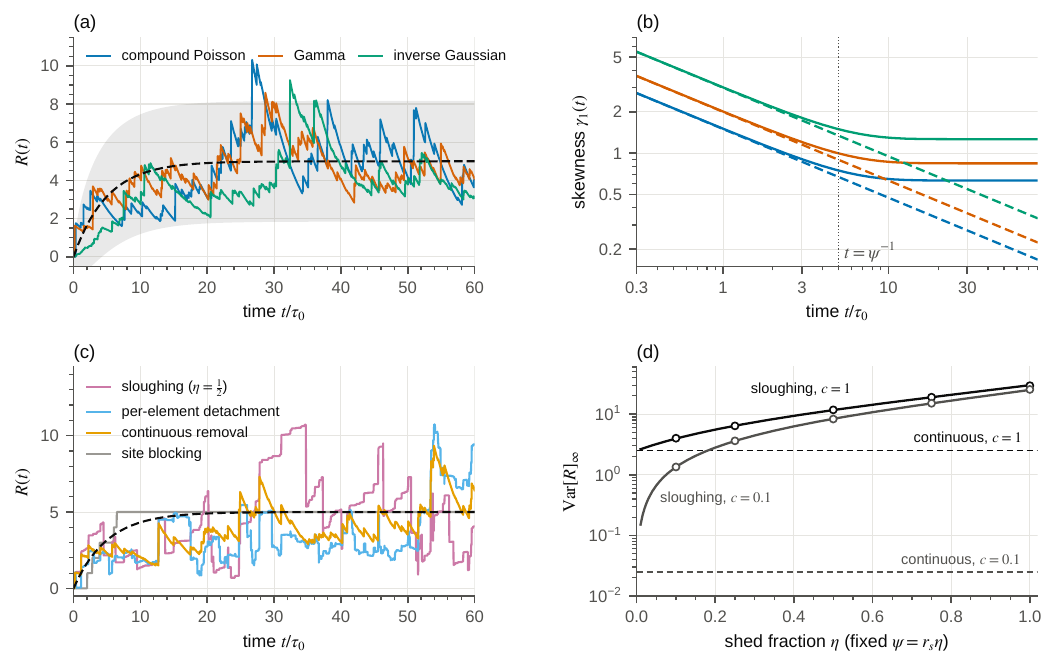}
\caption{Fluctuations under removal ($m=1$, $c=1$, $\tau_0=1$, $\psi=0.2$,
$R_\infty=5$). (a)~Sample paths of the L\'evy-driven Ornstein--Uhlenbeck
process for the three laws, with the mean and the $\pm2\sigma(t)$ band.
(b)~Exact skewness $\gamma_1(t)$ under removal (solid,
Eq.~\eqref{eq:ou_cumulants}) and the no-removal $t^{-1/2}$ reference (dashed).
(c)~One path each of continuous removal, per-element detachment, sloughing
($\eta=0.5$, $r_s=\psi/\eta$) and independent-site blocking (equal events of
size $\ell$), all with the same mean (dashed). (d)~Stationary variance against $\eta$ at
fixed $\psi$, Eq.~\eqref{eq:sloughing_variance} (lines) and Monte Carlo
(markers), for $c=1$ and $c=0.1$; dashed: the continuous-removal
baselines $s^2/2\psi$ (logarithmic scale).}
\label{fig:removal}
\end{figure*}

The fluctuations, however, differ. At equal mean and relaxation rate, the
stationary variance under sloughing is
\begin{equation}
  \operatorname{Var}[R]_\infty
  =\frac{s^2}{(2-\eta)\,\psi}+\frac{\eta}{2-\eta}\,R_\infty^2 ,
  \label{eq:sloughing_variance}
\end{equation}
to be compared with $s^2/2\psi$ for continuous removal
(Table~\ref{tab:plateau}). The second term does not depend on the deposition
statistics: it is the sawtooth generated by the random timing of the removal
events, and is present even if the deposition is perfectly smooth ($s=0$). Its
relative size is $\eta/(2-\eta)$. In the limit of frequent small events,
$\eta\to0$ at fixed $r_s\eta$, Eq.~\eqref{eq:sloughing_variance} reduces to
the continuous result, as expected for erosion by many small detachments. For
complete cleaning events ($\eta=1$), $\mathrm{CV}^2=\psi t_c+1\ge1$: the
fluctuations are of the order of the mean itself. Figure~\ref{fig:removal}(c)--(d)
compares one realization of each mechanism at matched mean and validates
the stationary variance against $\eta$, Eq.~\eqref{eq:sloughing_variance},
by Monte Carlo simulations. 
The moment comparison between the two mechanisms is collected in the Supplemental Material, Sec.~S4.

\begin{table*}[t]
\caption{Plateau statistics for three removal mechanisms reaching the same
plateau $R_\infty=m/\psi$ (for sloughing, $\psi=r_s\eta$); $t_c=s^2/m^2$. All
three share the autocorrelation $e^{-\psi|\tau|}$. Only sloughing retains a non-vanishing
plateau variance as the deposition becomes smooth, $c\to0$, because its second
term does not contain $s^2$.}
\label{tab:plateau}
\begin{ruledtabular}
\begin{tabular}{lcccc}
Mechanism & $R_\infty$ & $\operatorname{Var}[R]_\infty$ & $\mathrm{CV}_\infty^2$ & $\operatorname{Var}[R]_\infty$ as $c\to0$ \\
\hline
proportional erosion (continuous) & $\dfrac m\psi$ & $\dfrac{s^2}{2\psi}$ & $\dfrac{\psi t_c}{2}$ & $0$ \\[2mm]
per-element detachment & $\dfrac m\psi$ & $\dfrac{s^2}{\psi}$ & $\psi t_c$ & $0$ \\[2mm]
intermittent removal (sloughing) & $\dfrac m\psi$ & $\dfrac{s^2}{(2-\eta)\psi}+\dfrac{\eta R_\infty^2}{2-\eta}$ & $\dfrac{\psi t_c+\eta}{2-\eta}$ & $\dfrac{\eta R_\infty^2}{2-\eta}$ \\[2mm]

\end{tabular}
\end{ruledtabular}
\end{table*}

\subsection{Nonlinear removal}
\label{sec:removal_nonlinear}

For a removal law $g(R)$ that is neither constant nor linear, the mean
equation, which is still exact,
\begin{equation}
  \frac{d\mu}{dt}=m-\mathbb E[g(R)],\qquad\mu=\mathbb E[R],
  \label{eq:nonlinear_mean}
\end{equation}
no longer closes: $\mathbb E[g(R)]\neq g(\mu)$, and the mean depends on the
fluctuations. Expanding $g$ around $\mu$ gives, to second order,
\begin{equation}
  \frac{d\mu}{dt}\simeq m-g(\mu)-\tfrac12\,g''(\mu)\,V,\qquad
  \frac{dV}{dt}\simeq s^2-2\,g'(\mu)\,V,
  \label{eq:nonlinear_moments}
\end{equation}
with $V=\operatorname{Var}[R]$. The expansion is ordered in the relative
event size $1/N_\infty=\ell/R_\infty$; the third-cumulant terms it neglects
are smaller by a further factor $1/N_\infty$ (Supplemental Material,
Sec.~S4). To this order the deposition law enters only through
$s^2$, i.e.\ through $c$; the shape parameter $\alpha$ first appears
through $\kappa_3[R]$ at the next order in $1/N_\infty$. If $g$
is convex, fluctuations increase the mean removal rate and the mean fouling
lies below the deterministic (mean-field) prediction; for concave $g$ the
opposite holds. The universality of the mean is therefore lost, and the mean
curve now carries information on $c$.

The correction to the removal rate, $\tfrac12g''V/g\sim\mathrm{CV}^2\sim1/N_\infty$,
is small when the plateau holds many events, where the closure becomes
asymptotically exact, and the closure fails when it holds few.

In real exchangers nonlinear dependence on the deposit
arises, for instance, from the change of the flow cross-section and of the
wall shear stress with the deposit thickness (condition~(ii) of Sec.~\ref{sec:theory}); such
feedback also makes the deposition itself non-stationary, and lies outside the
scope of this work.

\paragraph*{Numerical verification.}
Every analytic result of this and the following sections is checked against
Monte Carlo simulation of Eq.~\eqref{eq:stochastic_balance}. Deposition
increments are drawn exactly from the three laws of Table~\ref{tab:three_laws};
continuous removal is applied as exponential decay over each step, sloughing
as Poisson-timed events, per-element detachment exactly by thinning the alive
deposit, and blocking by binomial filling of independent sites. Where a time
step is needed it is $0.005$--$0.02\,\tau_0$; ensembles range from $10^3$
records to $2\times10^5$ paths per law, with fixed seeds, and the code is
available (Data Availability). The simulations reproduce the transient and
stationary moments of Table~\ref{tab:plateau} to within sampling error, about
$0.5\,\%$ for the stationary variances (Supplemental Material, Fig.~S2), the
plateau skewness of Eq.~\eqref{eq:plateau_shape}, the record-length estimate
of Sec.~\ref{sec:observability} (Fig.~S5) and the first-passage statistics of
Sec.~\ref{sec:first_passage} (Fig.~S6).

\section{Identifiability of the deposition mechanism}
\label{sec:identifiability}
\subsection{What the mean cannot see}
\label{sec:removal_summary}

Linear removal and Poisson-timed sloughing, matched at $\psi=r_s\eta$, were
shown in Sec.~\ref{sec:removal} to share one mean curve while their
fluctuations remain distinguishable (Table~\ref{tab:plateau}).

Both mechanisms assume that every particle reaching the wall sticks, whereas
attachment efficiencies well below unity are the
rule in particulate deposition~\cite{Yao1971,Epstein1988}, and the reaction-engineering treatment of fouling writes the deposition rate as a foulant production rate times a sticking probability~\cite{PatersonFryer1988}. Writing $p$ for the sticking probability, two cases separate. A constant $p$ thins the event-rate density,
\begin{equation}
  \nu\to p\,\nu,\qquad m\to p\,m,\qquad c^{2}\to c^{2}/p,
  \label{eq:thinning}
\end{equation}
leaving $\ell$ and $\alpha$ unchanged: the tempered-stable family is closed
under the operation, and the only effect is a scaled net rate---a linear
mean curve of reduced slope, or, where linear removal is also present, the
same exponential saturation to a proportionally lower plateau.
This is the suppression reading of the Ebert--Panchal term
(Sec.~\ref{sec:theory}): with $p=1-\gamma\tau_w/m$ the thinned process
reproduces the linear mean of Table~\ref{tab:removal} exactly, without a
boundary at $R=0$, and with a variance $p\,s^2\,t$ smaller than that
of the erosion reading. If instead $p$
falls as the surface fills, $p(R)=1-R/R_\infty$, the Langmuir blocking law of
site-limited adsorption, the mechanism is not removal at all: nothing leaves
the surface and attachment becomes self-limiting. Each increment then depends
on how full the surface already is, so the process leaves the L\'evy class.
We take the blocking to be independent-site filling, for which the mean closes
exactly because $p$ is linear in $R$ and the variance vanishes at saturation;
random sequential adsorption with excluded volume is a different process,
approaching its jamming limit non-exponentially and retaining a variance
proportional to the area, so the two are themselves distinguishable by the
late-time scatter. The mean trajectory of independent-site blocking is
nonetheless the same exponential saturation to $R_\infty$.

Four mechanisms with nothing physically in common---continuous shear
removal, per-element detachment, intermittent sloughing and self-limiting
Langmuir adsorption---therefore reach the same saturating mean at matched net
rate [Fig.~\ref{fig:removal}(c)], while constant attachment thinning rescales the linear branch without
saturating it at all. Fitting an asymptotic curve returns $R_\infty$ and a
time constant and says nothing about which of the four produced it: a
Kern--Seaton fit is not evidence of shear removal.
Table~\ref{tab:plateau} collects the three that settle to a stationary
plateau; blocking is absent from it because its variance vanishes once the
sites are full, leaving no stationary fluctuation. The same difficulty, of
separating deposition, suppression, removal and consolidation in operating
data, is known within the mechanistic literature~\cite{Diaz-Bejarano24052017}.

The degeneracy is not even confined to the mechanism. Stationarity and a
finite first moment already give $\mathbb{E}[D(t)]=mt$, since expectations
add whether or not the increments are independent, and for a single record
rather than an ensemble it is enough that the increment sequence be ergodic;
independence is needed only from the second cumulant onward. A straight or
saturating mean curve is therefore not evidence even of the L\'evy structure
assumed in Sec.~\ref{sec:stochastic_formulation}. What such a curve identifies is the net mean
rate and, under linear removal, the relaxation rate---and nothing beyond
them.

\subsection{Variance and the deposition law}
\label{sec:variance_test}

The three laws of Sec.~\ref{sec:three_laws} are matched in their first two
cumulants by construction, so only the standardized third and fourth cumulants
carry information about the increment law. What the variance can test is
whether a \emph{granular} picture of deposition is quantitatively consistent at
all.

Suppose each attachment event is one particle of diameter $d_p$, depositing at solid
fraction $\phi_s$, and that arrivals form a Poisson
process. Averaged over the heat-transfer area $A$, a single attachment raises the fouling
resistance by
\begin{equation}
  J_1=\frac{\pi d_p^{3}/6}{A\,\phi_s\,k_f},
  \qquad\text{whence}\qquad
  c^{2}=\frac{J_1}{m\tau_0}.
  \label{eq:shot_noise}
\end{equation}
Equation~\eqref{eq:shot_noise} is not an external estimate: it is
Eq.~\eqref{eq:neff} evaluated at $\Delta=\tau_0$ in the monodisperse limit
$\alpha\to-\infty$ of the family, where every event deposits the same amount
and the size-biased event size $\ell$ collapses onto the jump $J_1$ itself.
Equivalently $c^{2}=1/(r\tau_0)$---the classical shot-noise result for Poisson-timed pulses of
equal weight~\cite{Rice1944,Rice1945}. No parameter is free: once the event
size is fixed, the mean rate fixes the event frequency through $r=m/J_1$. The
derivation is given in the Supplemental Material, Sec.~S5.
The resulting magnitudes are extreme. For an exchanger of $A=200~\mathrm{m^2}$
fouling at a rate of order $10^{-5}~\mathrm{m^{2}\,K\,W^{-1}}$ per sampling period, with deposit
properties $k_f\sim0.5~\mathrm{W\,m^{-1}K^{-1}}$ and $\phi_s\sim0.5$,
Eq.~\eqref{eq:shot_noise} gives $c\sim10^{-9}$ for $0.1~\mu\mathrm{m}$ particles
and $c\sim3\times10^{-5}$ at $100~\mu\mathrm{m}$. The transport mechanism
is thus invisible in the mean, which sees only the net rate
(Sec.~\ref{sec:removal_summary}), and invisible in the variance by smallness.

The scaling behind those numbers is the general statement. Because $c^{2}\propto J_1\propto A^{-1}$
at fixed mean rate, granular arrival noise is suppressed as $A^{-1/2}$ and attains order unity
only on the scale of a single particle. The same square-root suppression governs the coverage of a deposited
monolayer under random sequential adsorption, whose jamming count obeys a
central limit theorem with variance proportional to the system
size~\cite{Penrose2001,Loscar2003}. An area-averaged fouling resistance is therefore almost
perfectly deterministic with respect to the granularity of deposition, and any observed
coefficient of variation of order one must originate in a fluctuation that is coherent over a
length comparable to the exchanger itself.

That bound bears on assumption (b) of Sec.~\ref{sec:stochastic_formulation}: independent granular
arrivals cannot produce $c$ of order unity without single events equivalent to a uniform layer
several to tens of micrometres thick across the whole surface, which is not physical. A coherent fluctuation
is required instead: operating-condition drift ($c\approx E_a\Delta T/(R_gT_f^2)\approx0.1$ for a 5~K film-temperature
excursion at $E_a=50~\mathrm{kJ\,mol^{-1}}$, $T_f=500$~K), turbulent-burst intermittency in near-wall
transport~\cite{Vatistas1988,Nakagawa1978,Miele2025}, or a coherent measurement error. The first
two place the process outside the L\'evy class; the third leaves the subordinator description
intact and makes the observed $c$ partly a sensor parameter.

\subsection{Variance and the removal mechanism}
\label{sec:variance_removal}

For removal, linearising the variance equation gives
$\dot V\simeq s^2-2\,g'(\mu)\,V$ (Eq.~\eqref{eq:nonlinear_moments}), so what
matters is the \emph{slope} of the removal law at the operating point, not
whether removal is present. Pure deposition and Ebert--Panchal removal both
have $g'=0$ and therefore the same unbounded growth
(Table~\ref{tab:removal}; under the suppression reading the variance rate
is reduced by the factor $p$, but the growth remains unbounded); only
$g'\neq0$ makes the variance informative.

A saturating record is exactly the regime where the variance separates
mechanisms the mean cannot: sloughing retains a floor
$\eta R_\infty^2/(2-\eta)$ even as deposition becomes noise-free
(Eq.~\eqref{eq:sloughing_variance}), while every $s^2$-proportional term---including
per-element detachment's (Eq.~\eqref{eq:mginf_cumulants})---vanishes with $c$. At realistic $c$
the question is therefore not whether the plateau variance exceeds a baseline but whether it
fluctuates at all; blocking is separated from continuous removal
instead by its time course: with independent-site filling the variance is
$\ell R_\infty\,\theta(1-\theta)$ at coverage
$\theta=\mathbb E[R]/R_\infty=1-p$, rising to a maximum $\ell R_\infty/4$
at half coverage and returning to zero as the sites fill. The binomial
form holds for sites of a single capacity, each filled by one event of size
$J_1$, i.e.\ the monodisperse limit of Sec.~\ref{sec:variance_test} with
$\ell=J_1$; for polydisperse events the intermediate-time variance depends on
how events of different size share the sites, but it still vanishes as the
surface saturates. That maximum is set
by the event size, so unlike the sloughing floor it is a granular quantity and
inherits the smallness of Sec.~\ref{sec:variance_test}.

What this test costs is worth setting beside
Sec.~\ref{sec:observability}. The sloughing floor is a fixed fraction of the
plateau: at $\eta=\tfrac12$ it is $R_\infty^2/3$, so the discriminator asks
only that the measurement noise stay below $R_\infty/\sqrt3$, under
six-tenths of the plateau itself. What it does demand is duration, since the
state decorrelates on the relaxation time and the plateau variance must
therefore be estimated from a record spanning several multiples of
$\psi^{-1}$. The contrast with the deposition law is not one of degree.
Identifying the increment law needs both a better instrument and the record
length of Eq.~\eqref{eq:Nreq}; identifying the removal mechanism needs a
record that reaches its plateau and stays there.

\subsection{Measurement noise: observability and record length}
\label{sec:observability}

Temporal aggregation preserves the mean rate and suppresses every standardized
higher cumulant, Eq.~\eqref{eq:aggregation_scaling}: over a window $\Delta t$
the coefficient of variation falls as $c\sqrt{\tau_0/\Delta t}$ and the
skewness as $\gamma_1\sqrt{\tau_0/\Delta t}$. A smooth macroscopic fouling
curve is therefore compatible with strongly intermittent event-scale
deposition, and the question for a real record is whether any observation
window is short enough to retain that shape and long enough to clear the
measurement noise.

Let the state be observed at intervals $\Delta t=k\tau_0$ as
$Y_j=R(j\Delta t)+\varepsilon_j$ with
$\varepsilon_j\overset{\mathrm{iid}}{\sim}\mathcal{N}(0,\sigma^2)$, and take
disjoint increments $\Delta Y_j=Y_{2j+1}-Y_{2j}$ so that their measurement
errors are independent. Then
$\operatorname{Var}[\Delta Y_j]=c^2m^2\tau_0\Delta t+2\sigma^2$: the
deposition variance grows with the window, the differenced noise does not, and
everything that follows is governed by the single dimensionless group formed
from the two,
\begin{equation}
  \rho=\frac{2\sigma^2}{c^2m^2\tau_0^{2}},
  \label{eq:rho}
\end{equation}
the differenced noise variance divided by the deposition variance of one
reference window; equivalently, $\rho$ is the window length $k$ at which the
increment's signal-to-noise ratio $\mathrm{SNR}(k)=k/\rho$ reaches one, where
deposition and noise contribute equally.

\begin{figure}[ht]
\centering
\includegraphics[width=0.99\columnwidth]{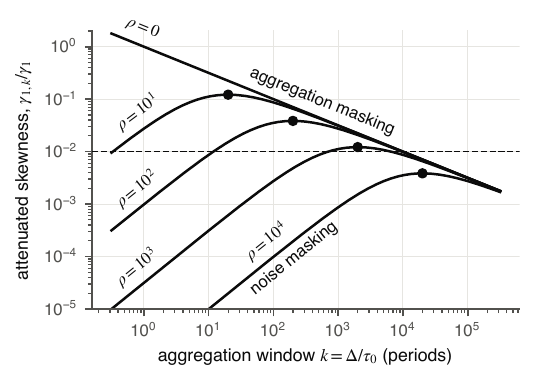}
\caption{Observability is governed by the noise-to-signal ratio $\rho$ alone.
The vertical axis is the attenuated skewness $\gamma_{1,k}$,
Eq.~\eqref{eq:attenuated_shape}, normalized by its noise-free value
$\gamma_1$, plotted against the aggregation window $k$ ($\rho=0$: the
noise-free $k^{-1/2}$ decay). The dot on each curve marks the best window
$k^\ast=2\rho$, with peak value $(2/3\sqrt3)\,\gamma_1/\sqrt\rho$: every curve has
the same shape, rescaled by $\rho$, so only this peak height falls as $\rho$
grows, clearing the dashed $1\,\%$ line for $\rho=10,10^2,10^3$ but falling
below it for any $\rho\gtrsim1.5\times10^{3}$.}
\label{fig:observability}
\end{figure}

Let $f=k/(k+\rho)=\mathrm{SNR}/(1+\mathrm{SNR})$ be the share of the observed
variance that is deposition. Gaussian noise leaves every cumulant above the
second unchanged and inflates $\kappa_2$ by $1/f$, so a standardized cumulant
of order $n$, which carries $\kappa_2^{-n/2}$, is attenuated by exactly
$f^{n/2}$:
\begin{align}
  \gamma_{1,k}&=\frac{\gamma_1}{\sqrt k}\,f^{3/2}
              =\gamma_1\,\frac{k}{(k+\rho)^{3/2}},\notag\\
  \gamma_{2,k}&=\frac{\gamma_2}{k}\,f^{2}
              =\gamma_2\,\frac{k}{(k+\rho)^{2}},
  \label{eq:attenuated_shape}
\end{align}
the first factor in each being the noise-free value at that window. The two
pull against each other: $f^{n/2}$ rises towards one as the window clears the
noise (noise masking) and can go no further, while the aggregation factor
(aggregation masking) falls without bound.
A bounded gain against an unbounded loss forces an interior maximum, at
$k^\ast=2\rho$ and $k^\ast=\rho$, with maxima $(2/3\sqrt3)\,\gamma_1/\sqrt\rho$ and
$\gamma_2/(4\rho)$, so that once $\rho\gg1$ no window reveals the shape of the
law: any window long enough to clear the noise has already aggregated the
shape away (Fig.~\ref{fig:observability}).

Detecting the shape is a different optimization from preserving it.
Each disjoint increment consumes $2k$ base periods, so a record of $N$
periods supplies $n_w=N/(2k)$ independent windows, on which the sample
skewness and sample excess kurtosis have standard errors $\sqrt{6/n_w}$ and
$\sqrt{24/n_w}$ in the Gaussian limit $\rho\gg c^2$. We define the
separability index $d'$ as the difference between the attenuated values of
two laws, $\Delta\gamma_{1,k}$ or $\Delta\gamma_{2,k}$, in units of that
standard error: the distance, in standard errors, between the estimate
obtained under one law and the value expected under the other. A longer
window retains more of the cumulant but leaves fewer independent windows in a
record of fixed length, and the two effects balance at $k=\rho/2$ for the
skewness and $k=\rho/3$ for the excess kurtosis---both shorter than the windows
that maximize the cumulants themselves, and both requiring $\rho\gtrsim3$
so that $k\ge1$. Requiring $d'=2$
between the Gamma and inverse Gaussian laws ($\Delta\gamma_1=c$) at the
better of the two statistics then demands a record of
\begin{equation}
  N\approx 324\,\frac{\rho^{2}}{c^{2}}
  \label{eq:Nreq}
\end{equation}
periods (Supplemental Material, Sec.~S6). This pair is the
easiest to separate; the Gamma and compound-Poisson laws differ by only
$\Delta\gamma_1=c/2$ and need four times as long a record. The accounting is
also conservative: when readings are available at every base period,
windows need only be separated by one period to keep their endpoint errors
independent, $n_w\simeq N/k$, which halves the prefactor, and using all
consecutive increments with their known first-order moving-average error
structure would gain a further constant factor. Neither changes the
$\rho^{2}$ scaling. The skewness is the cheaper
statistic by a factor $\rho/(14c^{2})$, so it, and not the excess kurtosis,
sets the scaling of the requirement with the noise. Since the required record grows as the
fourth power of the measurement noise,
improving the instrument is by far the more effective route to the event-scale
statistics.
The estimate is asymptotic and specific to moment-based estimation,
whereas the attenuation it rests on is not; its range of validity is given in
the Supplemental Material, Sec.~S6.

The numbers this implies can be estimated without reference to any particular
record. A thermally reconstructed fouling resistance is inferred from
temperature and flow measurements as a difference of reciprocals,
$R_f=A\,(1/UA-1/UA_{\mathrm{clean}})$, so that a relative error $\epsilon$ in
the inferred conductance gives an error $\epsilon/U$ in $R_f$, with $U=UA/A$. The
deposition increment of one period changes the conductance by the relative
amount $\zeta=U m\tau_0$, and Eq.~\eqref{eq:rho} becomes
\begin{equation}
  \rho=\frac{2}{c^{2}}\Bigl(\frac{\epsilon}{\zeta}\Bigr)^{2},
  \qquad
  N\approx\frac{1296}{c^{6}}\Bigl(\frac{\epsilon}{\zeta}\Bigr)^{4},
  \label{eq:rho_generic}
\end{equation}
the second from Eq.~\eqref{eq:Nreq}. For an exchanger with
$U\approx400~\mathrm{W\,m^{-2}K^{-1}}$ fouling by
$10^{-5}~\mathrm{m^{2}\,K\,W^{-1}}$ per three-day period, $\zeta\approx0.4\,\%$:
a conductance inferred to within $1\,\%$ already carries a noise comparable to
several periods of deposition (consistent with reported uncertainties of
$4.6\,\%$ for an industrial exchanger's overall
coefficient~\cite{Fguiri2020} and resistances uncertain below about
$10^{-5}~\mathrm{m^{2}\,K\,W^{-1}}$~\cite{Shen2015}).
Table~\ref{tab:record_length} evaluates
Eq.~\eqref{eq:rho_generic}. Only the combination of a conductance accurate to
about one per cent and strongly bursty deposition, $c\approx2$, brings the
requirement within years; at a few per cent, or for $c\lesssim1$ (the requirement
scales as $c^{-6}$), it runs to centuries and beyond, and for the granular values of $c$ found in
Sec.~\ref{sec:variance_test} it is unbounded for practical purposes. The
estimate is, moreover, favorable on two counts: it counts only the error of
the inferred conductance, not the property and heat-loss closures behind
the duty or the use of a design rather than a measured clean conductance, and
it assumes errors that average away rather than correlate in time. For
thermally reconstructed resistances, where a conductance error of a few per
cent already exceeds the deposition of several sampling periods, the
increment law is thus, in practice, unobservable.

\paragraph*{Measurement Uncertainty Effect.}
The error $\epsilon$ follows from the reconstruction of the conductance. With
$UA=Q/\Delta T_{\mathrm{lm}}$ and the duty $Q=\dot m c_p\,\Delta T$ taken from
one stream, first-order propagation gives, neglecting correlations through
shared sensors~\cite{Tatara2011,Shen2015},
\begin{equation}
  \epsilon^{2}\simeq\Bigl(\frac{\delta\dot m}{\dot m}\Bigr)^{2}
  +\Bigl(\frac{\delta(\Delta T)}{\Delta T}\Bigr)^{2}
  +\Bigl(\frac{\delta(\Delta T_{\mathrm{lm}})}{\Delta T_{\mathrm{lm}}}\Bigr)^{2}.
  \label{eq:epsilon_propagation}
\end{equation}
Only the random part of each error enters. A calibration offset of a
thermocouple or a flow meter is constant over the record and cancels in the
increments on which every statistic of this section is built, so that
$\epsilon$ is the precision error of a single reading: a better-calibrated
instrument does not help, only a less noisy one does. For a stream heated by
$\Delta T=20$~K, a log-mean temperature difference of $30$~K, a random
thermocouple error of $0.1$~K and a random flow-meter error of $0.5\,\%$, the
duty is uncertain to about $0.9\,\%$ and the log-mean difference to about
$0.3\,\%$, so that $\epsilon\approx1\,\%$ and $\epsilon/\zeta\approx2.5$, the
second row of Table~\ref{tab:record_length}. Averaging $n$ readings within a
sampling period reduces $\epsilon$ by $\sqrt n$ only if their errors are
independent; plant noise is usually correlated in time, and $\epsilon$ is
then the error of the averaged reading.

\begin{table}[tb]
\caption{Record length, in years, required to separate the Gamma and
inverse-Gaussian laws, Eq.~\eqref{eq:rho_generic}, against the relative error
$\epsilon$ of the inferred conductance, for a relative conductance change
$\zeta=0.4\,\%$ per sampling period $\tau_0=3$~days. The column
$\epsilon/\zeta$ expresses the noise in periods of deposition. Down a column
the requirement grows as $\epsilon^4$, across a row as $c^{-6}$.
Entries marked $^\dagger$ have $\rho<c^{2}$, outside the asymptotic regime of
Eq.~\eqref{eq:Nreq}, and are indicative only.}
\label{tab:record_length}
\vspace{4pt}\centering
\begin{tabular}{@{}cc@{\hspace{10pt}}cccc@{}}
\toprule
 & & \multicolumn{4}{c}{record length (years)}\\
\cmidrule(l){3-6}
$\epsilon$ & $\epsilon/\zeta$ & $c=2$ & $c=1$ & $c=0.5$ & $c=0.1$\\
\midrule
$0.5\,\%$ & $1.25$ & $0.4^{\dagger}$ & $26$ & $2\times10^{3}$ & $3\times10^{7}$\\
$1\,\%$ & $2.5$  & $6^{\dagger}$   & $4\times10^{2}$ & $3\times10^{4}$ & $4\times10^{8}$\\
$2\,\%$ & $5$    & $1\times10^{2}$ & $7\times10^{3}$ & $4\times10^{5}$ & $7\times10^{9}$\\
$3\,\%$ & $7.5$  & $5\times10^{2}$ & $3\times10^{4}$ & $2\times10^{6}$ & $3\times10^{10}$\\
$5\,\%$ & $12.5$ & $4\times10^{3}$ & $3\times10^{5}$ & $2\times10^{7}$ & $3\times10^{11}$\\
\bottomrule
\end{tabular}
\end{table}

The same competition between a noise floor at short lag and a degrading
signal at long lag governs parameter estimation for noisy degradation and
diffusion records generally~\cite{Whitmore1995,Bordes2015,Pulcini2016,Lanoiselee2018};
what is specific here is that the quantity estimated is a standardized higher
cumulant, which aggregation itself destroys, so the upper cutoff is a
property of the process rather than of the record length.

\section{First-passage consequences}
\label{sec:first_passage}

Consider deposition without removal, $R(t)=R_0+D(t)$, and a generic threshold
$x>R_0$. The continuous first-passage time is
\begin{equation}
  T_x=\inf\{t\geq0:R(t)\geq x\}.
  \label{eq:first_passage_definition}
\end{equation}
Because the process is nondecreasing,
\begin{equation}
  \Pr(T_x\leq t)=\Pr[D(t)\geq x-R_0].
  \label{eq:first_passage_identity}
\end{equation}
In an engineering application, setting $x$ equal to a cleaning or performance
limit makes $T_x$ equivalent to a remaining useful life.

We define the dimensionless threshold distance
\begin{equation}
  n_x=\frac{x-R_0}{m\tau_0}.
  \label{eq:threshold_distance}
\end{equation}
For $n_x\gg1$, and with time measured in sampling periods, the leading
passage-time moments are
\begin{align}
  \frac{\mathbb{E}[T_x]}{\tau_0}
  &=n_x+\frac{1+c^{2}}{2}+o(1),
  \label{eq:first_passage_moments}\\
  \frac{\operatorname{Var}(T_x)}{\tau_0^{2}}
  &=c^2 n_x\bigl[1+O(n_x^{-1})\bigr],
\end{align}
the expansion being in the threshold distance $n_x$ rather than in $\tau_0$.
Equation~\eqref{eq:first_passage_moments} is the mean for passage
\emph{detected at the sampling instants}; in continuous time the overshoot is
the stationary overshoot of the subordinator, of mean $\ell/2$, and the
correction is $c^{2}/2$ instead: the extra half-period is the cost of
discrete observation. To leading order,
\begin{equation}
  \operatorname{CV}(T_x)
  =\frac{c}{\sqrt{n_x}}+O(n_x^{-1}).
  \label{eq:first_passage_cv}
\end{equation}
The mean rate fixes the leading passage time, while $c$ fixes its leading
relative uncertainty. Cleaning-schedule optimization under uncertainty treats the
fouling law as deterministic with uncertain parameters or
inputs~\cite{AlIsmaili2019,DiPretoro2021}; for a distant threshold that uncertainty dominates, the intrinsic variability adding only a relative spread
$c/\sqrt{n_x}$.

\begin{figure}[ht]
\centering
\includegraphics[width=\columnwidth]{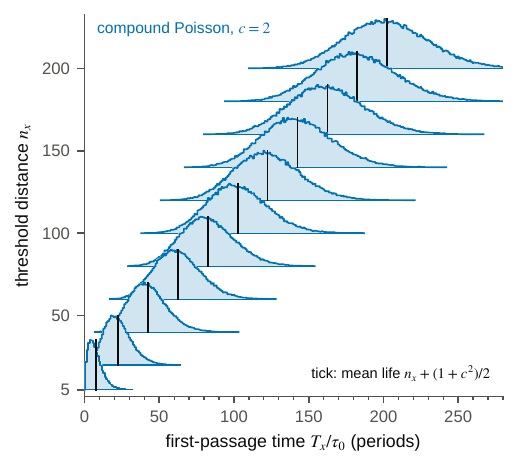}
\caption{Distribution of the remaining useful life for compound-Poisson
deposition at $c=2$ ($10^{5}$ paths, passage detected at integer sampling
periods), for thresholds $n_x=5,20,40,\dots,200$ increments away. Each ridge is
the probability mass of $T_x/\tau_0$, scaled to a common peak height and drawn
on a baseline at its $n_x$; the black tick marks the mean life
$n_x+(1+c^2)/2$, Eq.~\eqref{eq:first_passage_moments}. The distribution drifts
linearly with $n_x$, widens as $c\sqrt{n_x}$ and loses its right skew as
$(3c-\gamma_1)/\sqrt{n_x}$, Eq.~\eqref{eq:fpt_skew_main}.}
\label{fig:rul_distribution}
\end{figure}

The deposition law enters neither the mean nor the spread, but only the third
cumulant,
\begin{equation}
  \operatorname{skew}[T_x]=\frac{3c-\gamma_1}{\sqrt{n_x}},
  \label{eq:fpt_skew_main}
\end{equation}
so that the remaining life is positively skewed for the compound Poisson and
Gamma laws, whose increment skewness satisfies $\gamma_1<3c$, and symmetric to
this order for the inverse Gaussian law, whose increment skewness is exactly
$3c$ (Table~\ref{tab:three_laws}). Because the skew decays as $n_x^{-1/2}$
(Fig.~\ref{fig:rul_distribution}),
the law is visible in the life distribution only for a threshold a few
increments away. At a threshold $n_x=220$ 
increments away, the three laws agree on the mean
life to $0.2$ sampling periods and on every percentile from the $5$th to the $95$th to within 
$0.9\,\%$ (Supplemental Material, Sec.~S7); at $n_x=5$ the same percentiles spread by
up to three periods on a mean life of seven periods. For a distant
threshold the choice of law is immaterial, and what is not immaterial is $c$,
which sets the spread, and the uncertainty of $m$, which dominates it.

The asymptotic argument, overshoot qualification, and distinction
between continuous passage and passage first detected at discrete observation
times, together with the range of validity of
Eqs.~\eqref{eq:first_passage_moments}--\eqref{eq:fpt_skew_main} at small
$n_x$, are given in Supplemental Material, Sec.~S7. With removal, the process
is no longer monotone and Eq.~\eqref{eq:first_passage_identity} does not hold;
the passage problem must then be treated separately. Exact results are
nevertheless available in special cases: for constant-rate removal the
moments of the time to clear a deposit are known in closed form for compound
Poisson deposition \cite{Zacks2005}, and for infinite-activity
deposition the passage event can be simulated exactly
\cite{Chi2016,GonzalezCazares2023}.

The preceding results assume stationary operating conditions. A known,
deterministic variation in deposition intensity can be represented by the
operational clock
\begin{equation}
  \Lambda(t)=\int_0^t \chi(u)\,du,
  \label{eq:operational_clock}
\end{equation}
where $\chi(u)\geq0$ is a relative deposition intensity. Replacing $D(t)$ by
$D[\Lambda(t)]$ makes all cumulants scale with $\Lambda(t)$; in particular,
\begin{equation}
  \mathbb{E}\{D[\Lambda(t)]\}=m\Lambda(t).
  \label{eq:time_changed_mean}
\end{equation}
The stationary theory therefore remains valid on the operational clock.
Prediction in calendar time additionally requires future operating
conditions, whose uncertainty is distinct from the intrinsic deposition
variability represented by $c$ and $\alpha$.

\section{Implications and conclusions}
\label{sec:implications}
\label{sec:conclusions}

The classical linear and asymptotic fouling laws are universal first-moment
laws of stationary additive deposition with a finite mean rate, and this fixes
what a monitored fouling resistance can reveal. A linear or Kern--Seaton fit
returns a net deposition rate $m$ and, if the record saturates, a relaxation
rate $\psi$, or equivalently $R_\infty$ and $\psi^{-1}$; nothing more is in
the mean. A good fit is therefore not evidence that removal acts, nor of which
removal mechanism acts, nor of any deposition model, since every additive
mechanism with the same $m$ and $\psi$ fits equally well
(Sec.~\ref{sec:removal_summary}); the same caution applies to a best-fitting
blocking law in membrane
filtration~\cite{BoltonBoesch2006,CifuentesCabezas2023} and, once measurement
noise is included, to an increment law selected by
likelihood~\cite{ZhangRevie2016}. Even $m$ is a single combination: writing
deposition as transport to the wall followed by
attachment~\cite{PatersonFryer1988}, it constrains the product of a transport
coefficient, an attachment probability and the deposit density, not the
factors individually. How $m$ and $\psi$ depend on velocity, temperature and
composition remains the proper object of mechanistic modelling, but must be
read across operating points, not from the shape of one curve.

The information the fit discards is in the scatter about it, and for a
saturating record it can be read with three steps. First, subtract the
fitted mean curve. Second, estimate the variance of the residuals on the
plateau and subtract the measurement contribution $(\epsilon/U)^2$, with
$\epsilon$ from Eq.~\eqref{eq:epsilon_propagation}. Third, compare the excess
with the scales of Table~\ref{tab:plateau}. Continuous removal and
per-element detachment leave a plateau variance of order $\ell R_\infty$,
a relative scatter $\mathrm{CV}_\infty^2\sim\ell/R_\infty$ that is small
whenever the plateau holds many events. Sloughing leaves a floor of order
$\eta R_\infty^2$ that does not shrink as the deposition becomes smooth,
Eq.~\eqref{eq:sloughing_variance}. A plateau scatter well above the
measurement noise, and a fair fraction of $R_\infty$, therefore points to
intermittent removal; a scatter at the noise level is consistent with
continuous removal or with blocking. Blocking is told apart in time rather
than on the plateau: its variance peaks at half coverage and vanishes at
saturation (Sec.~\ref{sec:variance_removal}). The autocorrelation of the
residuals does not discriminate, since it decays as $e^{-\psi\tau}$ for all
three mechanisms of Table~\ref{tab:plateau}.

The deposition law itself, beyond its coefficient of variation $c$, cannot be
recovered from a thermally reconstructed resistance at realistic
measurement accuracy (Table~\ref{tab:record_length}); as
Sec.~\ref{sec:observability} showed, a better instrument helps far more
than a longer record, and a better-calibrated one not at all.

For a cleaning limit many sampling periods away, the time to reach it is set
by the mean rate, its uncertainty, and $c$ through the relative spread
$c/\sqrt{n_x}$, Eq.~\eqref{eq:first_passage_cv}; the deposition law enters
only through a skewness that fades with distance from the limit
(Sec.~\ref{sec:first_passage}). Schedules can therefore be planned on the
fitted rate, with an uncertainty budget dominated by that of $m$, and the
choice of law matters only when the limit is a few periods away.

Curve fitting and mechanism identification are thus different tasks: the
first is governed by universal mean laws, the second requires fluctuation
statistics, independent measurements, or observations across operating
conditions. The formulation is restricted to additive, non-decreasing
deposition with a removal rate that depends only on the current state;
processes that change the thermal resistance without changing the deposited
mass, such as ageing, consolidation or diffusion within wax
gels~\cite{Singh2000,Mehrotra2021}, and state-triggered detachment such as
spalling fall outside it. Extending the theory to these effects, and testing
the plateau-variance discriminator on records that visibly saturate, are the
natural next steps.

\section*{Acknowledgments}
This publication has been funded by the SFI NorwAI
(Centre for Research-based Innovation, 309834). The author gratefully
acknowledges the financial
support from the Research Council of Norway and the partners of the SFI NorwAI.

\section*{Data Availability}
Code producing figures and numerical checks is available at \texttt{github.com/\allowbreak simonecasolo/\allowbreak fouling-first-moment-laws}.

\clearpage
\setcounter{equation}{0}
\setcounter{figure}{0}
\setcounter{table}{0}
\setcounter{section}{0}
\renewcommand{\theequation}{S\arabic{equation}}
\renewcommand{\thesection}{S\arabic{section}}
\renewcommand{\thefigure}{S\arabic{figure}}
\renewcommand{\thetable}{S\arabic{table}}

\onecolumngrid
\begin{center}
{\large\bfseries Supplementary Material:\\[2pt]
Classical heat exchanger fouling-resistance laws as first-moment identities of additive stochastic deposition}\\[6pt]
Simone Casolo\\
{\small Cognite, Oslo, Norway}
\end{center}
\vspace{6pt}
\twocolumngrid

\section{Notation}
\label{app:notation}

The deposition $D(t)$ is a subordinator with L\'evy measure $\nu(\dd x)$ on
$(0,\infty)$, activity $r=\int\nu(\dd x)$ and cumulant rates
$q_n=\int_0^\infty x^n\,\nu(\dd x)$, so that $\kappa_n[D(t)]=q_nt$. Its law is
fixed by the Laplace exponent through
$\mathbb E[e^{-\theta D(t)}]=e^{-t\Phi(\theta)}$ with
\begin{equation}
  \Phi(\theta)=\int_0^\infty\bigl(1-e^{-\theta x}\bigr)\,\nu(\dd x),
  \qquad\theta\ge0 .
  \label{eq:S_lk}
\end{equation}
We write $m=q_1$ for the mean deposition rate, $s^2=q_2$ for the variance
rate, $c^2=s^2/(m^2\tau_0)$ for the squared coefficient of variation over a
reference window $\tau_0$, $t_c=s^2/m^2$ and $\ell=q_2/q_1=s^2/m=m\,t_c$ for
the size-biased event size. 

\section{Finite activity and the compound Poisson equivalence}
\label{app:cpp_equivalence}

This section proves the equivalence in Sec.~3.1: a
subordinator has finite activity, $r=\int\nu(dx)<\infty$, if and only if it
is a compound Poisson process~\cite{Sato1999,Applebaum2009}. 
It is worth separating the two directions:
finite activity is a statement about the L\'evy measure $\nu$, whereas
being a compound Poisson process is a statement about how the sample path
is assembled from discrete jumps. 

\paragraph*{From finite activity to a Poisson sum of jumps.}
Suppose $r<\infty$. Then $\nu$ splits into an overall rate and a shape,
\begin{equation}
  \nu(dx)=r\,F(dx),
\end{equation}
and $F$ is a probability distribution on $(0,\infty)$, since
$\int F(dx)=r^{-1}\int\nu(dx)=1$. This normalization step is only possible
because $r$ is finite.
 Substituting $\nu=rF$ into the L\'evy--Khintchine exponent, Eq.~\eqref{eq:S_lk}, 
\begin{equation}
  \Phi(\theta)=\int_0^\infty\bigl(1-e^{-\theta x}\bigr)\,r\,F(dx)
  =r\bigl[1-\widehat F(\theta)\bigr].
  \label{eq:cpp_phi}
\end{equation}
All dependence on the jump-size distribution is
now carried by $\widehat F$, where 
$\widehat F(\theta)=\int_0^\infty e^{-\theta x}\,F(dx)$ is the
Laplace transform of $F$.
Insert Eq.~\eqref{eq:cpp_phi} into $\mathbb E[e^{-\theta D(t)}]=e^{-t\Phi(\theta)}$
and expand the exponential in powers of $\widehat F(\theta)$:
\begin{equation}
  \mathbb E\bigl[e^{-\theta D(t)}\bigr]
  =e^{-rt}\,e^{\,rt\widehat F(\theta)}
  =\sum_{n=0}^\infty\underbrace{\frac{(rt)^n e^{-rt}}{n!}}_{\Pr[N(t)=n]}\,
    \widehat F(\theta)^n ,
  \label{eq:cpp_sum}
\end{equation}
where the coefficient of $\widehat F(\theta)^n$ is a
Poisson$(rt)$ probability. This is
exactly the Laplace transform one obtains by conditioning on the
number of jumps: for $N(t)\sim\mathrm{Poisson}(rt)$ and
$X_i\stackrel{\mathrm{iid}}{\sim}F$ independent of $N(t)$,
\begin{equation}
\begin{aligned}
  \mathbb E\Bigl[e^{-\theta\sum_{i=1}^{N(t)}X_i}\Bigr]
  &=\sum_{n=0}^\infty\Pr[N(t)=n]\,
     \mathbb E\Bigl[e^{-\theta\sum_{i=1}^nX_i}\Bigr]\\[2pt]
  &=\sum_{n=0}^\infty\Pr[N(t)=n]\,\widehat F(\theta)^n,
\end{aligned}
\end{equation}
which is term-by-term identical to Eq.~\eqref{eq:cpp_sum}. Since the
Laplace transform determines the law, $D(t)\stackrel{d}{=}\sum_{i=1}^{N(t)}X_i$:
$D$ is, by definition, a compound Poisson process of rate $r$ and jump law
$F=\nu/r$.

\paragraph*{From a Poisson sum of jumps back to finite activity.}
The converse needs no further calculation, only reading the definition in
the other direction. A compound Poisson process of rate $r$ and jump
law $F$ has a L\'evy--Khintchine representation with L\'evy measure
$\nu=r F$; by uniqueness of that representation, this is \emph{the}
L\'evy measure of the process, not merely one that reproduces its law.
Since $F$ is a probability measure, $\int\nu(dx)=r\int F(dx)=r<\infty$:
the process automatically has finite activity, equal to its jump rate.

Put together, the two directions say that the map
$\nu\leftrightarrow(r,F)$ is a bijection between L\'evy measures with
$r<\infty$ and (rate, jump-law) pairs: every finite-activity subordinator
is a compound Poisson process, and every compound Poisson process has
finite activity equal to its own rate. 

\section{Tempered-stable family: derivations}
\label{app:tempered}

This section carries out the five calculations that the tempered-stable
spectrum, Eq.~(13) of the main text, is built to support: fixing the two free scale parameters $B$ and $\lambda$ in terms of the physically meaningful triple $(m,c,\tau_0)$ at any shape $\alpha$; obtaining the
cumulant rates $q_n$, and from them the aggregation scaling of the main text; obtaining the Laplace exponent in
closed form; identifying the three named laws of Table~2 of the main text as three points on this family; and separating three notions---activity, path continuity, and observability---that are easy to conflate. Throughout, $\ell=q_2/q_1$ is the size-biased
event size of the main text: the typical event size as seen by the mass it deposits, rather than by a plain count of events. Only the first calculation, the matching relation Eq.~\eqref{eq:ts_matching}, is needed before any of the others, so a reader interested only in one named law can read that paragraph and then skip ahead.

\paragraph*{Cumulant rates, and matching to $(m,c,\tau_0)$.}
The spectrum $\nu(dx)$ gives the rate at which events of size $x$ occur; its moments $q_n=\int_0^\infty x^n\,\nu(dx)$ are therefore exactly the cumulant rates of the subordinator, with $q_1=m$ the mean deposition rate and $q_2$ the variance rate ($s^2=q_2$ in the notation of the main text). For the tempered-stable spectrum every such moment reduces to a single Gamma-function integral:
\begin{equation}
\begin{aligned}
  q_n&=\int_0^\infty x^n\nu(dx)
      =B\int_0^\infty x^{n-1-\alpha}e^{-\lambda x}\,dx\\[2pt]
     &=B\,\Gamma(n-\alpha)\,\lambda^{\alpha-n},
\end{aligned}
  \label{eq:ts_qn_general}
\end{equation}
valid for every $n\ge1$: the exponential tempering factor $e^{-\lambda x}$ ensures the integral converges at large $x$ regardless of $n$, and the restriction $\alpha<1$ ensures convergence at $x\to0$, which is what keeps even the first moment, $q_1=m$, finite.

Two of these moments are enough to fix both free parameters of the family, $B$ and $\lambda$. Imposing $q_1=m$ sets the overall deposition rate; imposing $q_2/q_1=\ell$ sets the size-biased event size rather than $q_2$ itself, because $\ell$ carries the same units as a jump size and is the more physically transparent quantity to specify. Using $\Gamma(2-\alpha)=(1-\alpha)\Gamma(1-\alpha)$ to simplify the resulting ratio, these two conditions solve to
\begin{equation}
  \lambda=\frac{1-\alpha}{\ell}=\frac{1-\alpha}{c^2m\tau_0},
  \qquad
  B=\frac{m\,\lambda^{1-\alpha}}{\Gamma(1-\alpha)}.
  \label{eq:ts_matching}
\end{equation}
Only two parameters are free in the family, so once $\lambda$ and $B$ are
fixed by $m$ and $\ell$ at a chosen $\alpha$, every higher $q_n$ is
determined by Eq.~\eqref{eq:ts_qn_general}; no further fitting is possible or needed, and $\alpha$ alone governs how the higher cumulants compare to the first two. This dependence can be made explicit without re-evaluating Gamma functions at every $n$. Taking the ratio of
consecutive cumulant rates in Eq.~\eqref{eq:ts_qn_general} gives
$q_n/q_{n-1}=(n-1-\alpha)/\lambda$, and substituting $\lambda=(1-\alpha)/\ell$
from Eq.~\eqref{eq:ts_matching} turns this into a one-step recursion, $q_n/q_{n-1}=(n-1-\alpha)\,\ell/(1-\alpha)$. Iterating the recursion down from $q_1=m$
telescopes it into the closed form
\begin{equation}
  q_n=m\,\ell^{\,n-1}\prod_{k=2}^{n-1}\frac{k-\alpha}{1-\alpha}
  \qquad(n\ge2);
  \label{eq:ts_qn_product}
\end{equation}
the product is empty (equal to $1$) for $n=2$, consistent with $q_2=m\ell$
by the definition of $\ell$ itself---a useful check on the algebra before
using the recursion for $n=3,4$ below.

\paragraph*{Skewness and kurtosis of the increments.}
The third and fourth cumulant rates are the ones most directly tied to data, since they set the skewness and excess kurtosis of the increment distribution over any observation window. Setting $n=3,4$ in
Equation~\eqref{eq:ts_qn_product} gives
\begin{equation}
  \frac{q_3}{q_2}=\frac{2-\alpha}{1-\alpha}\,\ell,
  \qquad
  \frac{q_4}{q_2}=\frac{(2-\alpha)(3-\alpha)}{(1-\alpha)^2}\,\ell^2 .
\end{equation}
These enter the standardized cumulants of a window $\Delta$,
$\gamma_1(\Delta)=q_3q_2^{-3/2}\Delta^{-1/2}$ and
$\gamma_2(\Delta)=q_4q_2^{-2}\Delta^{-1}$, only through the combination
$q_2\Delta=s^2\Delta$, the accumulated variance over the window; this is expected, since $\gamma_1$ and $\gamma_2$ are dimensionless cumulant ratios and $\Delta$ can only enter through how much variance has built up by the end of the window. Writing this out and using $c(\Delta)^2=\ell/(m\Delta)=s^2/(m^2\Delta)$
to trade $\ell$ and $\Delta$ for the single window-dependent coefficient of
variation $c(\Delta)$ reproduces the aggregation scaling of the main text
directly, i.e.\ it is this substitution that the main text's statement of how $\gamma_1$ and $\gamma_2$ shrink with the aggregation window rests on. A useful check that survives this substitution is the ratio
\begin{equation}
  \frac{\gamma_2(\Delta)}{\gamma_1(\Delta)^2}=\frac{q_4q_2}{q_3^{\,2}}=\frac{3-\alpha}{2-\alpha},
\end{equation}
which depends on $\alpha$ alone: $c$ and $\Delta$ cancel because both
$\gamma_1$ and $\gamma_2$ carry the same power of $c(\Delta)$ through
the aggregation scaling of the main text ($\gamma_1\propto c(\Delta)$,
$\gamma_2\propto c(\Delta)^2$), so this ratio is a pure shape diagnostic,
usable in principle to estimate $\alpha$ from data without first knowing $c$,
since it depends only on the shape of the increment distribution and not
on its overall scale.

\paragraph*{Laplace exponent.}
The Laplace exponent
\begin{equation}
  \Phi(\theta)=\int_0^\infty(1-e^{-\theta x})\,\nu(dx),
\end{equation}
already defined in the L\'evy--Khintchine formula, Eq.~\eqref{eq:S_lk}, is the generating function of the subordinator: it determines $\Delta D$'s distribution over any window through $\mathbb{E}[e^{-\theta\Delta D}]=e^{-\Delta\Phi(\theta)}$, and is the object matched below to identify the increment law of each named process. It cannot be obtained by integrating
$x^{-1-\alpha}e^{-\lambda x}$ alone: for $\alpha\ge0$ that integral diverges
at $x=0$, since a single power-law tail with $\alpha\ge0$ carries infinite
activity (infinitely many arbitrarily small events, Sec.~3.1 of the main
text). What converges is the \emph{difference} $1-e^{-\theta x}$ against
the spectrum, because this difference vanishes linearly as $x\to0$ and tames
the singularity that defeats direct integration. Since
$(1-e^{-\theta x})\,e^{-\lambda x}=e^{-\lambda x}-e^{-(\lambda+\theta)x}$, the
tempering already present in $\nu$ turns the integrand into a difference of
two exponentials, each divergent against $x^{-1-\alpha}$ at $x=0$ but with a
convergent difference. This reduces $\Phi$ to the one nontrivial integral
\begin{align}
  \int_0^\infty\bigl(e^{-\lambda x}-e^{-(\lambda+\theta)x}\bigr)\,x^{-1-\alpha}\,dx
  \notag\\
  =\Gamma(-\alpha)\bigl[\lambda^\alpha-(\lambda+\theta)^\alpha\bigr],
  \label{eq:ts_frullani}
\end{align}
a standard tempered-stable identity~\cite{Rosinski2007}, a generalization of
Frullani's integral to a power-law weight. Multiplying
Eq.~\eqref{eq:ts_frullani} by $B$ gives $\Phi(\theta)$ itself, so the only remaining step is to trade $B$ and
$\Gamma(-\alpha)$---both of which are awkward to evaluate directly for
general $\alpha$, the latter diverging at every nonnegative integer---for
the physically meaningful $m$ and $\lambda$ already fixed in
Eq.~\eqref{eq:ts_matching}. Using $\Gamma(1-\alpha)=-\alpha\,\Gamma(-\alpha)$
together with the matching relation $B\,\Gamma(1-\alpha)=m\lambda^{1-\alpha}$
from Eq.~\eqref{eq:ts_matching} eliminates $B$ and $\Gamma(-\alpha)$ in
favor of $m$ and $\lambda$ alone:
\begin{align}
  \Phi(\theta)&=\frac{m\lambda}{\alpha}
    \Bigl[\bigl(1+\theta/\lambda\bigr)^{\alpha}-1\Bigr],
    \qquad\alpha\neq0,
  \label{eq:ts_laplace_exponent}\\
  \Phi(\theta)&=m\lambda\,\ln\bigl(1+\theta/\lambda\bigr),
    \qquad\alpha=0,
  \label{eq:ts_laplace_exponent_gamma}
\end{align}
the $\alpha=0$ case following from the $\alpha\neq0$ one by L'H\^opital's
rule as $\alpha\to0$, since $[(1+\theta/\lambda)^\alpha-1]/\alpha\to\ln(1+\theta/\lambda)$
in that limit.

\begin{figure*}[t]
\centering
\includegraphics[width=\textwidth]{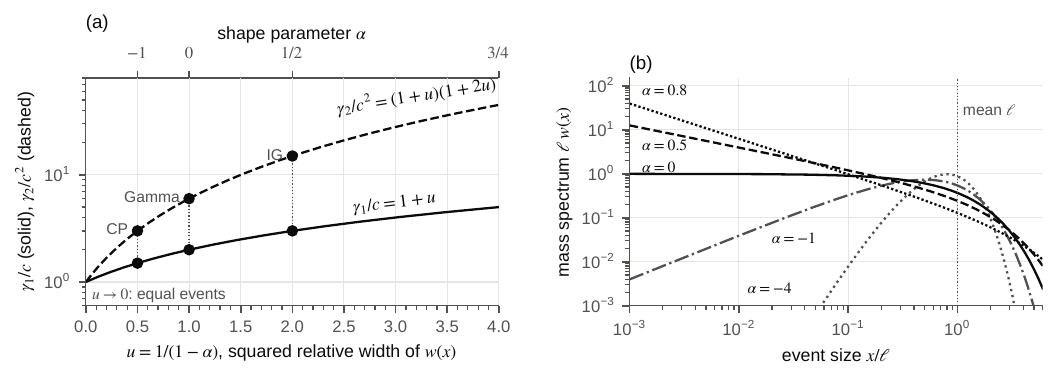}
\caption{The tempered-stable family as a continuum, at common $m$, $c$ and
$\tau_0$. (a)~Standardized skewness and excess kurtosis, in units of
$c(\Delta)$ and $c(\Delta)^2$, against $u=1/(1-\alpha)$ (top axis: $\alpha$):
$\gamma_1/c=1+u$, $\gamma_2/c^2=(1+u)(1+2u)$. Compound Poisson, Gamma and
inverse Gaussian ($u=\tfrac12,1,2$; markers) lie on these curves, and
$u\to0$ approaches the bound $\gamma_1\ge c$. (b)~Mass spectrum
$w(x)$ for five values of $\alpha$, all with mean $\ell$ (dotted): $\alpha$
redistributes the deposit between few large and many small events at fixed
mean and variance rate.}
\label{fig:S_alpha}
\end{figure*}

\paragraph*{Mass-weighted spectrum.}
The spectrum $\nu(x)$ weights every event by how often it occurs, not by
how much mass it carries, so a large rare event and a small common one can
have the same $\nu(x)$ while contributing very differently to the deposit.
To see the spectrum from the standpoint of mass rather than event count,
multiply the event-rate density by the size $x$ and normalize by the mean
rate $m$: this turns $\nu$ into a genuine probability density,
$w(x)=x\,\nu(x)/m$, giving the fraction of deposited mass that arrives in
events of size near $x$:
\begin{equation}
  w(x)=\frac{\lambda^{1-\alpha}}{\Gamma(1-\alpha)}\,x^{-\alpha}e^{-\lambda x},
\end{equation}
which is recognizable as the $\mathrm{Gamma}(1-\alpha,\lambda)$ density
--- shape $1-\alpha$, rate $\lambda$ --- with mean $(1-\alpha)/\lambda=\ell$
by construction (consistent with $\ell$ being defined as the size-biased,
i.e.\ mass-weighted, event size) and variance $(1-\alpha)/\lambda^2$, i.e.\
relative width (coefficient of variation) $(1-\alpha)^{-1/2}$. This relative
width is governed by $\alpha$ alone, independently of $m$ and $\ell$: the
mass spectrum narrows as $\alpha\to-\infty$, concentrating the deposit into
a few near-identical events, and broadens as $\alpha\to1$, spreading it over
an increasingly wide range of event sizes. This is the precise
sense in which $\alpha$ ``controls how the deposit is distributed over
event sizes'' in Sec.~3.2 of the main text.

\paragraph*{The three named laws.}
The tempered-stable family is a one-parameter continuum in $\alpha$, but
three particular values recover subordinators with well-known, independently
derived increment distributions. Evaluating
Eqs.~\eqref{eq:ts_matching}--\eqref{eq:ts_laplace_exponent_gamma}
at these three values of $\alpha$ recovers the three named processes,
identified by matching $\Phi$ above against their known Laplace exponents
term by term; once $\Phi$ matches, the increment law over any window
$\Delta$ follows immediately, since the Laplace exponent determines the
distribution uniquely. Beyond the algebra, the three laws correspond to
three qualitatively different pictures of how the deposit actually
accumulates, set by how far $\alpha$ has pushed the event-size spectrum
toward few-and-large or many-and-small events (Mass-weighted spectrum,
above); it is worth stating each picture in words before the matching
calculation that confirms it:
\begin{itemize}
  \item $\alpha=-1$ (compound Poisson). Physically, this is the law of a
    surface that sits idle most of the time and then, at random, unrelated
    instants, gains a lump of deposit of random size: nothing happens
    between events, and when something does happen, its size is drawn
    independently of everything else. Concretely, the waiting time between
    successive events is exponential (events occur at a constant rate, with
    no memory of how long it has already been quiet), and each event's size
    is itself exponentially distributed, independently of the waiting
    times and of all other events. It is this combination---a finite,
    constant rate of occurrence together with jumps that occur strictly one
    at a time---that produces the quiescent stretches visible in Fig.~2(d)
    of the main text: genuine flat intervals of the process, not merely
    small increments. Algebraically, substituting $\alpha=-1$ into
    Eq.~\eqref{eq:ts_matching} gives $\lambda=2/\ell$ and
    $B=m\lambda^2$, so by the finite-activity criterion of
    Sec.~3.1 of the main text the total event rate is
    $r=\int\nu(dx)=B/\lambda=2/t_c$: events of all sizes occur at a finite
    combined rate, as required for a list of discrete events to exist at
    all. This lets
    Sec.~\ref{app:cpp_equivalence} apply directly: the normalized
    jump law is $F(dx)=\nu(dx)/r=\lambda e^{-\lambda x}dx$, exponential of
    mean $1/\lambda=\ell/2$, so $\Delta D=\sum_{i=1}^{N}J_i$ with
    $N\sim\mathrm{Poisson}(r\Delta)$ and $J_i\stackrel{\mathrm{iid}}{\sim}\mathrm{Exp}(\mathrm{mean}=\ell/2)$,
    confirming in symbols the picture above: a Poisson-distributed number
    of events, each an independent draw from an exponential jump-size
    distribution. More generally, any
    $\alpha<0$---not only $\alpha=-1$---keeps $r<\infty$ and
    the same idle-then-jump picture, only with the jump size now
    $F$ a $\mathrm{Gamma}(\mathrm{shape}=-\alpha,\mathrm{rate}=\lambda)$-distributed
    draw rather than a plain exponential; writing $a=-\alpha$ for the shape, the aggregation scaling of the main text
    gives $\gamma_1/c=(a+2)/(a+1)$, which recovers $3/2$ at $a=1$
    ($\alpha=-1$) above.
  \item $\alpha=0$ (Gamma process). Physically, this is the opposite
    extreme from compound Poisson: instead of occasional large jumps
    separated by quiescent gaps, the surface receives a continuous rain of
    events so frequent and so small that no individual event can be
    pointed to, and no stretch of time is ever truly flat. The record
    still only ever increases (deposition cannot run backward), but it does
    so through an uninterrupted accumulation of infinitesimal contributions
    rather than through discrete, separately identifiable arrivals.
    Formally, $\lambda=1/\ell$ and $B=m\lambda$;
    the activity is infinite ($0\le\alpha<1$), so there is no finite list of
    discrete events to thin or enumerate---the process deposits through a
    dense set of arbitrarily small jumps rather than a countable sequence of
    them, which is the precise sense of ``continuous rain'' above. Even so, $\Phi(\theta)=m\lambda\ln(1+\theta/\lambda)$,
    Eq.~\eqref{eq:ts_laplace_exponent_gamma}, is exactly the Laplace
    exponent of a Gamma process of shape rate $B$ and scale $1/\lambda$,
    for which $\Delta D\sim\mathrm{Gamma}(\mathrm{shape}=B\Delta,\mathrm{scale}=1/\lambda)$:
    the match is read off by comparing this expression term by term against
    the textbook Laplace exponent of a Gamma process. Since $B\Delta=m\Delta/\ell=\Delta/t_c$ and $1/\lambda=\ell$, this is
    $\Delta D\sim\mathrm{Gamma}(\mathrm{shape}=\Delta/t_c,\ \mathrm{scale}=\ell)$.
  \item $\alpha=1/2$ (inverse Gaussian process). Physically, this law lies
    beyond the Gamma process: it shares its infinite activity,
    so there are no isolated events or quiescent gaps either, but its mass
    spectrum is broader (Mass-weighted spectrum, above), so a somewhat
    larger share of the deposit is carried by comparatively rarer,
    comparatively larger contributions riding on top of the same dense
    background of infinitesimal ones---a bursty, rather than perfectly
    smooth, continuous accumulation. (The inverse Gaussian process also has
    an independent, well-known construction as the first-passage time of a
    drifting Brownian motion through a fixed level, which is a separate
    fact about the law and not how it is derived here.) Formally,
    $\lambda=1/(2\ell)$
    and, from Eq.~\eqref{eq:ts_matching} with $\Gamma(1/2)=\sqrt\pi$,
    $B=m\sqrt{\lambda/\pi}$; by the same term-by-term matching,
    $\Phi(\theta)=2m\lambda[(1+\theta/\lambda)^{1/2}-1]$
    is exactly the Laplace exponent of an inverse Gaussian process, for
    which $\Delta D\sim\mathrm{IG}(\mu_\Delta=m\Delta,\lambda_{\mathrm{IG}}=m\Delta^2/t_c)$.
    As a consistency check, its variance $\mu_\Delta^3/\lambda_{\mathrm{IG}}=c^2m^2\tau_0\Delta$
    agrees with the variance rate of the main text, as it must since both
    were fixed by the same matching condition $q_2/q_1=\ell$.
\end{itemize}
Figure~\ref{fig:S_alpha} places the three
laws on one continuous family, showing both how their skewness and
kurtosis compare (panel a) and how the underlying mass spectrum $w(x)$
reshapes itself as $\alpha$ sweeps between them (panel b). Panel (a) shows
$\gamma_1/c$ approaching, but never crossing, the value $1$ as the mass
spectrum degenerates toward equal-sized events ($u\to0$, i.e.\
$\alpha\to-\infty$, where every event carries the same size). This
floor is not special to the tempered-stable family: it is a general
property of every subordinator, as Sec.~\ref{app:skew_bound} shows.

\paragraph*{Activity, path continuity and observability.}
\label{app:three_distinctions}
The results above are easy to over-interpret, because they blur together
three properties of the family that are logically independent and should
be kept apart.

\emph{Activity} is the event count: finite for $\alpha<0$, where the
process can be written as a genuine list of discrete jumps as in the
compound Poisson case above, and infinite for $\alpha\ge0$, where it
cannot. \emph{Path continuity} is a separate matter and does not track
activity: every member of the family, finite or infinite activity alike, is
a pure-jump process, so the Gamma and inverse-Gaussian paths---despite
having infinite activity---are not continuous but strictly increasing
through a dense set of arbitrarily small jumps. In particular, the Gamma
process is a continuum limit only in this measure-theoretic sense of
infinitely many infinitesimal jumps, and not in the sense of a deposit
thickening smoothly and uniformly the way a continuous function would; no
member of the family has continuous sample paths. \emph{Observability} is a
third, and independent, question again: whether any of the first two
properties can actually be resolved from a measured record depends on the
sampling interval and the measurement noise, and for an area-averaged
thermal resistance it cannot (main text, Sec.~5), regardless of the true
activity or path structure of the underlying process. The quiescent
stretches that distinguish compound Poisson in Fig.~2(d) of the main text
illustrate this: they are a statement about the model's event count
(activity), not a prediction that a measured record will show matching
flat intervals, since the measurement itself may not resolve them.

\subsection{Skewness bound}
\label{app:skew_bound}

For any subordinator, the Cauchy--Schwarz inequality
$q_2^2\le q_1q_3$ gives $\gamma_1(\Delta)\ge c(\Delta)$, with equality only
if all events have the same size (approached as $\alpha\to-\infty$ in the family above):
no matter how the event-size spectrum is shaped, the skewness of the
increments can never fall below the coefficient of variation itself.

\emph{Proof.}
With the window cumulants $\kappa_n=q_n\Delta$,
\begin{equation}
\begin{aligned}
  c(\Delta)&=\frac{\sqrt{q_2\Delta}}{q_1\Delta},\qquad
  \gamma_1(\Delta)=\frac{q_3\Delta}{(q_2\Delta)^{3/2}},\\
  \frac{\gamma_1(\Delta)}{c(\Delta)}&=\frac{q_1q_3}{q_2^{2}},
\end{aligned}
  \label{eq:S_skew_ratio}
\end{equation}
so the ratio is independent of the window and depends on the event-size
spectrum alone. Writing $x^2=x^{1/2}x^{3/2}$ and applying the
Cauchy--Schwarz inequality with respect to $\nu$,
\begin{equation}
\begin{aligned}
  q_2&=\int_0^\infty x^{1/2}x^{3/2}\,\nu(dx)\\
  &\le\Bigl(\int_0^\infty x\,\nu(dx)\Bigr)^{1/2}
     \Bigl(\int_0^\infty x^{3}\,\nu(dx)\Bigr)^{1/2}
  =\sqrt{q_1q_3},
\end{aligned}
\end{equation}
which is $\gamma_1\ge c$. Equality requires $x^{1/2}\propto x^{3/2}$ on the
support of $\nu$, i.e.\ a single event size: compound Poisson with fixed
jumps attains $\gamma_1=c$, and any spread of sizes raises $\gamma_1/c$
above $1$. In the tempered-stable family $\gamma_1/c=1+u$, i.e.\ $1.5$, $2$
and $3$ for the compound Poisson, Gamma and inverse Gaussian laws, and the
bound is approached only as $u\to0$ [Fig.~\ref{fig:S_alpha}(a)].

Physically, a deposit that grows only by adding non-negative pieces must
have increments at least as right-skewed as they are dispersed. A Gaussian
increment model, with zero skewness and finite spread, is therefore never
an exact additive law, only its large-window limit, in which both
$\gamma_1$ and $c$ vanish as $\Delta^{-1/2}$. Conversely, measured
increments whose skewness falls clearly below their coefficient of
variation cannot come from additive deposition alone: measurement noise,
removal or a non-additive mechanism must contribute.

\section{Removal: derivations}
\label{app:removal}

\label{app:linear_removal}
\paragraph*{Cumulants of the L\'evy-driven Ornstein--Uhlenbeck process.}
For a L\'evy process $D$ and a deterministic kernel $f$, the cumulants of a
stochastic integral against $D$ are
\begin{equation}
  \kappa_n\!\biggl[\int_0^tf(u)\,dD(u)\biggr]=q_n\int_0^tf(u)^n\,du.
\end{equation}
With $f(u)=e^{-\psi(t-u)}$ this gives
\begin{equation}
  \kappa_n[R(t)]=q_n\int_0^te^{-n\psi(t-u)}du=q_n\,\frac{1-e^{-n\psi t}}{n\psi},
\end{equation}
Eq.~(18) of the main text; the deterministic initial condition contributes
only to the mean. At stationarity, $\kappa_n=q_n/(n\psi)$.

\paragraph*{Stationary plateau.}
The mean obeys $d\mu/dt=m-\psi\mu$ (Table~1 of the main text), so
$\mu_\infty=R_\infty=m/\psi$. Taking $n=2$ above,
\begin{equation}
  \operatorname{Var}[R]_\infty=\kappa_2=\frac{q_2}{2\psi}=\frac{s^2}{2\psi},
\end{equation}
since $q_2=s^2$. The plateau accordingly contains
\begin{equation}
  N_\infty=\frac{R_\infty}{\ell}=\frac{m/\psi}{s^2/m}=\frac{m^2}{\psi s^2}=\frac1{\psi t_c}
\end{equation}
size-biased events (using $t_c=s^2/m^2=\ell/m$), and
\begin{equation*}
  \mathrm{CV}_\infty^2
  =\frac{\operatorname{Var}[R]_\infty}{R_\infty^2}
  =\frac{s^2/2\psi}{(m/\psi)^2}
  =\frac{\psi s^2}{2m^2}
  =\frac{\psi t_c}{2}
  =\frac1{2N_\infty},
\end{equation*}
the plateau relations of the main text. The shape of the plateau follows the
same route:
\begin{equation}
\begin{aligned}
  \gamma_{1,\infty}&=\frac{\kappa_3}{\kappa_2^{3/2}}
  =\frac{q_3/3\psi}{(q_2/2\psi)^{3/2}}
  =\frac{2\sqrt2}{3}\,\psi^{1/2}q_3q_2^{-3/2}\\
  &=\frac{2\sqrt2}{3}\,\gamma_1(\psi^{-1}),
\end{aligned}
\end{equation}
and $\gamma_{2,\infty}=\kappa_4/\kappa_2^2=\psi\,q_4q_2^{-2}=\gamma_2(\psi^{-1})$.
Using $\gamma_1(\Delta)=[(2-\alpha)/(1-\alpha)]c(\Delta)$,
$\gamma_2(\Delta)=[(2-\alpha)(3-\alpha)/(1-\alpha)^2]c(\Delta)^2$ and
$c(\psi^{-1})=\sqrt2\,\mathrm{CV}_\infty$ for the tempered-stable family gives
\begin{align*}
  \gamma_{1,\infty}&=\frac43\,\frac{2-\alpha}{1-\alpha}\,\mathrm{CV}_\infty,\\
  \gamma_{2,\infty}&=2\,\frac{(2-\alpha)(3-\alpha)}{(1-\alpha)^2}\,\mathrm{CV}_\infty^2,
\end{align*}
the plateau skewness and kurtosis of the main text.

\paragraph*{Per-element detachment.}
Write the deposition as $D(t)=\int_0^t\!\int_0^\infty x\,N_D(du,dx)$, with
$N_D$ a Poisson random measure of intensity $du\,\nu(dx)$, and mark each atom
with an independent lifetime $\tau\sim\mathrm{Exp}(\psi)$. The marked measure
$N(du,dx,d\tau)$ is again Poisson, with intensity
$du\,\nu(dx)\,\psi e^{-\psi\tau}d\tau$, and the deposit present at time $t$ is
$R(t)=\int\!\int\!\int x\,\mathbf 1_{\{t-u<\tau\}}\,N(du,dx,d\tau)$: each
event leaves as a single intact unit at the end of its lifetime. By Campbell's
theorem~\cite{Campbell1909} the cumulants of a Poisson integral are the integrals of the $n$th
power of the integrand against the intensity,
\begin{equation*}
\begin{aligned}
  \kappa_n[R(t)]&=\int_0^t\!du\int_0^\infty\!x^n\nu(dx)\,\Pr(\tau>t-u)\\
  &=q_n\int_0^te^{-\psi(t-u)}du=q_n\,\frac{1-e^{-\psi t}}{\psi},
\end{aligned}
\end{equation*}
which holds for finite and infinite activity alike, since only the moments
$q_n=\int x^n\nu(dx)$, $n\ge1$, enter. At stationarity
$\kappa_n=q_n/\psi$, $n$ times the Ornstein--Uhlenbeck value $q_n/(n\psi)$.
The difference lies in the integrand: under proportional erosion an event of
size $x$ contributes $x\,e^{-\psi(t-u)}$, whose $n$th power integrates to
$1/(n\psi)$, whereas under per-element detachment it contributes $x$ or nothing,
and $\mathbf 1^n=\mathbf 1$ integrates to $1/\psi$. An event made of many
sub-units detaching independently contributes, by the law of large numbers,
its surviving fraction $x\,e^{-\psi(t-u)}$, and so returns the
Ornstein--Uhlenbeck cumulants. The plateau shape follows as before:
\begin{equation}
\begin{aligned}
  \gamma_{1,\infty}&=\frac{q_3/\psi}{(q_2/\psi)^{3/2}}=\psi^{1/2}q_3q_2^{-3/2}=\gamma_1(\psi^{-1}),\\
  \gamma_{2,\infty}&=\psi q_4q_2^{-2}=\gamma_2(\psi^{-1});
\end{aligned}
\end{equation}
with $\mathrm{CV}_\infty^2=\psi q_2/m^2=\psi t_c=c(\psi^{-1})^2$ these are
\begin{equation*}
\begin{aligned}
  \gamma_{1,\infty}&=\frac{2-\alpha}{1-\alpha}\,\mathrm{CV}_\infty,\\
  \gamma_{2,\infty}&=\frac{(2-\alpha)(3-\alpha)}{(1-\alpha)^2}\,\mathrm{CV}_\infty^2.
\end{aligned}
\end{equation*}

\paragraph*{Autocorrelation.}
For continuous, per-element and intermittent removal alike the mean equation closes, and
it holds conditionally on the state at an earlier time:
\begin{equation}
  \mathbb E[R(t+\tau)\,|\,R(t)]=R_\infty+(R(t)-R_\infty)e^{-\psi\tau},\qquad\tau>0.
\end{equation}
Multiplying by $R(t)$ and averaging gives
\begin{equation}
  \operatorname{Cov}[R(t),R(t+\tau)]=\operatorname{Var}[R]\,e^{-\psi\tau}.
\end{equation}

\begin{figure*}[t]
\centering
\includegraphics[width=\textwidth]{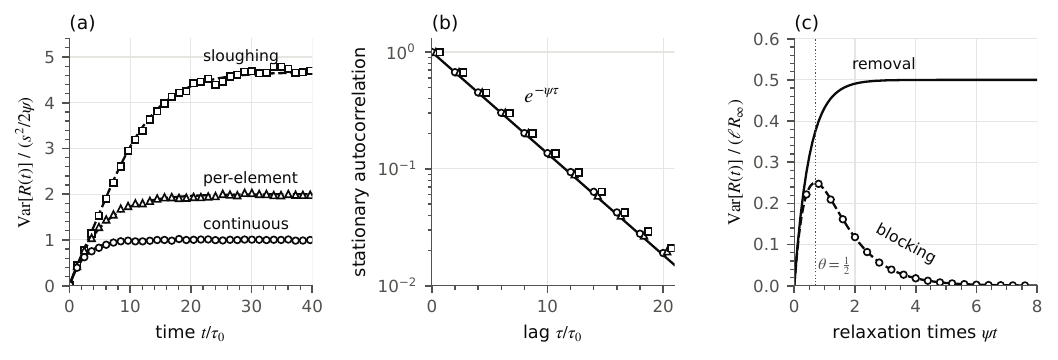}
\caption{Removal mechanisms with the same mean $R_\infty=m/\psi$ and
relaxation rate $\psi$ ($m=c=\tau_0=1$, $\psi=0.2$, Gamma deposition; lines
exact, markers Monte Carlo). (a)~Transient variance, normalized by the
continuous-removal plateau $s^2/2\psi$: continuous removal [Eq.~(18) of the
main text]; per-element detachment, $q_2(1-e^{-\psi t})/\psi$, saturating at
twice that value and half the rate; and sloughing ($\eta=\tfrac12$),
Eqs.~\eqref{eq:S_slough_m1}--\eqref{eq:S_slough_m2}, whose plateau carries a deposition-independent floor.
(b)~All three share
the stationary autocorrelation $e^{-\psi\tau}$, so neither mean nor
correlation time discriminates them. (c)~Independent-site blocking vs.\
continuous removal at the same $R_\infty,\psi$, for equal events of size
$\ell$: removal variance rises monotonically to $\ell R_\infty/2$, while
blocking variance $\ell R_\infty\theta(1-\theta)$ peaks at half coverage and
vanishes at saturation (markers: binomial site filling).}
\label{fig:S_removal}
\end{figure*}

\label{app:sloughing}
\paragraph*{Moments under sloughing.}
Applying the generator of the sloughing equation of the main text to $f(R)=R$ and $f(R)=R^2$,
\begin{align}
  \frac{d\mu_1}{dt}&=m-r_s\eta\,\mu_1,\label{eq:S_slough_m1}\\
  \frac{d\mu_2}{dt}&=2m\mu_1+s^2-r_s\eta(2-\eta)\,\mu_2,\label{eq:S_slough_m2}
\end{align}
where the deposition contributes $2R\,m+s^2$ and an event changes $R^2$ by
$-\eta(2-\eta)R^2$. At stationarity $\mu_1=m/(r_s\eta)=R_\infty$ and
$\mu_2=(2m\mu_1+s^2)/[r_s\eta(2-\eta)]$, so that
\begin{equation}
\begin{aligned}
  \operatorname{Var}&=\mu_2-\mu_1^2
   =\frac{2R_\infty^2}{2-\eta}+\frac{s^2}{(2-\eta)\psi}-R_\infty^2\\[2pt]
  &=\frac{s^2}{(2-\eta)\psi}+\frac{\eta}{2-\eta}R_\infty^2,
\end{aligned}
\end{equation}
with $\psi=r_s\eta$, which is the sloughing variance of the main text.
The transient moments of the three removal mechanisms, and of site
blocking, are compared in Fig.~\ref{fig:S_removal}.

\begin{figure*}[t]
\centering
\includegraphics[width=\textwidth]{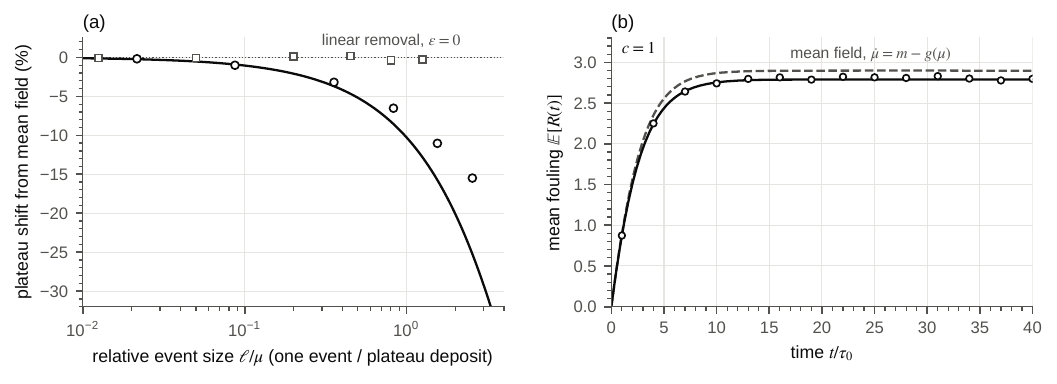}
\caption{Nonlinear removal, $g(R)=\psi R+\varepsilon R^2$ ($m=\tau_0=1$,
$\psi=0.2$, $\varepsilon=0.05$, Gamma deposition). (a)~Shift of the
stationary mean from the mean-field plateau $m=g(\mu)$, against the relative
event size $\ell/\mu$, varied through $c$: second-order closure, Eq.~(26) of
the main text (line), vs.\ Monte Carlo (circles). The closure captures the
shift while the plateau holds many events, and overestimates it once a
single event is comparable to the deposit ($\ell/\mu\gtrsim0.3$), having
dropped $-g''\kappa_3[R]$. Under linear removal ($\varepsilon=0$, squares)
the plateau does not move: only there is the mean universal. (b)~Mean
trajectory at $c=1$: fluctuations lower the plateau from $2.90$ (mean field,
dashed) to $2.81$ (Monte Carlo), against $2.79$ from the closure (solid).}
\label{fig:S_nonlinear}
\end{figure*}

\label{app:nonlinear_closure}
\paragraph*{Nonlinear removal.}
Starting from $dR=dD-g(R)\,dt$, the mean and the variance obey
\begin{equation}
  \frac{d\mu}{dt}=m-\mathbb E[g(R)],
  \qquad
  \frac{dV}{dt}=s^2-2\operatorname{Cov}[R,g(R)].
\end{equation}
Expanding $g(R)=g(\mu)+g'(\mu)\delta R+\tfrac12g''(\mu)\delta R^2+\dots$ around
the mean gives, to the order kept by the main text's second-order closure,
\begin{equation}
\begin{aligned}
  \mathbb E[g]&=g(\mu)+\tfrac12g''(\mu)V+O(\kappa_3),\\
  \operatorname{Cov}[R,g(R)]&=g'(\mu)V+O(\kappa_3);
\end{aligned}
\end{equation}
the neglected terms contain the third cumulant and carry the first
dependence on $\alpha$. Explicitly, the next order is
\begin{equation}
\begin{aligned}
  \operatorname{Cov}[R,g(R)]&=g'(\mu)V\\
  &\quad+\tfrac12g''(\mu)\kappa_3[R]+\dots,\\
  \mathbb E[g]&=g(\mu)+\tfrac12g''(\mu)V\\
  &\quad+\tfrac16g'''(\mu)\kappa_3[R]+\dots.
\end{aligned}
\end{equation}
The closure is ordered in $1/N_\infty=\ell/R_\infty$: near the plateau
$V=O(\ell R_\infty)$ and $\kappa_3[R]=O(\ell^2R_\infty)$, i.e.\
$\kappa_n[R]/\mu^n=O(N_\infty^{1-n})$, and with $g^{(j)}\sim g/\mu^{j}$ each
$\kappa_3$ term is smaller than the retained $V$ term of the same equation by
a factor $O(1/N_\infty)$. Dropping $g''\kappa_3$ in the variance equation
while keeping $g''V$ in the mean equation is therefore consistent at leading
order in $1/N_\infty$, although both terms carry $g''$.
Figure~\ref{fig:S_nonlinear} compares the closure with simulation.

\paragraph*{Numerical checks.}
Monte Carlo simulations ($m=c=\tau_0=1$, $\psi=0.2$, $2\times10^4$ paths,
run for $T=60=12\psi^{-1}$ to reach the stationary state) reproduce the stationary skewness of the plateau of the
main text for the three laws ($0.63$, $0.82$,
$1.27$ against $0.63$, $0.84$, $1.27$), the stationary variances of continuous removal, per-element detachment and
sloughing ($2.52$, $5.02$ and $11.61$ against $2.50$, $5.00$ and $11.67$ for
$\eta=0.5$, Gamma deposition, Fig.~\ref{fig:S_removal}; per-element
detachment is simulated exactly by thinning the alive deposit, and the
residuals are within sampling error), the exponential autocorrelation ($0.366$ against
$e^{-1}=0.368$ at $\tau=5$), and the nonlinear plateau of Fig.~\ref{fig:S_nonlinear}.

\section{Granular shot noise and the coefficient of variation}
\label{app:shot_noise}

This section derives the granular estimate of the coefficient of variation
quoted in Sec.~5.2 of the main text, and the area scaling that follows from
it.

\paragraph*{The monodisperse limit of the family.}
In the tempered-stable family, Eq.~(13) of the main text, the mass-weighted spectrum
has relative width $(1-\alpha)^{-1/2}$, so as $\alpha\to-\infty$ it collapses
onto a single size: deposition becomes a compound Poisson process in which
every event deposits the same amount $J_1$. Writing $r$ for the event rate,
the cumulant rates are $q_n=r\,J_1^{\,n}$, so that
\begin{equation}
  m=q_1=r J_1,\qquad s^2=q_2=r J_1^{2},\qquad
  \ell=\frac{q_2}{q_1}=J_1 ,
  \label{eq:S_mono}
\end{equation}
the size-biased event size coinciding with the jump itself, as it must when
all jumps are equal. Substituting $\ell=J_1$ into the squared coefficient of
variation over a window $\Delta$, $c(\Delta)^2=\ell/(m\Delta)$, and evaluating
it at the reference window $\Delta=\tau_0$,
\begin{equation}
  c^{2}=\frac{J_1}{m\,\tau_0}=\frac{1}{r\,\tau_0},
  \label{eq:S_cv_mono}
\end{equation}
one over the mean number of attachments per reference window. This is the
classical shot-noise result for a superposition of Poisson-timed pulses of
equal weight~\cite{Rice1944,Rice1945}; here it is not imported but obtained as
the $\alpha\to-\infty$ member of the family of the main text.

\paragraph*{The size of a single attachment.}
It remains to express $J_1$ for one attached particle. A sphere of diameter
$d_p$ has volume $\pi d_p^{3}/6$. Deposited into a porous layer of solid
fraction $\phi_s$ it occupies a layer volume $(\pi d_p^{3}/6)/\phi_s$;
spread over the heat-transfer area $A$ this raises the deposit thickness by
$\delta_1=(\pi d_p^{3}/6)/(A\phi_s)$, and with $R=\delta/k_f$ the
fouling resistance by
\begin{equation}
  J_1=\frac{\delta_1}{k_f}=\frac{\pi d_p^{3}/6}{A\,\phi_s\,k_f} .
  \label{eq:S_J1}
\end{equation}
Equations~\eqref{eq:S_cv_mono} and \eqref{eq:S_J1} together are the estimate
used in the main text. Nothing in it is adjustable: once $d_p$ and the deposit
properties are fixed, the observed mean rate fixes the event frequency through
$r=m/J_1$.

\paragraph*{Area scaling.}
Because $J_1\propto A^{-1}$ at fixed particle size, Eq.~\eqref{eq:S_cv_mono}
gives $c\propto A^{-1/2}$ at fixed mean rate: granular arrival noise in an
area-averaged observable is suppressed by the square root of the averaging
area. Setting $c=1$ identifies the area on which one sampling period deposits
a single event,
\begin{equation}
  A_1=\frac{\pi d_p^{3}/6}{\phi_s\,k_f\,m\,\tau_0} .
  \label{eq:S_area}
\end{equation}
For $d_p=1~\mu\mathrm{m}$, $k_f=0.5~\mathrm{W\,m^{-1}K^{-1}}$,
$\phi_s=0.5$ and a mean rate of order
$10^{-5}~\mathrm{m^{2}\,K\,W^{-1}}$ per sampling period,
$A_1\approx2\times10^{-13}~\mathrm{m^{2}}$, about a quarter of the
cross-section of the particle itself. That is the sense in which an
area-averaged fouling resistance is deterministic with respect to the
granularity of deposition: order-unity fluctuations would require averaging
over a surface no larger than one grain.

\section{Record length for law identification}
\label{app:record_length}

This section derives the noise attenuation of the standardized cumulants and
the record-length estimate quoted in the main text. Write
\begin{equation}
  \rho=\frac{2\sigma^2}{c^2m^2\tau_0^{2}}
  \label{eq:S_rho}
\end{equation}
for the noise-to-signal ratio. The observed increment over $k$
periods is the sum of the deposition increment, with cumulants
$\kappa_n[\Delta D]=q_nk\tau_0$, and two independent Gaussian measurement
errors of variance $\sigma^2$ each, which contribute only to $\kappa_2$.
Writing $\kappa_2[\Delta y]=\kappa_2[\Delta D](1+\rho/k)$ and substituting
into the standardized cumulants gives
\begin{equation}
  \gamma_{1,k}=\gamma_1\,\frac{k}{(k+\rho)^{3/2}},\qquad
  \gamma_{2,k}=\gamma_2\,\frac{k}{(k+\rho)^{2}}.
  \label{eq:S_atten}
\end{equation}

\begin{figure}[t]
\centering
\includegraphics[width=\columnwidth]{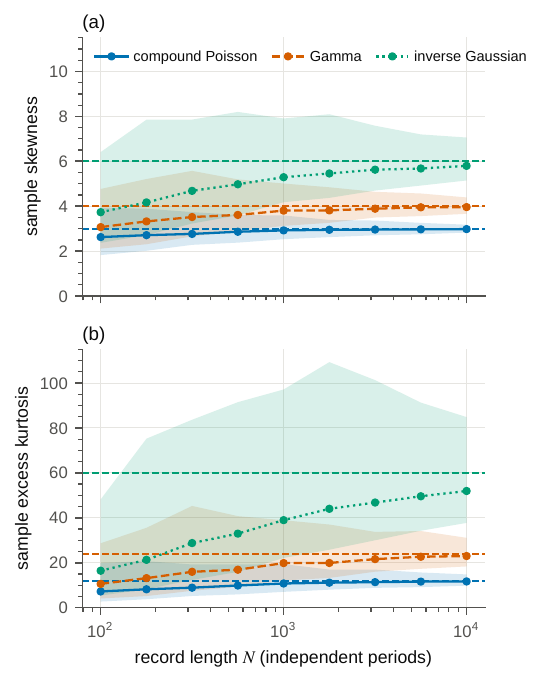}
\caption{Recovering the microscopic law from skewness and kurtosis:
finite-record convergence of the sample estimators, at $m$, $c=2$, $\tau_0$,
noise-free. For $N$ independent periods (300 Monte Carlo records per $N$),
markers/lines are the median sample skewness (a) and excess kurtosis (b),
shaded bands the $5$--$95\,\%$ range, dashed lines the exact values of
Table~2 of the main text. Both converge onto the correct, law-specific
value within a feasible record length ($N\sim10^2$--$10^3$), given clean
data; for an area-averaged thermal measurement, however, the same
estimators are attenuated by Eq.~\eqref{eq:S_atten}, so the required record
grows to the lengths estimated in Sec.~5.4 of the main text.}
\label{fig:S_recovery}
\end{figure}

The windows that best \emph{retain} the cumulants maximize
Eq.~\eqref{eq:S_atten} over $k$. For the skewness,
\begin{equation}
  \frac{d}{dk}\,\frac{k}{(k+\rho)^{3/2}}
  =\frac{(k+\rho)-\tfrac32k}{(k+\rho)^{5/2}}=0
  \quad\Rightarrow\quad k^\ast=2\rho,
\end{equation}
\begin{equation}
\max_k\gamma_{1,k}=\gamma_1\,\frac{2\rho}{(3\rho)^{3/2}}=(2/3\sqrt3)\,\frac{\gamma_1}{\sqrt\rho}.
\end{equation}
For the excess kurtosis,
$(k+\rho)-2k=0$ gives $k^\ast=\rho$ and
$\max_k\gamma_{2,k}=\gamma_2/(4\rho)$. 
Noise-free, the sample estimators numerically
converge on the law-specific values (Fig.~\ref{fig:S_recovery}).

\begin{figure}[t]
\centering
\includegraphics[width=\columnwidth]{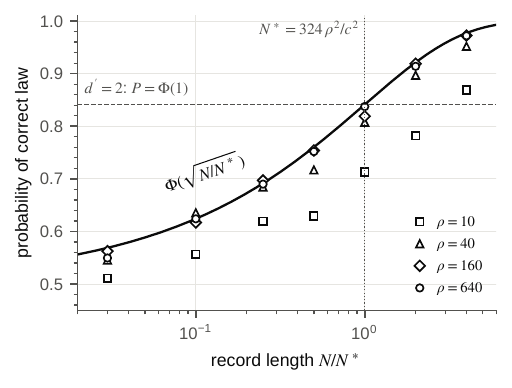}
\caption{Monte Carlo test of the record-length estimate,
Eq.~\eqref{eq:Nreq_general}, for the Gamma and inverse-Gaussian laws at $c=2$.
Each record of $N$ base periods is observed with Gaussian noise set by
$\rho$, Eq.~\eqref{eq:S_rho}, and differenced over $N/2k$ windows of
$k=\rho/2$ periods; the law is assigned by comparing sample skewness with the
midpoint of the two attenuated values, Eq.~\eqref{eq:S_atten}. Markers:
fraction of correct assignments in $1000$ records per law. Line: asymptotic
prediction $\Phi_{\mathrm G}(\sqrt{N/N^\ast})$, with $\Phi_{\mathrm G}$ the standard normal distribution function, i.e.\ $d'=2$ at $N=N^\ast=324\rho^2/c^2$,
which holds for $\rho\gtrsim10^2$ and is optimistic below (about $1.3\times$
longer needed at $\rho=40$, about $3\times$ at $\rho=10$, where $\rho$ is no
longer large compared with $c^2=4$).}
\label{fig:S_record}
\end{figure}

Detection requires a second ingredient---the sampling error of the
estimator---and therefore selects a different window. Let two laws differ,
before attenuation, by $\Delta\gamma_1$ or $\Delta\gamma_2$. Each window is
formed from a disjoint pair of observations $k$ periods apart and so consumes
$2k$ periods, giving $n_w=N/(2k)$ independent windows; the standard errors of
the sample skewness and sample excess kurtosis under approximate normality are
$\sqrt{6/n_w}$ and $\sqrt{24/n_w}$. We measure the separation of two
laws in units of the standard error of a \emph{single} estimate: the record is
tested against the value expected under the alternative law, so
$d'=\Delta\gamma_{1,k}/\sqrt{6/n_w}$ and
$d'=\Delta\gamma_{2,k}/\sqrt{24/n_w}$. Comparing two independent records,
the standard error of their difference is larger by $\sqrt2$ and $N$ doubles.
With the attenuated differences
$\Delta\gamma_{1,k}=\Delta\gamma_1\,k/(k+\rho)^{3/2}$ and
$\Delta\gamma_{2,k}=\Delta\gamma_2\,k/(k+\rho)^{2}$ of
Eq.~\eqref{eq:S_atten}, the separability indices are
\begin{equation}
\begin{aligned}
  d'_1&=\Delta\gamma_1\sqrt{\frac{N}{12}}\,\frac{k^{1/2}}{(k+\rho)^{3/2}},\\
  d'_2&=\Delta\gamma_2\sqrt{\frac{N}{48}}\,\frac{k^{1/2}}{(k+\rho)^{2}}.
\end{aligned}
  \label{eq:dprime}
\end{equation}
Setting the derivative of the $k$-dependent factor to zero gives
$\tfrac12(k+\rho)-\tfrac32k=0$, i.e.\ $k=\rho/2$, for the skewness, and
$\tfrac12(k+\rho)-2k=0$, i.e.\ $k=\rho/3$, for the excess kurtosis---both
shorter than $k^\ast$, because a longer window is also a rarer one. At these
windows the factors equal $(2/3\sqrt3)\,\rho^{-1}$ and
$(3\sqrt3/16)\,\rho^{-3/2}$, so that $d'=2$ requires
$N=12\cdot4\cdot(27/4)\,\rho^2/\Delta\gamma_1^2$ and
$N=48\cdot4\cdot(256/27)\,\rho^3/\Delta\gamma_2^2$, that is,
\begin{equation}
  N\gtrsim 324\,\frac{\rho^{2}}{\Delta\gamma_1^{2}},
  \qquad
  N\gtrsim 1820\,\frac{\rho^{3}}{\Delta\gamma_2^{2}}.
  \label{eq:Nreq_general}
\end{equation}
For the Gamma/inverse-Gaussian pair, $\Delta\gamma_1=c$ and
$\Delta\gamma_2=9c^2$ (Sec.~\ref{app:tempered}), giving
$N\approx324\,\rho^2/c^2$ and $N\approx22\,\rho^3/c^4$; the skewness is
therefore the cheaper statistic by a factor $\rho/(14c^{2})$, and it sets the
scaling of the requirement with the noise. Separating windows by one period
instead of $k$ gives $n_w\simeq N/k$ and halves both prefactors without
changing the powers of $\rho$. The estimates are asymptotic in $\rho$:
Fig.~\ref{fig:S_record} tests them by simulation and finds them accurate to
within a factor $1$--$2$ for $\rho\gtrsim10^{2}$ and optimistic below. They
are also specific to moment-based estimation, whereas the attenuation,
Eq.~\eqref{eq:S_atten}, is a property of the measurement and applies to any
estimator.
\section{First-passage time: exact distribution and asymptotics}
\label{app:fpt}

\paragraph*{Exact distribution.} Because $R(t)=R_0+D(t)$ is nondecreasing,
$\{T_x\le t\}=\{D(t)\ge x-R_0\}$, and $\Pr(T_x\le t)$ follows directly from the
law of $D(t)$ given in the main text. For the Gamma law,
$D(t)\sim\mathrm{Gamma}(t/(c^2\tau_0),\,c^2m\tau_0)$ and
$\Pr(T_x\le t)=Q\bigl(t/(c^2\tau_0),\,(x-R_0)/(c^2m\tau_0)\bigr)$, the
regularized upper incomplete gamma function. For the compound Poisson law,
conditioning on the number of events $N(t)=n$ gives a sum of $n$
exponential jumps, i.e.\ an Erlang$(n,\ell/2)$ variable, so
$\Pr(D(t)\ge x-R_0\mid N(t)=n)=Q(n,2(x-R_0)/\ell)$; weighting by the
Poisson probabilities of $N(t)$ and summing over $n\ge1$ gives
$\Pr(T_x\le t)$ in closed form. For the inverse Gaussian law,
$D(t)\sim\mathrm{IG}(\mu_t,\lambda_t)$ with $\mu_t=mt$,
$\lambda_t=mt^2/(c^2\tau_0)$, and the standard inverse-Gaussian distribution
function gives $\Pr(T_x\le t)=1-F_{\mathrm{IG}}(x-R_0;\mu_t,\lambda_t)$. All
three reduce, for $x-R_0$ large compared with the event scale $\ell$, to the
normal approximation $\Phi_{\mathrm G}\bigl[(mt-(x-R_0))/(cm\sqrt{\tau_0t})\bigr]$.

\begin{figure*}[t]
\centering
\includegraphics[width=\textwidth]{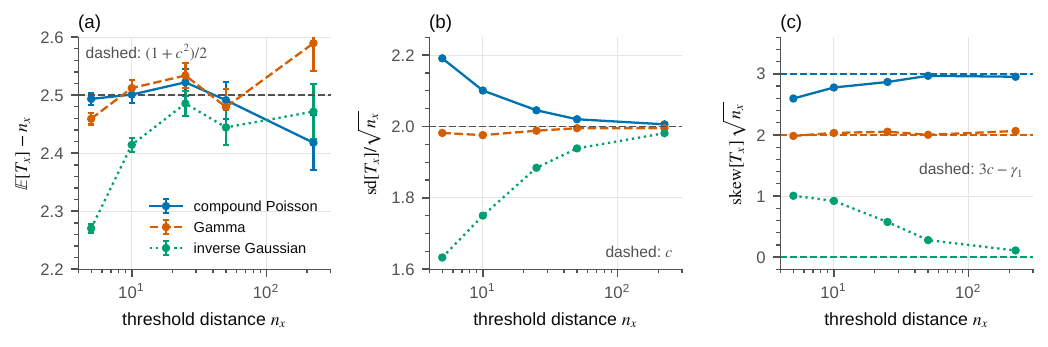}
\caption{Range of validity of the first-passage formulas, by Monte Carlo at
$c=2$ for the three matched laws. (a) $\mathbb E[T_x]-n_x$ vs.\ the predicted
$(1+c^2)/2$ (dashed; error bars $\pm1$ Monte Carlo s.e.). (b)
$\operatorname{sd}[T_x]/\sqrt{n_x}$ vs.\ predicted $c$. (c)
$\operatorname{skew}[T_x]\sqrt{n_x}$ vs.\ predicted $3c-\gamma_1$ (dashed;
zero for the inverse Gaussian law). Accurate at $n_x=220$, degrading towards
small $n_x$, the only regime where the deposition law would matter.}
\label{fig:S_fpt_validation}
\end{figure*}

\paragraph*{Asymptotic cumulants.} Write the deposition as a sum of
one-period increments $X_1,X_2,\dots$ with cumulant generating function
$K(\theta)=\log\mathbb E\,e^{\theta X_1}$, mean $\kappa_1=m\tau_0$, variance
$\kappa_2=c^2m^2\tau_0^2$ and third cumulant
$\kappa_3=\gamma_1c^3m^3\tau_0^3$, and let $T=T_x$ be the number of periods
at which the partial sum $S_T=\sum_{i\le T}X_i$ first reaches
$x-R_0$, with overshoot $O=S_T-(x-R_0)\ge0$. For $\theta$ in a neighbourhood
of $0$ where $K$ is finite, $M_n=\exp[\theta S_n-nK(\theta)]$ is a
martingale: since the increments are independent,
$\mathbb E[M_{n+1}\mid X_1,\dots,X_n]=M_n\,\mathbb E[e^{\theta X_{n+1}}]\,e^{-K(\theta)}=M_n$,
and $\mathbb E M_n=M_0=1$. The passage time $T$ is finite with exponential
moments for a subordinator of positive mean, so optional stopping applies and
$\mathbb E M_T=1$; with $S_T=(x-R_0)+O$ this reads
\begin{equation}
  \mathbb E\bigl[e^{\theta(x-R_0)+\theta O-TK(\theta)}\bigr]=1 .
\end{equation}
Choosing $\theta=\theta(u)$ to solve $K(\theta)=-u$ turns $-TK(\theta)$ into
$uT$, and moving the constant factor $e^{\theta(x-R_0)}$ to the right-hand
side gives Wald's identity
\begin{equation}
  \mathbb E\bigl[e^{uT+\theta(u)O}\bigr]=e^{-\theta(u)(x-R_0)}.
  \label{eq:wald_fpt}
\end{equation}
For a distant threshold the overshoot becomes asymptotically independent of
$T$ and converges to the stationary overshoot of renewal theory, with mean
$\mathbb E[O_\infty]=\mathbb E[X_1^2]/(2\kappa_1)=(\kappa_2+\kappa_1^2)/(2\kappa_1)$.
The expectation in Eq.~\eqref{eq:wald_fpt} then factorizes, and its logarithm
is $\log\mathbb E\,e^{uT}=-\theta(u)(x-R_0)-\log\mathbb E\,e^{\theta(u)O_\infty}$.
Since $\theta(u)=-u/\kappa_1+O(u^2)$, the last term is
$-\theta(u)\,\mathbb E[O_\infty]+O(u^2)=u\,\mathbb E[O_\infty]/\kappa_1+O(u^2)$,
so that
\begin{equation}
  \log\mathbb E\,e^{uT}
  =-\theta(u)\,(x-R_0)+\frac{u\,\mathbb E[O_\infty]}{\kappa_1}+O(u^2).
  \label{eq:cgf_fpt}
\end{equation}
Inverting $K(\theta)=-u$ order by order in $u$ (matching
$K(\theta)=\kappa_1\theta+\tfrac12\kappa_2\theta^2+\tfrac16\kappa_3\theta^3+\dots=-u$)
and reading off the coefficients of $u^n/n!$ in
$-\theta(u)(x-R_0)$ gives the cumulants of $T$ to leading order in
$x-R_0$:
\begin{equation}
\begin{aligned}
  \kappa_1(T)&=\frac{x-R_0}{\kappa_1}+O(1),\\[2pt]
  \kappa_2(T)&=\frac{(x-R_0)\,\kappa_2}{\kappa_1^3}+O(1),\\[2pt]
  \kappa_3(T)&=(x-R_0)\Bigl(\frac{3\kappa_2^2}{\kappa_1^5}
              -\frac{\kappa_3}{\kappa_1^4}\Bigr)+O(1).
\end{aligned}
\label{eq:fpt_cumulants}
\end{equation}
Substituting $\kappa_1=m\tau_0$, $\kappa_2=c^2m^2\tau_0^2$,
$\kappa_3=\gamma_1c^3m^3\tau_0^3$ and $n_x=(x-R_0)/(m\tau_0)$ reproduces the
expressions quoted in the main text for the mean and the coefficient of
variation of $T_x$ (in units of $\tau_0$),
and gives for the skewness
\begin{equation}
  \operatorname{skew}[T_x]=\frac{3c-\gamma_1}{\sqrt{n_x}},
  \label{eq:fpt_skewness}
\end{equation}
which is positive whenever $\gamma_1<3c$ (true for the compound Poisson and
Gamma laws at matched $(m,c)$, Sec.~\ref{app:tempered}) and vanishes for
the inverse Gaussian law, whose skewness is exactly $3c$. The $O(1)$
remainders in Eq.~\eqref{eq:fpt_cumulants} come from the overshoot and do
not grow with the distance to the threshold, which is why the leading-order
formulas improve as $n_x$ increases and are least reliable for $n_x$ of
order $10$ or smaller, the regime flagged in the main text.

\paragraph*{Range of validity.} Monte Carlo over $n_x\in\{5,10,25,50,220\}$
(Fig.~\ref{fig:S_fpt_validation}) recovers the mean to within $0.09$ sampling
periods for $n_x\ge10$, but at $n_x=5$ the inverse Gaussian law is biased by
$-0.23$ periods, its spread is $18\,\%$ below $c\sqrt{n_x}$, and its
predicted zero skewness is still $0.45$. At $n_x=220$ the three laws agree
on the mean life to $0.2$ periods and on the $5$th--$95$th percentiles to
$0.9\,\%$. For a threshold within a handful of
increments the life distribution should be simulated rather than taken from
Eqs.~\eqref{eq:fpt_cumulants}--\eqref{eq:fpt_skewness}.


%


\begin{thebibliography}{98}%
\makeatletter
\providecommand \@ifxundefined [1]{%
 \@ifx{#1\undefined}
}%
\providecommand \@ifnum [1]{%
 \ifnum #1\expandafter \@firstoftwo
 \else \expandafter \@secondoftwo
 \fi
}%
\providecommand \@ifx [1]{%
 \ifx #1\expandafter \@firstoftwo
 \else \expandafter \@secondoftwo
 \fi
}%
\providecommand \natexlab [1]{#1}%
\providecommand \enquote  [1]{``#1''}%
\providecommand \bibnamefont  [1]{#1}%
\providecommand \bibfnamefont [1]{#1}%
\providecommand \citenamefont [1]{#1}%
\providecommand \href@noop [0]{\@secondoftwo}%
\providecommand \href [0]{\begingroup \@sanitize@url \@href}%
\providecommand \@href[1]{\@@startlink{#1}\@@href}%
\providecommand \@@href[1]{\endgroup#1\@@endlink}%
\providecommand \@sanitize@url [0]{\catcode `\\12\catcode `\$12\catcode
  `\&12\catcode `\#12\catcode `\^12\catcode `\_12\catcode `\%12\relax}%
\providecommand \@@startlink[1]{}%
\providecommand \@@endlink[0]{}%
\providecommand \url  [0]{\begingroup\@sanitize@url \@url }%
\providecommand \@url [1]{\endgroup\@href {#1}{\urlprefix }}%
\providecommand \urlprefix  [0]{URL }%
\providecommand \Eprint [0]{\href }%
\providecommand \doibase [0]{https://doi.org/}%
\providecommand \selectlanguage [0]{\@gobble}%
\providecommand \bibinfo  [0]{\@secondoftwo}%
\providecommand \bibfield  [0]{\@secondoftwo}%
\providecommand \translation [1]{[#1]}%
\providecommand \BibitemOpen [0]{}%
\providecommand \bibitemStop [0]{}%
\providecommand \bibitemNoStop [0]{.\EOS\space}%
\providecommand \EOS [0]{\spacefactor3000\relax}%
\providecommand \BibitemShut  [1]{\csname bibitem#1\endcsname}%
\let\auto@bib@innerbib\@empty
\bibitem [{\citenamefont {Epstein}(1983)}]{Epstein1983}%
  \BibitemOpen
  \bibfield  {author} {\bibinfo {author} {\bibfnamefont {N.}~\bibnamefont
  {Epstein}},\ }\href {https://doi.org/10.1080/01457638108939594} {\bibfield
  {journal} {\bibinfo  {journal} {Heat Transfer Engineering}\ }\textbf
  {\bibinfo {volume} {4}},\ \bibinfo {pages} {43} (\bibinfo {year}
  {1983})}\BibitemShut {NoStop}%
\bibitem [{\citenamefont {Bott}(1995)}]{Bott1995}%
  \BibitemOpen
  \bibfield  {author} {\bibinfo {author} {\bibfnamefont {T.~R.}\ \bibnamefont
  {Bott}},\ }\href@noop {} {\emph {\bibinfo {title} {Fouling of Heat
  Exchangers}}},\ \bibinfo {series} {Chemical Engineering Monographs},
  Vol.~\bibinfo {volume} {26}\ (\bibinfo  {publisher} {Elsevier},\ \bibinfo
  {address} {Amsterdam},\ \bibinfo {year} {1995})\BibitemShut {NoStop}%
\bibitem [{\citenamefont {M{\"u}ller-Steinhagen}(2011)}]{MullerSteinhagen2011}%
  \BibitemOpen
  \bibfield  {author} {\bibinfo {author} {\bibfnamefont {H.}~\bibnamefont
  {M{\"u}ller-Steinhagen}},\ }\href
  {https://doi.org/10.1080/01457632.2010.505127} {\bibfield  {journal}
  {\bibinfo  {journal} {Heat Transfer Engineering}\ }\textbf {\bibinfo {volume}
  {32}},\ \bibinfo {pages} {1} (\bibinfo {year} {2011})}\BibitemShut {NoStop}%
\bibitem [{\citenamefont {Ishiyama}\ \emph
  {et~al.}(2011{\natexlab{a}})\citenamefont {Ishiyama}, \citenamefont
  {Paterson},\ and\ \citenamefont {Ian~Wilson}}]{IshiyamaCleaning2011}%
  \BibitemOpen
  \bibfield  {author} {\bibinfo {author} {\bibfnamefont {E.~M.}\ \bibnamefont
  {Ishiyama}}, \bibinfo {author} {\bibfnamefont {W.~R.}\ \bibnamefont
  {Paterson}},\ and\ \bibinfo {author} {\bibfnamefont {D.}~\bibnamefont
  {Ian~Wilson}},\ }\href {https://doi.org/10.1016/j.ces.2010.10.036} {\bibfield
   {journal} {\bibinfo  {journal} {Chemical Engineering Science}\ }\textbf
  {\bibinfo {volume} {66}},\ \bibinfo {pages} {604} (\bibinfo {year}
  {2011}{\natexlab{a}})}\BibitemShut {NoStop}%
\bibitem [{\citenamefont {Pogiatzis}\ \emph {et~al.}(2012)\citenamefont
  {Pogiatzis}, \citenamefont {Ishiyama}, \citenamefont {Paterson},
  \citenamefont {Vassiliadis},\ and\ \citenamefont {Wilson}}]{Pogiatzis2012}%
  \BibitemOpen
  \bibfield  {author} {\bibinfo {author} {\bibfnamefont {T.}~\bibnamefont
  {Pogiatzis}}, \bibinfo {author} {\bibfnamefont {E.~M.}\ \bibnamefont
  {Ishiyama}}, \bibinfo {author} {\bibfnamefont {W.~R.}\ \bibnamefont
  {Paterson}}, \bibinfo {author} {\bibfnamefont {V.~S.}\ \bibnamefont
  {Vassiliadis}},\ and\ \bibinfo {author} {\bibfnamefont {D.~I.}\ \bibnamefont
  {Wilson}},\ }\href {https://doi.org/10.1016/j.apenergy.2011.01.063}
  {\bibfield  {journal} {\bibinfo  {journal} {Applied Energy}\ }\textbf
  {\bibinfo {volume} {89}},\ \bibinfo {pages} {60} (\bibinfo {year}
  {2012})}\BibitemShut {NoStop}%
\bibitem [{\citenamefont {Tatara}\ and\ \citenamefont
  {Lupia}(2011)}]{Tatara2011}%
  \BibitemOpen
  \bibfield  {author} {\bibinfo {author} {\bibfnamefont {R.~A.}\ \bibnamefont
  {Tatara}}\ and\ \bibinfo {author} {\bibfnamefont {G.~M.}\ \bibnamefont
  {Lupia}},\ }\bibfield  {journal} {\bibinfo  {journal} {International Journal
  of Engineering, Science and Technology}\ }\textbf {\bibinfo {volume} {3}},\
  \href {https://doi.org/10.4314/ijest.v3i8.1} {10.4314/ijest.v3i8.1} (\bibinfo
  {year} {2011})\BibitemShut {NoStop}%
\bibitem [{\citenamefont {Shen}\ \emph {et~al.}(2015)\citenamefont {Shen},
  \citenamefont {Cirone},\ and\ \citenamefont {Wang}}]{Shen2015}%
  \BibitemOpen
  \bibfield  {author} {\bibinfo {author} {\bibfnamefont {C.}~\bibnamefont
  {Shen}}, \bibinfo {author} {\bibfnamefont {C.}~\bibnamefont {Cirone}},\ and\
  \bibinfo {author} {\bibfnamefont {X.}~\bibnamefont {Wang}},\ }\href
  {https://doi.org/10.1016/j.applthermaleng.2015.03.028} {\bibfield  {journal}
  {\bibinfo  {journal} {Applied Thermal Engineering}\ }\textbf {\bibinfo
  {volume} {85}},\ \bibinfo {pages} {148} (\bibinfo {year} {2015})}\BibitemShut
  {NoStop}%
\bibitem [{\citenamefont {Taborek}\ \emph {et~al.}(1972)\citenamefont
  {Taborek}, \citenamefont {Aoki}, \citenamefont {Ritter}, \citenamefont
  {Palen},\ and\ \citenamefont {Knudsen}}]{Taborek1972a}%
  \BibitemOpen
  \bibfield  {author} {\bibinfo {author} {\bibfnamefont {J.}~\bibnamefont
  {Taborek}}, \bibinfo {author} {\bibfnamefont {T.}~\bibnamefont {Aoki}},
  \bibinfo {author} {\bibfnamefont {R.~B.}\ \bibnamefont {Ritter}}, \bibinfo
  {author} {\bibfnamefont {J.~W.}\ \bibnamefont {Palen}},\ and\ \bibinfo
  {author} {\bibfnamefont {J.~G.}\ \bibnamefont {Knudsen}},\ }\href@noop {}
  {\bibfield  {journal} {\bibinfo  {journal} {Chem. Eng. Prog.}\ }\textbf
  {\bibinfo {volume} {68}},\ \bibinfo {pages} {59} (\bibinfo {year}
  {1972})}\BibitemShut {NoStop}%
\bibitem [{\citenamefont {Somerscales}(1990)}]{Somerscales1990}%
  \BibitemOpen
  \bibfield  {author} {\bibinfo {author} {\bibfnamefont {E.~F.~C.}\
  \bibnamefont {Somerscales}},\ }\href
  {https://doi.org/10.1080/01457639008939720} {\bibfield  {journal} {\bibinfo
  {journal} {Heat Transfer Engineering}\ }\textbf {\bibinfo {volume} {11}},\
  \bibinfo {pages} {19} (\bibinfo {year} {1990})}\BibitemShut {NoStop}%
\bibitem [{\citenamefont {Augustin}\ \emph {et~al.}(2025)\citenamefont
  {Augustin}, \citenamefont {Melo},\ and\ \citenamefont
  {Wilson}}]{AugustinMeloWilson2025}%
  \BibitemOpen
  \bibfield  {author} {\bibinfo {author} {\bibfnamefont {W.}~\bibnamefont
  {Augustin}}, \bibinfo {author} {\bibfnamefont {L.~F.}\ \bibnamefont {Melo}},\
  and\ \bibinfo {author} {\bibfnamefont {D.~I.}\ \bibnamefont {Wilson}},\
  }\href {https://doi.org/10.1007/s00231-025-03595-z} {\bibfield  {journal}
  {\bibinfo  {journal} {Heat and Mass Transfer}\ }\textbf {\bibinfo {volume}
  {61}},\ \bibinfo {pages} {72} (\bibinfo {year} {2025})}\BibitemShut {NoStop}%
\bibitem [{\citenamefont {Kern}\ and\ \citenamefont
  {Seaton}(1959)}]{KernSeaton1959}%
  \BibitemOpen
  \bibfield  {author} {\bibinfo {author} {\bibfnamefont {D.~Q.}\ \bibnamefont
  {Kern}}\ and\ \bibinfo {author} {\bibfnamefont {R.~E.}\ \bibnamefont
  {Seaton}},\ }\href@noop {} {\bibfield  {journal} {\bibinfo  {journal}
  {British Chemical Engineering}\ }\textbf {\bibinfo {volume} {4}},\ \bibinfo
  {pages} {258} (\bibinfo {year} {1959})}\BibitemShut {NoStop}%
\bibitem [{\citenamefont {Ebert}\ and\ \citenamefont
  {Panchal}(1997)}]{EbertPanchal1997}%
  \BibitemOpen
  \bibfield  {author} {\bibinfo {author} {\bibfnamefont {W.~A.}\ \bibnamefont
  {Ebert}}\ and\ \bibinfo {author} {\bibfnamefont {C.~B.}\ \bibnamefont
  {Panchal}},\ }in\ \href@noop {} {\emph {\bibinfo {booktitle} {Fouling
  Mitigation of Industrial Heat-Exchange Equipment}}},\ \bibinfo {editor}
  {edited by\ \bibinfo {editor} {\bibfnamefont {T.~R.}\ \bibnamefont {Bott}},
  \bibinfo {editor} {\bibfnamefont {L.~F.}\ \bibnamefont {Melo}}, \bibinfo
  {editor} {\bibfnamefont {C.~B.}\ \bibnamefont {Panchal}},\ and\ \bibinfo
  {editor} {\bibfnamefont {E.~F.~C.}\ \bibnamefont {Somerscales}}}\ (\bibinfo
  {publisher} {Begell House},\ \bibinfo {address} {New York},\ \bibinfo {year}
  {1997})\ pp.\ \bibinfo {pages} {451--460}\BibitemShut {NoStop}%
\bibitem [{\citenamefont {Yao}\ \emph {et~al.}(1971)\citenamefont {Yao},
  \citenamefont {Habibian},\ and\ \citenamefont {O’Melia}}]{Yao1971}%
  \BibitemOpen
  \bibfield  {author} {\bibinfo {author} {\bibfnamefont {K.-M.}\ \bibnamefont
  {Yao}}, \bibinfo {author} {\bibfnamefont {M.~T.}\ \bibnamefont {Habibian}},\
  and\ \bibinfo {author} {\bibfnamefont {C.~R.}\ \bibnamefont {O’Melia}},\
  }\href {https://doi.org/10.1021/es60058a005} {\bibfield  {journal} {\bibinfo
  {journal} {Environmental Science \& Technology}\ }\textbf {\bibinfo {volume}
  {5}},\ \bibinfo {pages} {1105} (\bibinfo {year} {1971})}\BibitemShut
  {NoStop}%
\bibitem [{\citenamefont {Epstein}(1988)}]{Epstein1988}%
  \BibitemOpen
  \bibfield  {author} {\bibinfo {author} {\bibfnamefont {N.}~\bibnamefont
  {Epstein}},\ }\bibinfo {title} {Particulate fouling of heat transfer
  surfaces: Mechanisms and models},\ in\ \href
  {https://doi.org/10.1007/978-94-009-2813-8_10} {\emph {\bibinfo {booktitle}
  {Fouling Science and Technology}}}\ (\bibinfo  {publisher} {Springer
  Netherlands},\ \bibinfo {year} {1988})\ pp.\ \bibinfo {pages}
  {143--164}\BibitemShut {NoStop}%
\bibitem [{\citenamefont {Bohnet}(1987)}]{Bohnet1987}%
  \BibitemOpen
  \bibfield  {author} {\bibinfo {author} {\bibfnamefont {M.}~\bibnamefont
  {Bohnet}},\ }\href {https://doi.org/10.1002/ceat.270100115} {\bibfield
  {journal} {\bibinfo  {journal} {Chemical Engineering \& Technology}\ }\textbf
  {\bibinfo {volume} {10}},\ \bibinfo {pages} {113} (\bibinfo {year}
  {1987})}\BibitemShut {NoStop}%
\bibitem [{\citenamefont {Kostoglou}\ and\ \citenamefont
  {Karabelas}(1998)}]{Kostoglou1998}%
  \BibitemOpen
  \bibfield  {author} {\bibinfo {author} {\bibfnamefont {M.}~\bibnamefont
  {Kostoglou}}\ and\ \bibinfo {author} {\bibfnamefont {A.~J.}\ \bibnamefont
  {Karabelas}},\ }\href {https://doi.org/10.1021/ie970559g} {\bibfield
  {journal} {\bibinfo  {journal} {Industrial \& Engineering Chemistry
  Research}\ }\textbf {\bibinfo {volume} {37}},\ \bibinfo {pages} {1536}
  (\bibinfo {year} {1998})}\BibitemShut {NoStop}%
\bibitem [{\citenamefont {Brahim}\ \emph {et~al.}(2003)\citenamefont {Brahim},
  \citenamefont {Augustin},\ and\ \citenamefont {Bohnet}}]{Brahim2003}%
  \BibitemOpen
  \bibfield  {author} {\bibinfo {author} {\bibfnamefont {F.}~\bibnamefont
  {Brahim}}, \bibinfo {author} {\bibfnamefont {W.}~\bibnamefont {Augustin}},\
  and\ \bibinfo {author} {\bibfnamefont {M.}~\bibnamefont {Bohnet}},\ }\href
  {https://doi.org/10.1016/s1290-0729(02)00021-2} {\bibfield  {journal}
  {\bibinfo  {journal} {International Journal of Thermal Sciences}\ }\textbf
  {\bibinfo {volume} {42}},\ \bibinfo {pages} {323} (\bibinfo {year}
  {2003})}\BibitemShut {NoStop}%
\bibitem [{\citenamefont {Han}\ \emph {et~al.}(2019)\citenamefont {Han},
  \citenamefont {Xu},\ and\ \citenamefont {Yu}}]{Han2019}%
  \BibitemOpen
  \bibfield  {author} {\bibinfo {author} {\bibfnamefont {Z.}~\bibnamefont
  {Han}}, \bibinfo {author} {\bibfnamefont {Z.}~\bibnamefont {Xu}},\ and\
  \bibinfo {author} {\bibfnamefont {X.}~\bibnamefont {Yu}},\ }\href
  {https://doi.org/10.1016/j.ijheatmasstransfer.2019.07.078} {\bibfield
  {journal} {\bibinfo  {journal} {International Journal of Heat and Mass
  Transfer}\ }\textbf {\bibinfo {volume} {144}},\ \bibinfo {pages} {118428}
  (\bibinfo {year} {2019})}\BibitemShut {NoStop}%
\bibitem [{\citenamefont {Maddahi}\ \emph {et~al.}(2019)\citenamefont
  {Maddahi}, \citenamefont {Hatamipour},\ and\ \citenamefont
  {Jamialahmadi}}]{Maddahi2019}%
  \BibitemOpen
  \bibfield  {author} {\bibinfo {author} {\bibfnamefont {M.}~\bibnamefont
  {Maddahi}}, \bibinfo {author} {\bibfnamefont {M.}~\bibnamefont
  {Hatamipour}},\ and\ \bibinfo {author} {\bibfnamefont {M.}~\bibnamefont
  {Jamialahmadi}},\ }\href {https://doi.org/10.1016/j.ijthermalsci.2019.106017}
  {\bibfield  {journal} {\bibinfo  {journal} {International Journal of Thermal
  Sciences}\ }\textbf {\bibinfo {volume} {145}},\ \bibinfo {pages} {106017}
  (\bibinfo {year} {2019})}\BibitemShut {NoStop}%
\bibitem [{\citenamefont {Akbarzadeh}\ and\ \citenamefont
  {Zougari}(2008)}]{Akbarzadeh2008}%
  \BibitemOpen
  \bibfield  {author} {\bibinfo {author} {\bibfnamefont {K.}~\bibnamefont
  {Akbarzadeh}}\ and\ \bibinfo {author} {\bibfnamefont {M.}~\bibnamefont
  {Zougari}},\ }\href {https://doi.org/10.1021/ie0711325} {\bibfield  {journal}
  {\bibinfo  {journal} {Industrial \& Engineering Chemistry Research}\ }\textbf
  {\bibinfo {volume} {47}},\ \bibinfo {pages} {953} (\bibinfo {year}
  {2008})}\BibitemShut {NoStop}%
\bibitem [{\citenamefont {Watkinson}\ and\ \citenamefont
  {Wilson}(1997)}]{WatkinsonWilson1997}%
  \BibitemOpen
  \bibfield  {author} {\bibinfo {author} {\bibfnamefont {A.}~\bibnamefont
  {Watkinson}}\ and\ \bibinfo {author} {\bibfnamefont {D.}~\bibnamefont
  {Wilson}},\ }\href {https://doi.org/10.1016/s0894-1777(96)00138-0} {\bibfield
   {journal} {\bibinfo  {journal} {Experimental Thermal and Fluid Science}\
  }\textbf {\bibinfo {volume} {14}},\ \bibinfo {pages} {361} (\bibinfo {year}
  {1997})}\BibitemShut {NoStop}%
\bibitem [{\citenamefont {Rammerstorfer}\ \emph {et~al.}(2020)\citenamefont
  {Rammerstorfer}, \citenamefont {Karner},\ and\ \citenamefont
  {Siebenhofer}}]{EP6}%
  \BibitemOpen
  \bibfield  {author} {\bibinfo {author} {\bibfnamefont {E.}~\bibnamefont
  {Rammerstorfer}}, \bibinfo {author} {\bibfnamefont {T.}~\bibnamefont
  {Karner}},\ and\ \bibinfo {author} {\bibfnamefont {M.}~\bibnamefont
  {Siebenhofer}},\ }\href {https://doi.org/10.1080/01457632.2018.1564202}
  {\bibfield  {journal} {\bibinfo  {journal} {Heat Transfer Engineering}\
  }\textbf {\bibinfo {volume} {41}},\ \bibinfo {pages} {691} (\bibinfo {year}
  {2020})}\BibitemShut {NoStop}%
\bibitem [{\citenamefont {Panchal}\ \emph {et~al.}(1999)\citenamefont
  {Panchal}, \citenamefont {Kuru}, \citenamefont {Liao}, \citenamefont
  {Ebert},\ and\ \citenamefont {Palen}}]{EP4}%
  \BibitemOpen
  \bibfield  {author} {\bibinfo {author} {\bibfnamefont {C.~B.}\ \bibnamefont
  {Panchal}}, \bibinfo {author} {\bibfnamefont {W.~C.}\ \bibnamefont {Kuru}},
  \bibinfo {author} {\bibfnamefont {C.~F.}\ \bibnamefont {Liao}}, \bibinfo
  {author} {\bibfnamefont {W.~A.}\ \bibnamefont {Ebert}},\ and\ \bibinfo
  {author} {\bibfnamefont {J.~W.}\ \bibnamefont {Palen}},\ }in\ \href@noop {}
  {\emph {\bibinfo {booktitle} {Understanding Heat Exchanger Fouling and its
  Mitigation}}},\ \bibinfo {editor} {edited by\ \bibinfo {editor}
  {\bibfnamefont {T.~R.}\ \bibnamefont {Bott}}, \bibinfo {editor}
  {\bibfnamefont {L.~F.}\ \bibnamefont {Melo}}, \bibinfo {editor}
  {\bibfnamefont {C.~B.}\ \bibnamefont {Panchal}},\ and\ \bibinfo {editor}
  {\bibfnamefont {E.~F.~C.}\ \bibnamefont {Somerscales}}}\ (\bibinfo
  {publisher} {Begell House},\ \bibinfo {year} {1999})\ pp.\ \bibinfo {pages}
  {273--279}\BibitemShut {NoStop}%
\bibitem [{\citenamefont {Polley}\ \emph {et~al.}(2002)\citenamefont {Polley},
  \citenamefont {Wilson}, \citenamefont {Yeap},\ and\ \citenamefont
  {Pugh}}]{EP3}%
  \BibitemOpen
  \bibfield  {author} {\bibinfo {author} {\bibfnamefont {G.~T.}\ \bibnamefont
  {Polley}}, \bibinfo {author} {\bibfnamefont {D.~I.}\ \bibnamefont {Wilson}},
  \bibinfo {author} {\bibfnamefont {B.~L.}\ \bibnamefont {Yeap}},\ and\
  \bibinfo {author} {\bibfnamefont {S.~J.}\ \bibnamefont {Pugh}},\ }\href
  {https://doi.org/10.1016/S1359-4311(02)00023-6} {\bibfield  {journal}
  {\bibinfo  {journal} {Applied Thermal Engineering}\ }\textbf {\bibinfo
  {volume} {22}},\ \bibinfo {pages} {777} (\bibinfo {year} {2002})}\BibitemShut
  {NoStop}%
\bibitem [{\citenamefont {Yeap}\ \emph {et~al.}(2004)\citenamefont {Yeap},
  \citenamefont {Wilson}, \citenamefont {Polley},\ and\ \citenamefont
  {Pugh}}]{EP2}%
  \BibitemOpen
  \bibfield  {author} {\bibinfo {author} {\bibfnamefont {B.~L.}\ \bibnamefont
  {Yeap}}, \bibinfo {author} {\bibfnamefont {D.~I.}\ \bibnamefont {Wilson}},
  \bibinfo {author} {\bibfnamefont {G.~T.}\ \bibnamefont {Polley}},\ and\
  \bibinfo {author} {\bibfnamefont {S.~J.}\ \bibnamefont {Pugh}},\ }\href
  {https://doi.org/10.1205/026387604772803070} {\bibfield  {journal} {\bibinfo
  {journal} {Chemical Engineering Research and Design}\ }\textbf {\bibinfo
  {volume} {82}},\ \bibinfo {pages} {53} (\bibinfo {year} {2004})}\BibitemShut
  {NoStop}%
\bibitem [{\citenamefont {Nasr}\ and\ \citenamefont {Givi}(2006)}]{EP1}%
  \BibitemOpen
  \bibfield  {author} {\bibinfo {author} {\bibfnamefont {M.~R.~J.}\
  \bibnamefont {Nasr}}\ and\ \bibinfo {author} {\bibfnamefont {M.~M.}\
  \bibnamefont {Givi}},\ }\href
  {https://doi.org/10.1016/j.applthermaleng.2005.12.001} {\bibfield  {journal}
  {\bibinfo  {journal} {Applied Thermal Engineering}\ }\textbf {\bibinfo
  {volume} {26}},\ \bibinfo {pages} {1572} (\bibinfo {year}
  {2006})}\BibitemShut {NoStop}%
\bibitem [{\citenamefont {Wilson}\ \emph {et~al.}(2017)\citenamefont {Wilson},
  \citenamefont {Ishiyama},\ and\ \citenamefont {Polley}}]{Wilson24052017}%
  \BibitemOpen
  \bibfield  {author} {\bibinfo {author} {\bibfnamefont {D.~I.}\ \bibnamefont
  {Wilson}}, \bibinfo {author} {\bibfnamefont {E.~M.}\ \bibnamefont
  {Ishiyama}},\ and\ \bibinfo {author} {\bibfnamefont {G.~T.}\ \bibnamefont
  {Polley}},\ }\href {https://doi.org/10.1080/01457632.2016.1206407} {\bibfield
   {journal} {\bibinfo  {journal} {Heat Transfer Engineering}\ }\textbf
  {\bibinfo {volume} {38}},\ \bibinfo {pages} {669} (\bibinfo {year}
  {2017})}\BibitemShut {NoStop}%
\bibitem [{\citenamefont {Geddert}\ \emph {et~al.}(2009)\citenamefont
  {Geddert}, \citenamefont {Bialuch}, \citenamefont {Augustin},\ and\
  \citenamefont {Scholl}}]{Geddert2009}%
  \BibitemOpen
  \bibfield  {author} {\bibinfo {author} {\bibfnamefont {T.}~\bibnamefont
  {Geddert}}, \bibinfo {author} {\bibfnamefont {I.}~\bibnamefont {Bialuch}},
  \bibinfo {author} {\bibfnamefont {W.}~\bibnamefont {Augustin}},\ and\
  \bibinfo {author} {\bibfnamefont {S.}~\bibnamefont {Scholl}},\ }\href
  {https://doi.org/10.1080/01457630902753789} {\bibfield  {journal} {\bibinfo
  {journal} {Heat Transfer Engineering}\ }\textbf {\bibinfo {volume} {30}},\
  \bibinfo {pages} {868} (\bibinfo {year} {2009})}\BibitemShut {NoStop}%
\bibitem [{\citenamefont {Mayer}\ \emph {et~al.}(2013)\citenamefont {Mayer},
  \citenamefont {Augustin},\ and\ \citenamefont {Scholl}}]{Mayer2013}%
  \BibitemOpen
  \bibfield  {author} {\bibinfo {author} {\bibfnamefont {M.}~\bibnamefont
  {Mayer}}, \bibinfo {author} {\bibfnamefont {W.}~\bibnamefont {Augustin}},\
  and\ \bibinfo {author} {\bibfnamefont {S.}~\bibnamefont {Scholl}},\ }\href
  {https://doi.org/10.1007/s00231-013-1183-2} {\bibfield  {journal} {\bibinfo
  {journal} {Heat and Mass Transfer}\ }\textbf {\bibinfo {volume} {49}},\
  \bibinfo {pages} {1419} (\bibinfo {year} {2013})}\BibitemShut {NoStop}%
\bibitem [{\citenamefont {Yang}(2020)}]{YangCFD2020}%
  \BibitemOpen
  \bibfield  {author} {\bibinfo {author} {\bibfnamefont {J.}~\bibnamefont
  {Yang}},\ }\href {https://doi.org/10.1016/j.ijheatmasstransfer.2020.120129}
  {\bibfield  {journal} {\bibinfo  {journal} {International Journal of Heat and
  Mass Transfer}\ }\textbf {\bibinfo {volume} {159}},\ \bibinfo {pages}
  {120129} (\bibinfo {year} {2020})}\BibitemShut {NoStop}%
\bibitem [{\citenamefont {Coletti}\ \emph {et~al.}(2010)\citenamefont
  {Coletti}, \citenamefont {Ishiyama}, \citenamefont {Paterson}, \citenamefont
  {Wilson},\ and\ \citenamefont {Macchietto}}]{EP5}%
  \BibitemOpen
  \bibfield  {author} {\bibinfo {author} {\bibfnamefont {F.}~\bibnamefont
  {Coletti}}, \bibinfo {author} {\bibfnamefont {E.~M.}\ \bibnamefont
  {Ishiyama}}, \bibinfo {author} {\bibfnamefont {W.~R.}\ \bibnamefont
  {Paterson}}, \bibinfo {author} {\bibfnamefont {D.~I.}\ \bibnamefont
  {Wilson}},\ and\ \bibinfo {author} {\bibfnamefont {S.}~\bibnamefont
  {Macchietto}},\ }\href {https://doi.org/https://doi.org/10.1002/aic.12221}
  {\bibfield  {journal} {\bibinfo  {journal} {AIChE Journal}\ }\textbf
  {\bibinfo {volume} {56}},\ \bibinfo {pages} {3257} (\bibinfo {year}
  {2010})}\BibitemShut {NoStop}%
\bibitem [{\citenamefont {Ishiyama}\ \emph
  {et~al.}(2011{\natexlab{b}})\citenamefont {Ishiyama}, \citenamefont
  {Paterson},\ and\ \citenamefont {Ian~Wilson}}]{Ishiyama2011}%
  \BibitemOpen
  \bibfield  {author} {\bibinfo {author} {\bibfnamefont {E.~M.}\ \bibnamefont
  {Ishiyama}}, \bibinfo {author} {\bibfnamefont {W.~R.}\ \bibnamefont
  {Paterson}},\ and\ \bibinfo {author} {\bibfnamefont {D.}~\bibnamefont
  {Ian~Wilson}},\ }\href {https://doi.org/10.1002/aic.12514} {\bibfield
  {journal} {\bibinfo  {journal} {AIChE Journal}\ }\textbf {\bibinfo {volume}
  {57}},\ \bibinfo {pages} {3199} (\bibinfo {year}
  {2011}{\natexlab{b}})}\BibitemShut {NoStop}%
\bibitem [{\citenamefont {Ishiyama}\ \emph {et~al.}(2013)\citenamefont
  {Ishiyama}, \citenamefont {Paterson},\ and\ \citenamefont
  {Wilson}}]{IshiyamaAgeing2013}%
  \BibitemOpen
  \bibfield  {author} {\bibinfo {author} {\bibfnamefont {E.~M.}\ \bibnamefont
  {Ishiyama}}, \bibinfo {author} {\bibfnamefont {W.~R.}\ \bibnamefont
  {Paterson}},\ and\ \bibinfo {author} {\bibfnamefont {D.~I.}\ \bibnamefont
  {Wilson}},\ }\href {https://doi.org/10.1080/01457632.2013.825192} {\bibfield
  {journal} {\bibinfo  {journal} {Heat Transfer Engineering}\ }\textbf
  {\bibinfo {volume} {35}},\ \bibinfo {pages} {311} (\bibinfo {year}
  {2013})}\BibitemShut {NoStop}%
\bibitem [{\citenamefont {Coletti}\ and\ \citenamefont
  {Macchietto}(2011)}]{ColettiMacchietto2011}%
  \BibitemOpen
  \bibfield  {author} {\bibinfo {author} {\bibfnamefont {F.}~\bibnamefont
  {Coletti}}\ and\ \bibinfo {author} {\bibfnamefont {S.}~\bibnamefont
  {Macchietto}},\ }\href {https://doi.org/10.1021/ie901991g} {\bibfield
  {journal} {\bibinfo  {journal} {Industrial \& Engineering Chemistry
  Research}\ }\textbf {\bibinfo {volume} {50}},\ \bibinfo {pages} {4515}
  (\bibinfo {year} {2011})}\BibitemShut {NoStop}%
\bibitem [{\citenamefont {Diaz‐Bejarano}\ \emph {et~al.}(2015)\citenamefont
  {Diaz‐Bejarano}, \citenamefont {Coletti},\ and\ \citenamefont
  {Macchietto}}]{DiazBejarano2015}%
  \BibitemOpen
  \bibfield  {author} {\bibinfo {author} {\bibfnamefont {E.}~\bibnamefont
  {Diaz‐Bejarano}}, \bibinfo {author} {\bibfnamefont {F.}~\bibnamefont
  {Coletti}},\ and\ \bibinfo {author} {\bibfnamefont {S.}~\bibnamefont
  {Macchietto}},\ }\href {https://doi.org/10.1002/aic.15036} {\bibfield
  {journal} {\bibinfo  {journal} {AIChE Journal}\ }\textbf {\bibinfo {volume}
  {62}},\ \bibinfo {pages} {90} (\bibinfo {year} {2015})}\BibitemShut {NoStop}%
\bibitem [{\citenamefont {Coletti}\ \emph {et~al.}(2015)\citenamefont
  {Coletti}, \citenamefont {Crittenden},\ and\ \citenamefont
  {Macchietto}}]{ColettiCrittenden2015}%
  \BibitemOpen
  \bibfield  {author} {\bibinfo {author} {\bibfnamefont {F.}~\bibnamefont
  {Coletti}}, \bibinfo {author} {\bibfnamefont {B.}~\bibnamefont
  {Crittenden}},\ and\ \bibinfo {author} {\bibfnamefont {S.}~\bibnamefont
  {Macchietto}},\ }\bibinfo {title} {Basic science of the fouling process},\
  in\ \href {https://doi.org/10.1016/b978-0-12-801256-7.00002-6} {\emph
  {\bibinfo {booktitle} {Crude Oil Fouling}}}\ (\bibinfo  {publisher}
  {Elsevier},\ \bibinfo {year} {2015})\ pp.\ \bibinfo {pages}
  {23--50}\BibitemShut {NoStop}%
\bibitem [{\citenamefont {Diaz-Bejarano}\ \emph {et~al.}(2017)\citenamefont
  {Diaz-Bejarano}, \citenamefont {Coletti},\ and\ \citenamefont
  {Macchietto}}]{Diaz-Bejarano24052017}%
  \BibitemOpen
  \bibfield  {author} {\bibinfo {author} {\bibfnamefont {E.}~\bibnamefont
  {Diaz-Bejarano}}, \bibinfo {author} {\bibfnamefont {F.}~\bibnamefont
  {Coletti}},\ and\ \bibinfo {author} {\bibfnamefont {S.}~\bibnamefont
  {Macchietto}},\ }\href {https://doi.org/10.1080/01457632.2016.1206408}
  {\bibfield  {journal} {\bibinfo  {journal} {Heat Transfer Engineering}\
  }\textbf {\bibinfo {volume} {38}},\ \bibinfo {pages} {681} (\bibinfo {year}
  {2017})}\BibitemShut {NoStop}%
\bibitem [{\citenamefont {Loyola-Fuentes}\ \emph {et~al.}(2019)\citenamefont
  {Loyola-Fuentes}, \citenamefont {Jobson},\ and\ \citenamefont
  {Smith}}]{LoyolaFuentes2019}%
  \BibitemOpen
  \bibfield  {author} {\bibinfo {author} {\bibfnamefont {J.}~\bibnamefont
  {Loyola-Fuentes}}, \bibinfo {author} {\bibfnamefont {M.}~\bibnamefont
  {Jobson}},\ and\ \bibinfo {author} {\bibfnamefont {R.}~\bibnamefont
  {Smith}},\ }\href {https://doi.org/10.1021/acs.iecr.9b00457} {\bibfield
  {journal} {\bibinfo  {journal} {Industrial \& Engineering Chemistry
  Research}\ }\textbf {\bibinfo {volume} {58}},\ \bibinfo {pages} {10418}
  (\bibinfo {year} {2019})}\BibitemShut {NoStop}%
\bibitem [{\citenamefont {Loyola-Fuentes}\ and\ \citenamefont
  {Smith}(2019)}]{LoyolaFuentesSmith2019}%
  \BibitemOpen
  \bibfield  {author} {\bibinfo {author} {\bibfnamefont {J.}~\bibnamefont
  {Loyola-Fuentes}}\ and\ \bibinfo {author} {\bibfnamefont {R.}~\bibnamefont
  {Smith}},\ }\href {https://doi.org/10.1016/j.energy.2019.06.119} {\bibfield
  {journal} {\bibinfo  {journal} {Energy}\ }\textbf {\bibinfo {volume} {183}},\
  \bibinfo {pages} {368} (\bibinfo {year} {2019})}\BibitemShut {NoStop}%
\bibitem [{\citenamefont {Markowski}\ and\ \citenamefont
  {Trzcinski}(2025)}]{Markowski2025}%
  \BibitemOpen
  \bibfield  {author} {\bibinfo {author} {\bibfnamefont {M.}~\bibnamefont
  {Markowski}}\ and\ \bibinfo {author} {\bibfnamefont {P.}~\bibnamefont
  {Trzcinski}},\ }\href {https://doi.org/10.1038/s41598-025-26069-y} {\bibfield
   {journal} {\bibinfo  {journal} {Scientific Reports}\ }\textbf {\bibinfo
  {volume} {15}},\ \bibinfo {pages} {42078} (\bibinfo {year}
  {2025})}\BibitemShut {NoStop}%
\bibitem [{\citenamefont {Sundar}\ \emph {et~al.}(2020)\citenamefont {Sundar},
  \citenamefont {Rajagopal}, \citenamefont {Zhao}, \citenamefont {Kuntumalla},
  \citenamefont {Meng}, \citenamefont {Chang}, \citenamefont {Shao},
  \citenamefont {Ferreira}, \citenamefont {Miljkovic}, \citenamefont {Sinha},\
  and\ \citenamefont {Salapaka}}]{Sundar2020}%
  \BibitemOpen
  \bibfield  {author} {\bibinfo {author} {\bibfnamefont {S.}~\bibnamefont
  {Sundar}}, \bibinfo {author} {\bibfnamefont {M.~C.}\ \bibnamefont
  {Rajagopal}}, \bibinfo {author} {\bibfnamefont {H.}~\bibnamefont {Zhao}},
  \bibinfo {author} {\bibfnamefont {G.}~\bibnamefont {Kuntumalla}}, \bibinfo
  {author} {\bibfnamefont {Y.}~\bibnamefont {Meng}}, \bibinfo {author}
  {\bibfnamefont {H.~C.}\ \bibnamefont {Chang}}, \bibinfo {author}
  {\bibfnamefont {C.}~\bibnamefont {Shao}}, \bibinfo {author} {\bibfnamefont
  {P.}~\bibnamefont {Ferreira}}, \bibinfo {author} {\bibfnamefont
  {N.}~\bibnamefont {Miljkovic}}, \bibinfo {author} {\bibfnamefont
  {S.}~\bibnamefont {Sinha}},\ and\ \bibinfo {author} {\bibfnamefont
  {S.}~\bibnamefont {Salapaka}},\ }\href
  {https://doi.org/10.1016/j.ijheatmasstransfer.2020.120112} {\bibfield
  {journal} {\bibinfo  {journal} {International Journal of Heat and Mass
  Transfer}\ }\textbf {\bibinfo {volume} {159}},\ \bibinfo {pages} {120112}
  (\bibinfo {year} {2020})}\BibitemShut {NoStop}%
\bibitem [{\citenamefont {Zubair}\ \emph
  {et~al.}(1997{\natexlab{a}})\citenamefont {Zubair}, \citenamefont {Sheikh},
  \citenamefont {Budair}, \citenamefont {Haq}, \citenamefont {Quddus},\ and\
  \citenamefont {Ashiru}}]{Zubair1997CaCO3}%
  \BibitemOpen
  \bibfield  {author} {\bibinfo {author} {\bibfnamefont {S.~M.}\ \bibnamefont
  {Zubair}}, \bibinfo {author} {\bibfnamefont {A.~K.}\ \bibnamefont {Sheikh}},
  \bibinfo {author} {\bibfnamefont {M.~O.}\ \bibnamefont {Budair}}, \bibinfo
  {author} {\bibfnamefont {M.~U.}\ \bibnamefont {Haq}}, \bibinfo {author}
  {\bibfnamefont {A.}~\bibnamefont {Quddus}},\ and\ \bibinfo {author}
  {\bibfnamefont {O.~A.}\ \bibnamefont {Ashiru}},\ }\href
  {https://doi.org/10.1115/1.2824145} {\bibfield  {journal} {\bibinfo
  {journal} {Journal of Heat Transfer}\ }\textbf {\bibinfo {volume} {119}},\
  \bibinfo {pages} {581} (\bibinfo {year} {1997}{\natexlab{a}})}\BibitemShut
  {NoStop}%
\bibitem [{\citenamefont {Zubair}\ \emph {et~al.}(1992)\citenamefont {Zubair},
  \citenamefont {Sheikh},\ and\ \citenamefont {Shaik}}]{Zubair1992}%
  \BibitemOpen
  \bibfield  {author} {\bibinfo {author} {\bibfnamefont {S.~M.}\ \bibnamefont
  {Zubair}}, \bibinfo {author} {\bibfnamefont {A.~K.}\ \bibnamefont {Sheikh}},\
  and\ \bibinfo {author} {\bibfnamefont {M.~N.}\ \bibnamefont {Shaik}},\ }\href
  {https://doi.org/10.1016/0360-5442(92)90120-O} {\bibfield  {journal}
  {\bibinfo  {journal} {Energy}\ }\textbf {\bibinfo {volume} {17}},\ \bibinfo
  {pages} {769} (\bibinfo {year} {1992})}\BibitemShut {NoStop}%
\bibitem [{\citenamefont {Zubair}\ \emph
  {et~al.}(1997{\natexlab{b}})\citenamefont {Zubair}, \citenamefont {Sheikh},
  \citenamefont {Budair},\ and\ \citenamefont {Badar}}]{Zubair1997}%
  \BibitemOpen
  \bibfield  {author} {\bibinfo {author} {\bibfnamefont {S.~M.}\ \bibnamefont
  {Zubair}}, \bibinfo {author} {\bibfnamefont {A.~K.}\ \bibnamefont {Sheikh}},
  \bibinfo {author} {\bibfnamefont {M.~O.}\ \bibnamefont {Budair}},\ and\
  \bibinfo {author} {\bibfnamefont {M.~A.}\ \bibnamefont {Badar}},\ }\href
  {https://doi.org/10.1115/1.2824144} {\bibfield  {journal} {\bibinfo
  {journal} {Journal of Heat Transfer}\ }\textbf {\bibinfo {volume} {119}},\
  \bibinfo {pages} {575} (\bibinfo {year} {1997}{\natexlab{b}})}\BibitemShut
  {NoStop}%
\bibitem [{\citenamefont {Zubair}\ \emph {et~al.}(2000)\citenamefont {Zubair},
  \citenamefont {Sheikh}, \citenamefont {Younas},\ and\ \citenamefont
  {Budair}}]{Zubair2000}%
  \BibitemOpen
  \bibfield  {author} {\bibinfo {author} {\bibfnamefont {S.~M.}\ \bibnamefont
  {Zubair}}, \bibinfo {author} {\bibfnamefont {A.~K.}\ \bibnamefont {Sheikh}},
  \bibinfo {author} {\bibfnamefont {M.}~\bibnamefont {Younas}},\ and\ \bibinfo
  {author} {\bibfnamefont {M.~O.}\ \bibnamefont {Budair}},\ }\href@noop {}
  {\bibfield  {journal} {\bibinfo  {journal} {Energy}\ }\textbf {\bibinfo
  {volume} {25}},\ \bibinfo {pages} {427} (\bibinfo {year} {2000})}\BibitemShut
  {NoStop}%
\bibitem [{\citenamefont {Sheikh}\ \emph {et~al.}(2001)\citenamefont {Sheikh},
  \citenamefont {Zubair}, \citenamefont {Younas},\ and\ \citenamefont
  {Budair}}]{Sheikh2001}%
  \BibitemOpen
  \bibfield  {author} {\bibinfo {author} {\bibfnamefont {A.~K.}\ \bibnamefont
  {Sheikh}}, \bibinfo {author} {\bibfnamefont {S.~M.}\ \bibnamefont {Zubair}},
  \bibinfo {author} {\bibfnamefont {M.}~\bibnamefont {Younas}},\ and\ \bibinfo
  {author} {\bibfnamefont {M.~O.}\ \bibnamefont {Budair}},\ }\href
  {https://doi.org/10.1243/0954408011530532} {\bibfield  {journal} {\bibinfo
  {journal} {Proceedings of the Institution of Mechanical Engineers, Part E:
  Journal of Process Mechanical Engineering}\ }\textbf {\bibinfo {volume}
  {215}},\ \bibinfo {pages} {331} (\bibinfo {year} {2001})}\BibitemShut
  {NoStop}%
\bibitem [{\citenamefont {Vatistas}(1988)}]{Vatistas1988}%
  \BibitemOpen
  \bibfield  {author} {\bibinfo {author} {\bibfnamefont {N.}~\bibnamefont
  {Vatistas}},\ }\bibinfo {title} {Stochastic model for the induction step on
  particulate fouling},\ in\ \href
  {https://doi.org/10.1007/978-94-009-2813-8_11} {\emph {\bibinfo {booktitle}
  {Fouling Science and Technology}}}\ (\bibinfo  {publisher} {Springer
  Netherlands},\ \bibinfo {year} {1988})\ pp.\ \bibinfo {pages}
  {165--171}\BibitemShut {NoStop}%
\bibitem [{\citenamefont {Flora}(1993)}]{Flora1993}%
  \BibitemOpen
  \bibfield  {author} {\bibinfo {author} {\bibfnamefont {J.~R.}\ \bibnamefont
  {Flora}},\ }\href {https://doi.org/10.1016/0376-7388(93)87008-y} {\bibfield
  {journal} {\bibinfo  {journal} {Journal of Membrane Science}\ }\textbf
  {\bibinfo {volume} {76}},\ \bibinfo {pages} {85} (\bibinfo {year}
  {1993})}\BibitemShut {NoStop}%
\bibitem [{\citenamefont {Wessling}(2001)}]{Wessling2001}%
  \BibitemOpen
  \bibfield  {author} {\bibinfo {author} {\bibfnamefont {M.}~\bibnamefont
  {Wessling}},\ }\href {https://doi.org/10.1016/s1383-5866(01)00138-1}
  {\bibfield  {journal} {\bibinfo  {journal} {Separation and Purification
  Technology}\ }\textbf {\bibinfo {volume} {24}},\ \bibinfo {pages} {375}
  (\bibinfo {year} {2001})}\BibitemShut {NoStop}%
\bibitem [{\citenamefont {Liu}\ \emph {et~al.}(2020)\citenamefont {Liu},
  \citenamefont {Huang}, \citenamefont {Ji}, \citenamefont {Wang},
  \citenamefont {Tang},\ and\ \citenamefont {Leckie}}]{Liu2020}%
  \BibitemOpen
  \bibfield  {author} {\bibinfo {author} {\bibfnamefont {J.}~\bibnamefont
  {Liu}}, \bibinfo {author} {\bibfnamefont {T.}~\bibnamefont {Huang}}, \bibinfo
  {author} {\bibfnamefont {R.}~\bibnamefont {Ji}}, \bibinfo {author}
  {\bibfnamefont {Z.}~\bibnamefont {Wang}}, \bibinfo {author} {\bibfnamefont
  {C.~Y.}\ \bibnamefont {Tang}},\ and\ \bibinfo {author} {\bibfnamefont
  {J.~O.}\ \bibnamefont {Leckie}},\ }\href
  {https://doi.org/10.1021/acs.est.0c04165} {\bibfield  {journal} {\bibinfo
  {journal} {Environmental Science \& Technology}\ }\textbf {\bibinfo {volume}
  {54}},\ \bibinfo {pages} {12703} (\bibinfo {year} {2020})}\BibitemShut
  {NoStop}%
\bibitem [{\citenamefont {Shaikh}\ \emph {et~al.}(2023)\citenamefont {Shaikh},
  \citenamefont {Newaz}, \citenamefont {Zubir}, \citenamefont {Wong},
  \citenamefont {Khan}, \citenamefont {Abdullah}, \citenamefont {Alam},\ and\
  \citenamefont {Sugumaran}}]{Shaikh2023}%
  \BibitemOpen
  \bibfield  {author} {\bibinfo {author} {\bibfnamefont {K.}~\bibnamefont
  {Shaikh}}, \bibinfo {author} {\bibfnamefont {K.~M.~S.}\ \bibnamefont
  {Newaz}}, \bibinfo {author} {\bibfnamefont {M.~N.~M.}\ \bibnamefont {Zubir}},
  \bibinfo {author} {\bibfnamefont {K.~H.}\ \bibnamefont {Wong}}, \bibinfo
  {author} {\bibfnamefont {W.~A.}\ \bibnamefont {Khan}}, \bibinfo {author}
  {\bibfnamefont {S.}~\bibnamefont {Abdullah}}, \bibinfo {author}
  {\bibfnamefont {M.~S.}\ \bibnamefont {Alam}},\ and\ \bibinfo {author}
  {\bibfnamefont {L.}~\bibnamefont {Sugumaran}},\ }\bibfield  {journal}
  {\bibinfo  {journal} {Journal of Thermal Analysis and Calorimetry}\ }\href
  {https://doi.org/10.1007/s10973-023-12544-z} {10.1007/s10973-023-12544-z}
  (\bibinfo {year} {2023})\BibitemShut {NoStop}%
\bibitem [{\citenamefont {Fguiri}\ \emph {et~al.}(2020)\citenamefont {Fguiri},
  \citenamefont {Jradi}, \citenamefont {Marvillet},\ and\ \citenamefont
  {Jeday}}]{Fguiri2020}%
  \BibitemOpen
  \bibfield  {author} {\bibinfo {author} {\bibfnamefont {A.}~\bibnamefont
  {Fguiri}}, \bibinfo {author} {\bibfnamefont {R.}~\bibnamefont {Jradi}},
  \bibinfo {author} {\bibfnamefont {C.}~\bibnamefont {Marvillet}},\ and\
  \bibinfo {author} {\bibfnamefont {M.~R.}\ \bibnamefont {Jeday}},\ }\bibfield
  {journal} {\bibinfo  {journal} {Heat and Mass Transfer}\ }\href
  {https://doi.org/10.1007/s00231-020-02858-1} {10.1007/s00231-020-02858-1}
  (\bibinfo {year} {2020})\BibitemShut {NoStop}%
\bibitem [{\citenamefont {Crittenden}\ \emph {et~al.}(1992)\citenamefont
  {Crittenden}, \citenamefont {Kolaczkowski},\ and\ \citenamefont
  {Downey}}]{Crittenden1992}%
  \BibitemOpen
  \bibfield  {author} {\bibinfo {author} {\bibfnamefont {B.~D.}\ \bibnamefont
  {Crittenden}}, \bibinfo {author} {\bibfnamefont {S.~T.}\ \bibnamefont
  {Kolaczkowski}},\ and\ \bibinfo {author} {\bibfnamefont {I.~L.}\ \bibnamefont
  {Downey}},\ }\href@noop {} {\bibfield  {journal} {\bibinfo  {journal} {Trans.
  IChemE, Part A, Chem. Eng. Res. Des.}\ }\textbf {\bibinfo {volume} {70}},\
  \bibinfo {pages} {547} (\bibinfo {year} {1992})}\BibitemShut {NoStop}%
\bibitem [{\citenamefont {Zhang}\ and\ \citenamefont
  {Revie}(2016)}]{ZhangRevie2016}%
  \BibitemOpen
  \bibfield  {author} {\bibinfo {author} {\bibfnamefont {M.}~\bibnamefont
  {Zhang}}\ and\ \bibinfo {author} {\bibfnamefont {M.}~\bibnamefont {Revie}},\
  }in\ \href@noop {} {\emph {\bibinfo {booktitle} {Risk, Reliability and
  Safety: Innovating Theory and Practice (Proceedings of ESREL 2016)}}}\
  (\bibinfo  {publisher} {CRC/Taylor \& Francis},\ \bibinfo {address}
  {London},\ \bibinfo {year} {2016})\BibitemShut {NoStop}%
\bibitem [{\citenamefont {Rehman}\ \emph {et~al.}(2023)\citenamefont {Rehman},
  \citenamefont {Ramasamy}, \citenamefont {Mohammad~Rozali}, \citenamefont
  {Mahadzir}, \citenamefont {Ghumman},\ and\ \citenamefont
  {Qureshi}}]{Rehman2023}%
  \BibitemOpen
  \bibfield  {author} {\bibinfo {author} {\bibfnamefont {O.~u.}\ \bibnamefont
  {Rehman}}, \bibinfo {author} {\bibfnamefont {M.~G.}\ \bibnamefont
  {Ramasamy}}, \bibinfo {author} {\bibfnamefont {N.~E.}\ \bibnamefont
  {Mohammad~Rozali}}, \bibinfo {author} {\bibfnamefont {S.}~\bibnamefont
  {Mahadzir}}, \bibinfo {author} {\bibfnamefont {A.~S.~M.}\ \bibnamefont
  {Ghumman}},\ and\ \bibinfo {author} {\bibfnamefont {A.~H.}\ \bibnamefont
  {Qureshi}},\ }\href {https://doi.org/10.3390/pr11041036} {\bibfield
  {journal} {\bibinfo  {journal} {Processes}\ }\textbf {\bibinfo {volume}
  {11}},\ \bibinfo {pages} {1036} (\bibinfo {year} {2023})}\BibitemShut
  {NoStop}%
\bibitem [{\citenamefont {Crittenden}\ and\ \citenamefont
  {Alderman}(1988)}]{Crittenden1988}%
  \BibitemOpen
  \bibfield  {author} {\bibinfo {author} {\bibfnamefont {B.}~\bibnamefont
  {Crittenden}}\ and\ \bibinfo {author} {\bibfnamefont {N.}~\bibnamefont
  {Alderman}},\ }\href {https://doi.org/10.1016/0009-2509(88)80078-2}
  {\bibfield  {journal} {\bibinfo  {journal} {Chemical Engineering Science}\
  }\textbf {\bibinfo {volume} {43}},\ \bibinfo {pages} {829} (\bibinfo {year}
  {1988})}\BibitemShut {NoStop}%
\bibitem [{\citenamefont {Sato}(1999)}]{Sato1999}%
  \BibitemOpen
  \bibfield  {author} {\bibinfo {author} {\bibfnamefont {K.-i.}\ \bibnamefont
  {Sato}},\ }\href@noop {} {\emph {\bibinfo {title} {L{\'e}vy Processes and
  Infinitely Divisible Distributions}}},\ \bibinfo {series} {Cambridge Studies
  in Advanced Mathematics}, Vol.~\bibinfo {volume} {68}\ (\bibinfo  {publisher}
  {Cambridge University Press},\ \bibinfo {address} {Cambridge},\ \bibinfo
  {year} {1999})\BibitemShut {NoStop}%
\bibitem [{\citenamefont {Applebaum}(2009)}]{Applebaum2009}%
  \BibitemOpen
  \bibfield  {author} {\bibinfo {author} {\bibfnamefont {D.}~\bibnamefont
  {Applebaum}},\ }\href {https://doi.org/10.1017/CBO9780511809781} {\emph
  {\bibinfo {title} {L{\'e}vy Processes and Stochastic Calculus}}},\ \bibinfo
  {edition} {2nd}\ ed.,\ \bibinfo {series} {Cambridge Studies in Advanced
  Mathematics}, Vol.\ \bibinfo {volume} {116}\ (\bibinfo  {publisher}
  {Cambridge University Press},\ \bibinfo {address} {Cambridge},\ \bibinfo
  {year} {2009})\BibitemShut {NoStop}%
\bibitem [{\citenamefont {Campbell}(1909)}]{Campbell1909}%
  \BibitemOpen
  \bibfield  {author} {\bibinfo {author} {\bibfnamefont {N.}~\bibnamefont
  {Campbell}},\ }\href@noop {} {\bibfield  {journal} {\bibinfo  {journal}
  {Proceedings of the Cambridge Philosophical Society}\ }\textbf {\bibinfo
  {volume} {15}},\ \bibinfo {pages} {117} (\bibinfo {year} {1909})}\BibitemShut
  {NoStop}%
\bibitem [{\citenamefont {Rosi{\'n}ski}(2007)}]{Rosinski2007}%
  \BibitemOpen
  \bibfield  {author} {\bibinfo {author} {\bibfnamefont {J.}~\bibnamefont
  {Rosi{\'n}ski}},\ }\href {https://doi.org/10.1016/j.spa.2006.10.003}
  {\bibfield  {journal} {\bibinfo  {journal} {Stochastic Processes and their
  Applications}\ }\textbf {\bibinfo {volume} {117}},\ \bibinfo {pages} {677}
  (\bibinfo {year} {2007})}\BibitemShut {NoStop}%
\bibitem [{\citenamefont {Jørgensen}(1987)}]{Jorgensen1987}%
  \BibitemOpen
  \bibfield  {author} {\bibinfo {author} {\bibfnamefont {B.}~\bibnamefont
  {Jørgensen}},\ }\href {https://doi.org/10.1111/j.2517-6161.1987.tb01685.x}
  {\bibfield  {journal} {\bibinfo  {journal} {Journal of the Royal Statistical
  Society Series B: Statistical Methodology}\ }\textbf {\bibinfo {volume}
  {49}},\ \bibinfo {pages} {127} (\bibinfo {year} {1987})}\BibitemShut
  {NoStop}%
\bibitem [{\citenamefont {Hougaard}(1986)}]{Hougaard1986}%
  \BibitemOpen
  \bibfield  {author} {\bibinfo {author} {\bibfnamefont {P.}~\bibnamefont
  {Hougaard}},\ }\href {https://doi.org/10.1093/biomet/73.2.387} {\bibfield
  {journal} {\bibinfo  {journal} {Biometrika}\ }\textbf {\bibinfo {volume}
  {73}},\ \bibinfo {pages} {387} (\bibinfo {year} {1986})}\BibitemShut
  {NoStop}%
\bibitem [{\citenamefont {Abdel-Hameed}(1975)}]{AbdelHameed1975}%
  \BibitemOpen
  \bibfield  {author} {\bibinfo {author} {\bibfnamefont {M.}~\bibnamefont
  {Abdel-Hameed}},\ }\href {https://doi.org/10.1109/tr.1975.5215123} {\bibfield
   {journal} {\bibinfo  {journal} {IEEE Transactions on Reliability}\ }\textbf
  {\bibinfo {volume} {R-24}},\ \bibinfo {pages} {152} (\bibinfo {year}
  {1975})}\BibitemShut {NoStop}%
\bibitem [{\citenamefont {Lawless}\ and\ \citenamefont
  {Crowder}(2004)}]{LawlessCrowder2004}%
  \BibitemOpen
  \bibfield  {author} {\bibinfo {author} {\bibfnamefont {J.}~\bibnamefont
  {Lawless}}\ and\ \bibinfo {author} {\bibfnamefont {M.}~\bibnamefont
  {Crowder}},\ }\href {https://doi.org/10.1023/b:lida.0000036389.14073.dd}
  {\bibfield  {journal} {\bibinfo  {journal} {Lifetime Data Analysis}\ }\textbf
  {\bibinfo {volume} {10}},\ \bibinfo {pages} {213} (\bibinfo {year}
  {2004})}\BibitemShut {NoStop}%
\bibitem [{\citenamefont {van Noortwijk}(2009)}]{vanNoortwijk2009}%
  \BibitemOpen
  \bibfield  {author} {\bibinfo {author} {\bibfnamefont {J.}~\bibnamefont {van
  Noortwijk}},\ }\href {https://doi.org/10.1016/j.ress.2007.03.019} {\bibfield
  {journal} {\bibinfo  {journal} {Reliability Engineering \& System Safety}\
  }\textbf {\bibinfo {volume} {94}},\ \bibinfo {pages} {2} (\bibinfo {year}
  {2009})}\BibitemShut {NoStop}%
\bibitem [{\citenamefont {Esary}\ \emph {et~al.}(1973)\citenamefont {Esary},
  \citenamefont {Marshall},\ and\ \citenamefont {Proschan}}]{Esary1973}%
  \BibitemOpen
  \bibfield  {author} {\bibinfo {author} {\bibfnamefont {J.~D.}\ \bibnamefont
  {Esary}}, \bibinfo {author} {\bibfnamefont {A.~W.}\ \bibnamefont
  {Marshall}},\ and\ \bibinfo {author} {\bibfnamefont {F.}~\bibnamefont
  {Proschan}},\ }\href {https://doi.org/10.1214/aop/1176996891} {\bibfield
  {journal} {\bibinfo  {journal} {The Annals of Probability}\ }\textbf
  {\bibinfo {volume} {1}},\ \bibinfo {pages} {627} (\bibinfo {year}
  {1973})}\BibitemShut {NoStop}%
\bibitem [{\citenamefont {Wang}\ and\ \citenamefont {Xu}(2010)}]{WangXu2010}%
  \BibitemOpen
  \bibfield  {author} {\bibinfo {author} {\bibfnamefont {X.}~\bibnamefont
  {Wang}}\ and\ \bibinfo {author} {\bibfnamefont {D.}~\bibnamefont {Xu}},\
  }\href {https://doi.org/10.1198/tech.2009.08197} {\bibfield  {journal}
  {\bibinfo  {journal} {Technometrics}\ }\textbf {\bibinfo {volume} {52}},\
  \bibinfo {pages} {188} (\bibinfo {year} {2010})}\BibitemShut {NoStop}%
\bibitem [{\citenamefont {Ye}\ and\ \citenamefont {Chen}(2014)}]{YeChen2014}%
  \BibitemOpen
  \bibfield  {author} {\bibinfo {author} {\bibfnamefont {Z.-S.}\ \bibnamefont
  {Ye}}\ and\ \bibinfo {author} {\bibfnamefont {N.}~\bibnamefont {Chen}},\
  }\href {https://doi.org/10.1080/00401706.2013.830074} {\bibfield  {journal}
  {\bibinfo  {journal} {Technometrics}\ }\textbf {\bibinfo {volume} {56}},\
  \bibinfo {pages} {302} (\bibinfo {year} {2014})}\BibitemShut {NoStop}%
\bibitem [{\citenamefont {Ye}\ and\ \citenamefont {Xie}(2015)}]{YeXie2015}%
  \BibitemOpen
  \bibfield  {author} {\bibinfo {author} {\bibfnamefont {Z.-S.}\ \bibnamefont
  {Ye}}\ and\ \bibinfo {author} {\bibfnamefont {M.}~\bibnamefont {Xie}},\
  }\href {https://doi.org/10.1002/asmb.2063} {\bibfield  {journal} {\bibinfo
  {journal} {Applied Stochastic Models in Business and Industry}\ }\textbf
  {\bibinfo {volume} {31}},\ \bibinfo {pages} {16} (\bibinfo {year}
  {2015})}\BibitemShut {NoStop}%
\bibitem [{\citenamefont {Si}\ \emph {et~al.}(2011)\citenamefont {Si},
  \citenamefont {Wang}, \citenamefont {Hu},\ and\ \citenamefont
  {Zhou}}]{Si2011}%
  \BibitemOpen
  \bibfield  {author} {\bibinfo {author} {\bibfnamefont {X.-S.}\ \bibnamefont
  {Si}}, \bibinfo {author} {\bibfnamefont {W.}~\bibnamefont {Wang}}, \bibinfo
  {author} {\bibfnamefont {C.-H.}\ \bibnamefont {Hu}},\ and\ \bibinfo {author}
  {\bibfnamefont {D.-H.}\ \bibnamefont {Zhou}},\ }\href
  {https://doi.org/10.1016/j.ejor.2010.11.018} {\bibfield  {journal} {\bibinfo
  {journal} {European Journal of Operational Research}\ }\textbf {\bibinfo
  {volume} {213}},\ \bibinfo {pages} {1} (\bibinfo {year} {2011})}\BibitemShut
  {NoStop}%
\bibitem [{\citenamefont {Kahle}\ \emph {et~al.}(2016)\citenamefont {Kahle},
  \citenamefont {Mercier},\ and\ \citenamefont {Paroissin}}]{Kahle2016}%
  \BibitemOpen
  \bibfield  {author} {\bibinfo {author} {\bibfnamefont {W.}~\bibnamefont
  {Kahle}}, \bibinfo {author} {\bibfnamefont {S.}~\bibnamefont {Mercier}},\
  and\ \bibinfo {author} {\bibfnamefont {C.}~\bibnamefont {Paroissin}},\ }\href
  {https://doi.org/10.1002/9781119307488} {\emph {\bibinfo {title} {Degradation
  Processes in Reliability}}}\ (\bibinfo  {publisher} {Wiley},\ \bibinfo {year}
  {2016})\BibitemShut {NoStop}%
\bibitem [{\citenamefont {{\c{C}}inlar}(1977)}]{Cinlar1977}%
  \BibitemOpen
  \bibfield  {author} {\bibinfo {author} {\bibfnamefont {E.}~\bibnamefont
  {{\c{C}}inlar}},\ }\bibinfo {title} {Shock and wear models and {Markov}
  additive processes},\ in\ \href
  {https://doi.org/10.1016/b978-0-12-702101-0.50016-0} {\emph {\bibinfo
  {booktitle} {The Theory and Applications of Reliability with Emphasis on
  Bayesian and Nonparametric Methods}}}\ (\bibinfo  {publisher} {Elsevier},\
  \bibinfo {year} {1977})\ pp.\ \bibinfo {pages} {193--214}\BibitemShut
  {NoStop}%
\bibitem [{\citenamefont {Abdel-Hameed}(1984)}]{AbdelHameed1984}%
  \BibitemOpen
  \bibfield  {author} {\bibinfo {author} {\bibfnamefont {M.}~\bibnamefont
  {Abdel-Hameed}},\ }\href {https://doi.org/10.1287/moor.9.4.606} {\bibfield
  {journal} {\bibinfo  {journal} {Mathematics of Operations Research}\ }\textbf
  {\bibinfo {volume} {9}},\ \bibinfo {pages} {606} (\bibinfo {year}
  {1984})}\BibitemShut {NoStop}%
\bibitem [{\citenamefont {Shu}\ \emph {et~al.}(2016)\citenamefont {Shu},
  \citenamefont {Feng}, \citenamefont {Kao},\ and\ \citenamefont
  {Liu}}]{Shu2016}%
  \BibitemOpen
  \bibfield  {author} {\bibinfo {author} {\bibfnamefont {Y.}~\bibnamefont
  {Shu}}, \bibinfo {author} {\bibfnamefont {Q.}~\bibnamefont {Feng}}, \bibinfo
  {author} {\bibfnamefont {E.~P.}\ \bibnamefont {Kao}},\ and\ \bibinfo {author}
  {\bibfnamefont {H.}~\bibnamefont {Liu}},\ }\href
  {https://doi.org/10.1080/0740817x.2016.1172743} {\bibfield  {journal}
  {\bibinfo  {journal} {IIE Transactions}\ }\textbf {\bibinfo {volume} {48}},\
  \bibinfo {pages} {993} (\bibinfo {year} {2016})}\BibitemShut {NoStop}%
\bibitem [{\citenamefont {Collett}\ \emph {et~al.}(2026)\citenamefont
  {Collett}, \citenamefont {Stasik}, \citenamefont {Casolo},\ and\
  \citenamefont {Riemer-S{\o}rensen}}]{Collet2026}%
  \BibitemOpen
  \bibfield  {author} {\bibinfo {author} {\bibfnamefont {P.}~\bibnamefont
  {Collett}}, \bibinfo {author} {\bibfnamefont {A.~J.}\ \bibnamefont {Stasik}},
  \bibinfo {author} {\bibfnamefont {S.}~\bibnamefont {Casolo}},\ and\ \bibinfo
  {author} {\bibfnamefont {S.}~\bibnamefont {Riemer-S{\o}rensen}},\ }\href@noop
  {} {\bibinfo {title} {Fast {B}ayesian equipment condition monitoring via
  simulation based inference: applications to heat exchanger health}},\
  \bibinfo {howpublished} {arXiv preprint} (\bibinfo {year} {2026}),\ \Eprint
  {https://arxiv.org/abs/2604.20735} {arXiv:2604.20735 [cs.LG]} \BibitemShut
  {NoStop}%
\bibitem [{\citenamefont {Barndorff-Nielsen}\ and\ \citenamefont
  {Shephard}(2001)}]{BarndorffNielsenShephard2001}%
  \BibitemOpen
  \bibfield  {author} {\bibinfo {author} {\bibfnamefont {O.~E.}\ \bibnamefont
  {Barndorff-Nielsen}}\ and\ \bibinfo {author} {\bibfnamefont {N.}~\bibnamefont
  {Shephard}},\ }\href {https://doi.org/10.1111/1467-9868.00282} {\bibfield
  {journal} {\bibinfo  {journal} {Journal of the Royal Statistical Society
  Series B: Statistical Methodology}\ }\textbf {\bibinfo {volume} {63}},\
  \bibinfo {pages} {167} (\bibinfo {year} {2001})}\BibitemShut {NoStop}%
\bibitem [{\citenamefont {Stewart}(1993)}]{Stewart1993}%
  \BibitemOpen
  \bibfield  {author} {\bibinfo {author} {\bibfnamefont {P.~S.}\ \bibnamefont
  {Stewart}},\ }\href {https://doi.org/10.1002/bit.260410115} {\bibfield
  {journal} {\bibinfo  {journal} {Biotechnology and Bioengineering}\ }\textbf
  {\bibinfo {volume} {41}},\ \bibinfo {pages} {111} (\bibinfo {year}
  {1993})}\BibitemShut {NoStop}%
\bibitem [{\citenamefont {Bohn}\ \emph {et~al.}(2007)\citenamefont {Bohn},
  \citenamefont {Zippel}, \citenamefont {Almeida},\ and\ \citenamefont
  {Xavier}}]{Bohn2007}%
  \BibitemOpen
  \bibfield  {author} {\bibinfo {author} {\bibfnamefont {A.}~\bibnamefont
  {Bohn}}, \bibinfo {author} {\bibfnamefont {B.}~\bibnamefont {Zippel}},
  \bibinfo {author} {\bibfnamefont {J.}~\bibnamefont {Almeida}},\ and\ \bibinfo
  {author} {\bibfnamefont {J.}~\bibnamefont {Xavier}},\ }\href
  {https://doi.org/10.2166/wst.2007.266} {\bibfield  {journal} {\bibinfo
  {journal} {Water Science and Technology}\ }\textbf {\bibinfo {volume} {55}},\
  \bibinfo {pages} {257} (\bibinfo {year} {2007})}\BibitemShut {NoStop}%
\bibitem [{\citenamefont {Paterson}\ and\ \citenamefont
  {Fryer}(1988)}]{PatersonFryer1988}%
  \BibitemOpen
  \bibfield  {author} {\bibinfo {author} {\bibfnamefont {W.~R.}\ \bibnamefont
  {Paterson}}\ and\ \bibinfo {author} {\bibfnamefont {P.~J.}\ \bibnamefont
  {Fryer}},\ }\href {https://doi.org/10.1016/0009-2509(88)85166-2} {\bibfield
  {journal} {\bibinfo  {journal} {Chem. Eng. Sci.}\ }\textbf {\bibinfo {volume}
  {43}},\ \bibinfo {pages} {1714} (\bibinfo {year} {1988})}\BibitemShut
  {NoStop}%
\bibitem [{\citenamefont {Rice}(1944)}]{Rice1944}%
  \BibitemOpen
  \bibfield  {author} {\bibinfo {author} {\bibfnamefont {S.~O.}\ \bibnamefont
  {Rice}},\ }\href {https://doi.org/10.1002/j.1538-7305.1944.tb00874.x}
  {\bibfield  {journal} {\bibinfo  {journal} {Bell System Technical Journal}\
  }\textbf {\bibinfo {volume} {23}},\ \bibinfo {pages} {282–332} (\bibinfo
  {year} {1944})}\BibitemShut {NoStop}%
\bibitem [{\citenamefont {Rice}(1945)}]{Rice1945}%
  \BibitemOpen
  \bibfield  {author} {\bibinfo {author} {\bibfnamefont {S.~O.}\ \bibnamefont
  {Rice}},\ }\href {https://doi.org/10.1002/j.1538-7305.1945.tb00453.x}
  {\bibfield  {journal} {\bibinfo  {journal} {Bell System Technical Journal}\
  }\textbf {\bibinfo {volume} {24}},\ \bibinfo {pages} {46–156} (\bibinfo
  {year} {1945})}\BibitemShut {NoStop}%
\bibitem [{\citenamefont {Penrose}(2001)}]{Penrose2001}%
  \BibitemOpen
  \bibfield  {author} {\bibinfo {author} {\bibfnamefont {M.~D.}\ \bibnamefont
  {Penrose}},\ }\href {https://doi.org/10.1007/s002200100387} {\bibfield
  {journal} {\bibinfo  {journal} {Communications in Mathematical Physics}\
  }\textbf {\bibinfo {volume} {218}},\ \bibinfo {pages} {153} (\bibinfo {year}
  {2001})}\BibitemShut {NoStop}%
\bibitem [{\citenamefont {Loscar}\ \emph {et~al.}(2003)\citenamefont {Loscar},
  \citenamefont {Borzi},\ and\ \citenamefont {Albano}}]{Loscar2003}%
  \BibitemOpen
  \bibfield  {author} {\bibinfo {author} {\bibfnamefont {E.~S.}\ \bibnamefont
  {Loscar}}, \bibinfo {author} {\bibfnamefont {R.~A.}\ \bibnamefont {Borzi}},\
  and\ \bibinfo {author} {\bibfnamefont {E.~V.}\ \bibnamefont {Albano}},\
  }\href {https://doi.org/10.1103/physreve.68.041106} {\bibfield  {journal}
  {\bibinfo  {journal} {Physical Review E}\ }\textbf {\bibinfo {volume} {68}},\
  \bibinfo {pages} {041106} (\bibinfo {year} {2003})}\BibitemShut {NoStop}%
\bibitem [{\citenamefont {Nakagawa}\ and\ \citenamefont
  {Nezu}(1978)}]{Nakagawa1978}%
  \BibitemOpen
  \bibfield  {author} {\bibinfo {author} {\bibfnamefont {H.}~\bibnamefont
  {Nakagawa}}\ and\ \bibinfo {author} {\bibfnamefont {I.}~\bibnamefont
  {Nezu}},\ }\href@noop {} {\bibfield  {journal} {\bibinfo  {journal} {Memoirs
  of the Faculty of Engineering, Kyoto University}\ }\textbf {\bibinfo {volume}
  {40}},\ \bibinfo {pages} {213} (\bibinfo {year} {1978})}\BibitemShut
  {NoStop}%
\bibitem [{\citenamefont {Miele}\ \emph {et~al.}(2025)\citenamefont {Miele},
  \citenamefont {Bordoloi}, \citenamefont {Dentz}, \citenamefont {Tabuteau},
  \citenamefont {Morales},\ and\ \citenamefont {de~Anna}}]{Miele2025}%
  \BibitemOpen
  \bibfield  {author} {\bibinfo {author} {\bibfnamefont {F.}~\bibnamefont
  {Miele}}, \bibinfo {author} {\bibfnamefont {A.~D.}\ \bibnamefont {Bordoloi}},
  \bibinfo {author} {\bibfnamefont {M.}~\bibnamefont {Dentz}}, \bibinfo
  {author} {\bibfnamefont {H.}~\bibnamefont {Tabuteau}}, \bibinfo {author}
  {\bibfnamefont {V.~L.}\ \bibnamefont {Morales}},\ and\ \bibinfo {author}
  {\bibfnamefont {P.}~\bibnamefont {de~Anna}},\ }\href@noop {} {\bibfield
  {journal} {\bibinfo  {journal} {arXiv preprint arXiv:2505.02689}\ } (\bibinfo
  {year} {2025})}\BibitemShut {NoStop}%
\bibitem [{\citenamefont {Whitmore}(1995)}]{Whitmore1995}%
  \BibitemOpen
  \bibfield  {author} {\bibinfo {author} {\bibfnamefont {G.~A.}\ \bibnamefont
  {Whitmore}},\ }\href {https://doi.org/10.1007/bf00985762} {\bibfield
  {journal} {\bibinfo  {journal} {Lifetime Data Analysis}\ }\textbf {\bibinfo
  {volume} {1}},\ \bibinfo {pages} {307} (\bibinfo {year} {1995})}\BibitemShut
  {NoStop}%
\bibitem [{\citenamefont {Bordes}\ \emph {et~al.}(2015)\citenamefont {Bordes},
  \citenamefont {Paroissin},\ and\ \citenamefont {Salami}}]{Bordes2015}%
  \BibitemOpen
  \bibfield  {author} {\bibinfo {author} {\bibfnamefont {L.}~\bibnamefont
  {Bordes}}, \bibinfo {author} {\bibfnamefont {C.}~\bibnamefont {Paroissin}},\
  and\ \bibinfo {author} {\bibfnamefont {A.}~\bibnamefont {Salami}},\ }\href
  {https://doi.org/10.1080/03610926.2014.892133} {\bibfield  {journal}
  {\bibinfo  {journal} {Communications in Statistics - Theory and Methods}\
  }\textbf {\bibinfo {volume} {45}},\ \bibinfo {pages} {2730} (\bibinfo {year}
  {2015})}\BibitemShut {NoStop}%
\bibitem [{\citenamefont {Pulcini}(2016)}]{Pulcini2016}%
  \BibitemOpen
  \bibfield  {author} {\bibinfo {author} {\bibfnamefont {G.}~\bibnamefont
  {Pulcini}},\ }\href {https://doi.org/10.1016/j.ress.2016.03.024} {\bibfield
  {journal} {\bibinfo  {journal} {Reliability Engineering \& System Safety}\
  }\textbf {\bibinfo {volume} {152}},\ \bibinfo {pages} {296} (\bibinfo {year}
  {2016})}\BibitemShut {NoStop}%
\bibitem [{\citenamefont {Lanoisel{\'e}e}\ \emph {et~al.}(2018)\citenamefont
  {Lanoisel{\'e}e}, \citenamefont {Grebenkov}, \citenamefont {Sikora},
  \citenamefont {Grzesiek},\ and\ \citenamefont
  {Wy{\l}oma{\'n}ska}}]{Lanoiselee2018}%
  \BibitemOpen
  \bibfield  {author} {\bibinfo {author} {\bibfnamefont {Y.}~\bibnamefont
  {Lanoisel{\'e}e}}, \bibinfo {author} {\bibfnamefont {D.~S.}\ \bibnamefont
  {Grebenkov}}, \bibinfo {author} {\bibfnamefont {G.}~\bibnamefont {Sikora}},
  \bibinfo {author} {\bibfnamefont {A.}~\bibnamefont {Grzesiek}},\ and\
  \bibinfo {author} {\bibfnamefont {A.}~\bibnamefont {Wy{\l}oma{\'n}ska}},\
  }\href@noop {} {\bibfield  {journal} {\bibinfo  {journal} {Physical Review
  E}\ }\textbf {\bibinfo {volume} {98}},\ \bibinfo {pages} {062139} (\bibinfo
  {year} {2018})},\ \bibinfo {note} {arXiv:1808.00412}\BibitemShut {NoStop}%
\bibitem [{\citenamefont {Al~Ismaili}\ \emph {et~al.}(2019)\citenamefont
  {Al~Ismaili}, \citenamefont {Lee}, \citenamefont {Wilson},\ and\
  \citenamefont {Vassiliadis}}]{AlIsmaili2019}%
  \BibitemOpen
  \bibfield  {author} {\bibinfo {author} {\bibfnamefont {R.}~\bibnamefont
  {Al~Ismaili}}, \bibinfo {author} {\bibfnamefont {M.~W.}\ \bibnamefont {Lee}},
  \bibinfo {author} {\bibfnamefont {D.~I.}\ \bibnamefont {Wilson}},\ and\
  \bibinfo {author} {\bibfnamefont {V.~S.}\ \bibnamefont {Vassiliadis}},\
  }\href@noop {} {\bibfield  {journal} {\bibinfo  {journal} {Computers \&
  Chemical Engineering}\ }\textbf {\bibinfo {volume} {121}},\ \bibinfo {pages}
  {409} (\bibinfo {year} {2019})}\BibitemShut {NoStop}%
\bibitem [{\citenamefont {Di~Pretoro}\ \emph {et~al.}(2021)\citenamefont
  {Di~Pretoro}, \citenamefont {D'Iglio},\ and\ \citenamefont
  {Manenti}}]{DiPretoro2021}%
  \BibitemOpen
  \bibfield  {author} {\bibinfo {author} {\bibfnamefont {A.}~\bibnamefont
  {Di~Pretoro}}, \bibinfo {author} {\bibfnamefont {F.}~\bibnamefont
  {D'Iglio}},\ and\ \bibinfo {author} {\bibfnamefont {F.}~\bibnamefont
  {Manenti}},\ }\href {https://doi.org/10.3390/pr9010093} {\bibfield  {journal}
  {\bibinfo  {journal} {Processes}\ }\textbf {\bibinfo {volume} {9}},\ \bibinfo
  {pages} {93} (\bibinfo {year} {2021})}\BibitemShut {NoStop}%
\bibitem [{\citenamefont {Zacks}(2005)}]{Zacks2005}%
  \BibitemOpen
  \bibfield  {author} {\bibinfo {author} {\bibfnamefont {S.}~\bibnamefont
  {Zacks}},\ }\href {https://doi.org/10.1016/j.jspi.2003.05.003} {\bibfield
  {journal} {\bibinfo  {journal} {Journal of Statistical Planning and
  Inference}\ }\textbf {\bibinfo {volume} {130}},\ \bibinfo {pages} {95}
  (\bibinfo {year} {2005})}\BibitemShut {NoStop}%
\bibitem [{\citenamefont {Chi}(2016)}]{Chi2016}%
  \BibitemOpen
  \bibfield  {author} {\bibinfo {author} {\bibfnamefont {Z.}~\bibnamefont
  {Chi}},\ }\href {https://doi.org/10.1016/j.spa.2015.11.001} {\bibfield
  {journal} {\bibinfo  {journal} {Stochastic Processes and their Applications}\
  }\textbf {\bibinfo {volume} {126}},\ \bibinfo {pages} {1124} (\bibinfo {year}
  {2016})}\BibitemShut {NoStop}%
\bibitem [{\citenamefont {Gonz{\'a}lez~C{\'a}zares}\ \emph
  {et~al.}(2025)\citenamefont {Gonz{\'a}lez~C{\'a}zares}, \citenamefont {Lin},\
  and\ \citenamefont {Mijatovi{\'c}}}]{GonzalezCazares2023}%
  \BibitemOpen
  \bibfield  {author} {\bibinfo {author} {\bibfnamefont {J.~I.}\ \bibnamefont
  {Gonz{\'a}lez~C{\'a}zares}}, \bibinfo {author} {\bibfnamefont
  {F.}~\bibnamefont {Lin}},\ and\ \bibinfo {author} {\bibfnamefont
  {A.}~\bibnamefont {Mijatovi{\'c}}},\ }\href
  {https://doi.org/10.1287/stsy.2023.0014} {\bibfield  {journal} {\bibinfo
  {journal} {Stochastic Systems}\ }\textbf {\bibinfo {volume} {15}},\ \bibinfo
  {pages} {50} (\bibinfo {year} {2025})}\BibitemShut {NoStop}%
\bibitem [{\citenamefont {Bolton}\ \emph {et~al.}(2006)\citenamefont {Bolton},
  \citenamefont {Boesch},\ and\ \citenamefont {Lazzara}}]{BoltonBoesch2006}%
  \BibitemOpen
  \bibfield  {author} {\bibinfo {author} {\bibfnamefont {G.~R.}\ \bibnamefont
  {Bolton}}, \bibinfo {author} {\bibfnamefont {A.~W.}\ \bibnamefont {Boesch}},\
  and\ \bibinfo {author} {\bibfnamefont {M.~J.}\ \bibnamefont {Lazzara}},\
  }\href {https://doi.org/10.1016/j.memsci.2005.12.057} {\bibfield  {journal}
  {\bibinfo  {journal} {Journal of Membrane Science}\ }\textbf {\bibinfo
  {volume} {279}},\ \bibinfo {pages} {625} (\bibinfo {year}
  {2006})}\BibitemShut {NoStop}%
\bibitem [{\citenamefont {Cifuentes-Cabezas}\ \emph {et~al.}(2023)\citenamefont
  {Cifuentes-Cabezas}, \citenamefont {Boh{\'o}rquez-Zurita}, \citenamefont
  {Gil-Herrero}, \citenamefont {Vincent-Vela}, \citenamefont {Mendoza-Roca},\
  and\ \citenamefont {{\'A}lvarez-Blanco}}]{CifuentesCabezas2023}%
  \BibitemOpen
  \bibfield  {author} {\bibinfo {author} {\bibfnamefont {M.}~\bibnamefont
  {Cifuentes-Cabezas}}, \bibinfo {author} {\bibfnamefont {J.~L.}\ \bibnamefont
  {Boh{\'o}rquez-Zurita}}, \bibinfo {author} {\bibfnamefont {S.}~\bibnamefont
  {Gil-Herrero}}, \bibinfo {author} {\bibfnamefont {M.~C.}\ \bibnamefont
  {Vincent-Vela}}, \bibinfo {author} {\bibfnamefont {J.~A.}\ \bibnamefont
  {Mendoza-Roca}},\ and\ \bibinfo {author} {\bibfnamefont {S.}~\bibnamefont
  {{\'A}lvarez-Blanco}},\ }\href {https://doi.org/10.1007/s11947-023-03033-0}
  {\bibfield  {journal} {\bibinfo  {journal} {Food and Bioprocess Technology}\
  }\textbf {\bibinfo {volume} {16}},\ \bibinfo {pages} {2126} (\bibinfo {year}
  {2023})}\BibitemShut {NoStop}%
\bibitem [{\citenamefont {Singh}\ \emph {et~al.}(2000)\citenamefont {Singh},
  \citenamefont {Venkatesan}, \citenamefont {Fogler},\ and\ \citenamefont
  {Nagarajan}}]{Singh2000}%
  \BibitemOpen
  \bibfield  {author} {\bibinfo {author} {\bibfnamefont {P.}~\bibnamefont
  {Singh}}, \bibinfo {author} {\bibfnamefont {R.}~\bibnamefont {Venkatesan}},
  \bibinfo {author} {\bibfnamefont {H.~S.}\ \bibnamefont {Fogler}},\ and\
  \bibinfo {author} {\bibfnamefont {N.}~\bibnamefont {Nagarajan}},\ }\href
  {https://doi.org/10.1002/aic.690460517} {\bibfield  {journal} {\bibinfo
  {journal} {AIChE Journal}\ }\textbf {\bibinfo {volume} {46}},\ \bibinfo
  {pages} {1059} (\bibinfo {year} {2000})}\BibitemShut {NoStop}%
\bibitem [{\citenamefont {Mehrotra}\ \emph {et~al.}(2021)\citenamefont
  {Mehrotra}, \citenamefont {Haj-Shafiei},\ and\ \citenamefont
  {Ehsani}}]{Mehrotra2021}%
  \BibitemOpen
  \bibfield  {author} {\bibinfo {author} {\bibfnamefont {A.~K.}\ \bibnamefont
  {Mehrotra}}, \bibinfo {author} {\bibfnamefont {S.}~\bibnamefont
  {Haj-Shafiei}},\ and\ \bibinfo {author} {\bibfnamefont {S.}~\bibnamefont
  {Ehsani}},\ }\href {https://doi.org/10.1016/j.jpse.2021.09.001} {\bibfield
  {journal} {\bibinfo  {journal} {Journal of Pipeline Science and Engineering}\
  }\textbf {\bibinfo {volume} {1}},\ \bibinfo {pages} {428} (\bibinfo {year}
  {2021})}\BibitemShut {NoStop}%
\end{thebibliography}
\end{document}